\documentclass[11pt]{article}
\usepackage{graphicx,psfrag}
\usepackage[utf8]{inputenc}
\usepackage{tabularx}
\usepackage[DIV13]{typearea}
\usepackage{ragged2e}
\usepackage{amsmath,amsfonts,bbm,mathrsfs,amssymb}
\usepackage{slashed,cancel}
\usepackage[usenames,dvipsnames]{xcolor}
\usepackage{cite}
\usepackage{amsmath}
\usepackage{bm} 
\usepackage{soul}
\usepackage{hyperref}
\hypersetup{colorlinks=true,urlcolor=blue,anchorcolor=blue,citecolor=blue,filecolor=blue,
            linkcolor=Magenta,menucolor=blue, linktocpage=true,pdfproducer=medialab}

\usepackage[english]{babel}
\usepackage{catchfile}
\usepackage{indentfirst}
\usepackage{graphicx}
\usepackage{float}
\usepackage{enumerate}
\usepackage{subcaption}
\usepackage{multirow}
\usepackage{physics}
\usepackage{makecell}
\usepackage{subcaption}

\usepackage[normalem]{ulem}
\usepackage{tcolorbox}\usepackage{adjustbox}
\usepackage[compat=1.1.0]{tikz-feynman}

\newcommand{\email}[1]{\href{mailto:#1}{\tt #1}}

\numberwithin{equation}{section}

\definecolor{vierde}{rgb}{0.0, 0.5, 0.0}
\definecolor{OliveGreen}{rgb}{0,0.6,0}

\newcommand{\be}{\begin{equation}}
\newcommand{\ee}{\end{equation}}
\newcommand{\bea}{\begin{eqnarray}}
\newcommand{\eea}{\end{eqnarray}}

\def\ba{\begin{eqnarray}}
\def\ea{\end{eqnarray}}

\def\bt{\bar{t}}
\def\bp{\bar{p}}

\def\bs{\bar{s}}
\def\btheta{\bar{\theta}}
\def\bphi{\bar{\phi}}
\def\bh{\bar{h}}

\begin{document}
\renewcommand*{\thefootnote}{\fnsymbol{footnote}}
\begin{titlepage}

\vspace*{-1cm}

\begin{center}
\bf\LARGE{
Quantum detection of CP violation in the $t\bar{t}$ system: tomography\\[4mm]}
\centering
\vskip .3cm
\end{center}
\vskip 0.5  cm
\begin{center}
{\bf Priyanka Lamba}${}^{a}$~\footnote{\email{priyanka.lamba@uclouvain.be}},
{\bf Fabio Maltoni}${}^{a,b}$~\footnote{\email{fabio.maltoni@unibo.it}},
{\bf Olimpia Miniati}${}^{b}$~\footnote{\email{olimpia.miniati2@unibo.it}}
and
{\bf Eleni Vryonidou}${}^{c,d}$~\footnote{\email{vryonidou.eleni@ucy.ac.cy}}
\vskip .5cm
{\small
\vskip .2cm
${}^{a)}$ Centre for Cosmology, Particle Physics and Phenomenology (CP3), \\Universit\'{e} Catholique de Louvain, B-1348 Louvain-la-Neuve, Belgium\\
${}^{b)}$ Dipartimento di Fisica e Astronomia, Universit\`{a} di Bologna and INFN, Sezione di Bologna,\\ Via Irnerio 46, 40126 Bologna, Italy\\
${}^{c)}$ {Department of Physics and Astronomy, University of Manchester,\\ Oxford Road, Manchester M13~9PL, United Kingdom}\\
${}^{d)}$ {Department of Physics, University of Cyprus, P.O. Box 20537, 1678 Nicosia, Cyprus}

}
\end{center}
\vskip 2cm
\begin{abstract}
We develop a quantum-tomographic framework for determining whether possible
CP-odd effects in $t\bar t$ events originate in production, in decay, or in
both. In the narrow-width approximation, the process
$I\to t\bar t\to b\ell^+\nu\,\bar b\ell^-\bar\nu$ factorises into a
production density matrix and top and antitop decay density matrices. We
extend the standard tomography procedure to a general anomalous $Wtb$ vertex
and derive the corresponding angular distributions. Polar-angle
distributions retain their usual tomographic form, up to modifications of
the spin-analysing powers, and can therefore be used to reconstruct the
production density matrix and test its CP properties. By contrast, dedicated
azimuthal observables involving the $b$--lepton decay planes contain
characteristic sine modulations that provide linear probes of possible new
CP-violating interactions in the decay vertex. Combining the two classes of
observables gives a systematic strategy for separating sources of CP violation in production and in
decay. We illustrate the resulting angular
signatures for representative $t\bar t$ production scenarios at hadron and
lepton colliders.
\end{abstract}
\vskip 3cm

\end{titlepage}
\setcounter{footnote}{0}

\renewcommand*{\thefootnote}{\arabic{footnote}}

\tableofcontents
\newpage 


\section{Introduction}
\label{sec:intro}

The top quark provides direct access to spin physics at colliders. Since it decays before hadronisation can wash out its spin information, the angular distributions of its decay products retain memory of the spin state in which it was produced. In top--antitop production this makes the \(t\bar t\) system an experimentally accessible pair of spin-\(1/2\) particles, whose state can be described by a two-qubit density matrix. The study of top polarisation and spin-correlations has a long history, both theoretically~\cite{Barger:1988jj,Mahlon:1995zn,Stelzer:1995gc,Parke:1996pr,Mahlon:1996pn,Mahlon:1997uc,Brandenburg:2002xr} and experimentally at the Tevatron~\cite{D0:2015kta} and the LHC~\cite{CMS:2013roq,ATLAS:2014aus,CMS:2016piu,ATLAS:2016bac,CMS:2019nrx,ATLAS:2019zrq}. More recently, the observation of entanglement in top-quark pairs by ATLAS and CMS has given this programme a new interpretation in the language of quantum information~\cite{ATLAS:2023fsd,CMS:2024pts,CMS:2024zkc}.

In this language, the spin state of the pair is encoded in a density matrix whose Fano--Bloch decomposition contains the top and antitop polarisation vectors, \(\mathbf B\) and \(\mathbf{\bar{B}}\), and the spin-correlation matrix \(\mathbf C\)~\cite{Fano:1957zz,Fano:1983zz}. These are the same quantities that appear in the conventional spin-correlation formalism, but the density-matrix viewpoint makes it possible to connect collider observables with quantum-information concepts such as entanglement, Bell inequalities, discord, entropy and magic~\cite{2003.02280,Fabbrichesi:2021npl,Severi:2021cnj,Severi:2022qjy,Aoude:2022imd,Afik:2022kwm,Aguilar-Saavedra:2022uye,Afik:2022dgh,Cheng:2023qmz,Han:2023fci,Dong:2023xiw,Cheng:2024btk,Aguilar-Saavedra:2024hwd,Aguilar-Saavedra:2023lwb,DelGratta:2025qyp,Lamba:2026qnk,Afik:2025ejh,A:2026pug,LoChiatto:2024dmx,Aguilar-Saavedra:2024fig,Cheng:2024rxi,Aguilar-Saavedra:2024vpd,Han:2024ugl,Maltoni:2024tul,Maltoni:2024csn,Aoude:2025jzc,Fabbrichesi:2025psr,Altakach:2026fpl,Choi:2026omc,Arai:2026jtc,Fang:2026ddi,Antozzi:2026vdi,Afik:2026pxv,Guo:2026yhz,Aoude:2026eeg}. For the purposes of this work, however, the essential point is more operational: the density matrix is not measured directly. It is reconstructed from decay products, and this reconstruction may itself be affected by new physics in the decay.

This issue is especially relevant in searches for possible new sources of CP violation. In the Standard Model, CP violation originates from the CKM phase and is consistent with current accelerator data, but it is insufficient to explain the observed matter--antimatter asymmetry of the Universe. The top sector is a natural place to search for additional CP-violating interactions, both because of its large coupling to the electroweak symmetry-breaking sector and because several SMEFT operators can modify top production and decay~\cite{Kane:1991bg,Atwood:1991ka,Bernreuther:1992be,Atwood:2000tu,Zhang:2010dr,Aguilar-Saavedra:2008nuh,Gupta:2009wu,Bernreuther:2013aga,Bernreuther:2015yna,Cirigliano:2016nyn,Cirigliano:2016njn,Aguilar-Saavedra:2018ksv,deBeurs:2018pvs,Degrande:2021zpv,Bernreuther:2024ltu}. Quantum-state tomography has also been advocated as a useful framework for probing CP violation in collider processes~\cite{Afik:2022dgh,Altakach:2022ywa,Fabbrichesi:2025igr}.

This paper is the second part of a two-paper study of CP violation in the \(t\bar t\) quantum state. In the companion paper~\cite{Lamba:2026jfm}, the focus is on production. There, the production density matrix is derived for several benchmark mechanisms, including spin-zero decays, \(e^+e^-\) annihilation, \(\gamma\gamma\) collisions, and the partonic channels \(q\bar q\to t\bar t\) and \(gg\to t\bar t\). In the spin basis used in both papers, CP invariance of the production density matrix implies
\[
\Delta\mathbf{B}=0,
\qquad
\mathbf C^A=0,
\]
where $\Delta\mathbf{B}=(\mathbf{B}-\bar{\mathbf B})/2$, i.e., half of the difference of the polarisation vectors and $\mathbf   C^{A}=\left(\mathbf C-\mathbf C^T\right)/2$, i.e., the antisymmetric part of the spin-correlation matrix ~\cite{Lamba:2026jfm}. The companion paper quantifies how several
quantum-information observables respond to these structures at the level of
the produced \(t\bar t\) state.

The question addressed here is how these production-level structures can be
reconstructed experimentally when the top and antitop decay vertices may
themselves contain non-standard interactions. An observed CP-odd angular
dependence can originate in production, in decay, or in both. We formulate
the problem in the narrow-width approximation, in which the process
\begin{equation}
    I\to t\bar t
    \to b\ell^+\nu_\ell\,\bar b\ell^-\bar\nu_\ell ,
\end{equation}
with $I=e^+e^-,\gamma\gamma,pp,\ldots$, factorises into a production density
matrix and top and antitop decay density matrices. This factorised
spin-density-matrix treatment, together with the use of decay products as
spin analysers, follows the general strategy of
Refs.~\cite{Godbole:2006tq,Boudjema:2009fz,Rahaman:2021fcz}.

We first review the standard tomographic reconstruction for SM top decays. At tree level, the charged leptons are optimal spin analysers, and polar-angle distributions give direct access to the Fano--Bloch coefficients, up to the corresponding spin-analysing powers~\cite{Godbole:2006tq,Aguilar-Saavedra:2022uye}. We then extend the analysis to a general anomalous \(Wtb\) vertex. Non-standard decay interactions modify the decay density matrices and can shift the spin-analysing powers~\cite{Aguilar-Saavedra:2006qvv,Zhang:2014rja,Aguilar-Saavedra:2010ljg}. Nevertheless, the polar-angle distributions retain the same tomographic form. Therefore, the usual reconstruction of \(\mathbf{B}\), \(\mathbf{\bar{B}}\), and \(\mathbf C\) remains valid in structure, although the extracted normalisations can be biased if anomalous decay effects are neglected.

Non-standard decay interactions also generate angular structures that are
absent in the SM. In particular, CP-odd contributions to the $Wtb$ vertex
induce sine modulations in relative azimuthal angles associated with the
$b$--lepton decay planes, whereas CP-even contributions generate cosine
modulations. Such azimuthal correlations are established probes of spin and
anomalous decay structures
~\cite{Boudjema:2009fz,Baumgart:2012ay,Fischer:2018lme}. We classify the
corresponding angular distributions according to whether their sensitivity
to new physics in the decay is linear or quadratic, and distinguish them from
distributions whose functional form remains SM-like. In particular, mixed
polar--azimuthal distributions can retain linear sensitivity to CP-odd
decay interactions through the $t\bar t$ spin-correlations even when the
individual top and antitop polarisation vectors vanish.

The outcome is a strategy with two complementary components. Polar-angle tomography reconstructs the $t\bar t$ density matrix and tests  CP violation in the production through $\Delta\mathbf{B}$ and ${\mathbf C}^A$. Dedicated azimuthal observables involving the $b$-lepton decay planes instead diagnose CP violation in the decay vertex. Combining the two makes it possible to determine whether a CP-odd signal originates in production, in decay, or in both. In this sense, the present work supplies the tomographic and decay-level counterpart of the production analysis performed in the companion paper.

The paper is organised as follows. In Sec.~\ref{sec:general} we introduce the factorised spin-density-matrix description of production and decay in the narrow-width approximation. In Sec.~\ref{sec:decay} we derive the top and antitop decay density matrices, first in the SM and then for a general anomalous $Wtb$ vertex. In Sec.~\ref{sec:tomography} we derive the angular distributions used for quantum tomography and discuss how anomalous decays affect the reconstruction of $\mathbf{B}$, $\mathbf{\bar{B}}$, and \(\mathbf C\). In Sec.~\ref{sec:discussion} we classify the distributions according to their sensitivity to CP violation in the decay. In Sec.~\ref{sec:strategy} we summarise the strategy for separating CP-odd effects in production and in decay. We conclude in Sec.~\ref{sec:conclusions}.


\section{Spin-density-matrix factorisation of production and decay} \label{sec:general}

In this section we introduce the general framework used to describe the production of a $t\bar t$ pair and its subsequent decay into a dileptonic final state. 

We consider the production of a $t\bar t$ pair from an initial state $I(a_i)$, where the set of variables $a_i$ denotes the independent parameters needed to characterise the production process. Since the top quarks decay rapidly, almost exclusively through the $Wb$ channel, and the $W$ bosons can in turn decay leptonically, we consider the full process
\begin{equation}
I \to t \bar t \to b\,\ell^+\nu_\ell\,\bar b\,\ell^-\,\bar\nu_\ell \,.
\end{equation}
Our purpose is to separate the production and decay contributions while
retaining their complete spin-correlations. We leave the hard-production
interaction and the two $Wtb$ decay vertices unspecified at this stage,
while assuming the leptonic $W\ell\nu$ interaction to be SM-like.

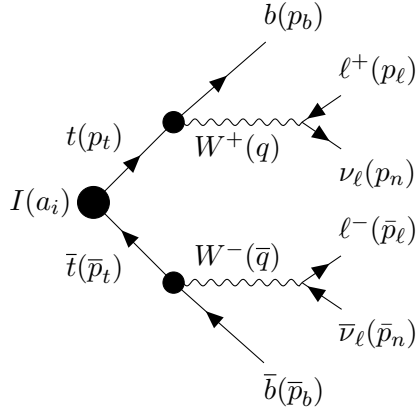
\begin{figure}[t!]
    \centering
    \begin{tikzpicture}
\begin{feynman}[]
\vertex (a) {$I(a_i)$};
\vertex [right=0.7 cm of a] (b);
\vertex [above right=of b] (f1); 
\vertex [below right=of b] (c);
\vertex[right=1.7cm of f1] (wp);
\vertex[right= 1.7cm of c] (wm);
\vertex [above right=0.5 cm of wp] (lp) {$\ell^+(p_\ell)$};
\vertex [below right=0.5 cm of wp] (n) {$\nu_{\ell}(p_n)$};
\vertex [above right= of f1] (bb) {$b(p_b)$};
\vertex [below right= of c] (bbb) {$\overline{b}(\overline{p}_b)$};
\vertex [above right=0.5 cm of wm] (llp) {$\ell^-(\bar{p}_\ell)$};
\vertex [below right=0.5 cm of wm] (nn) {$\overline{\nu}_{\ell}(\bar{p}_n)$};
\diagram* {
(a) -- [scalar, draw=white] (b) -- [fermion, edge label={\(t(p_t)\)}] (f1),
(b) -- [anti fermion, edge label'=\(\overline t( \overline p_t)\)] (c),
(f1) -- [boson, edge label'=\(W^{+}(q)\)](wp),
(c) -- [boson, edge label=\(W^{-}(\overline q)\)](wm), 
(wp) --  [anti fermion] (lp),
(wp) -- [fermion] (n),
(f1) -- [fermion] (bb),
(wm)-- [anti fermion] (nn),
(wm)-- [fermion] (llp),
(c)-- [anti fermion] (bbb),
};
\node[fill=black, draw=black, circle, minimum size=12pt, inner sep=0pt] at (b) {};
\node[fill=black, draw=black, circle, minimum size=8pt, inner sep=0pt] at (f1) {};
\node[fill=black, draw=black, circle, minimum size=8pt, inner sep=0pt] at (c) {};
\end{feynman}
\end{tikzpicture}
\caption{Diagrammatic representation of the process
$I\to t\bar t\to b\,\ell^+\nu_\ell\,\bar b\,\ell^-\bar\nu_\ell$.
All final-state leptons are taken to be massless, while $m_t$, $m_b$ and
$m_W$ denote the top-quark, bottom-quark and $W$-boson masses. In our
notation, barred four-momenta $\bar p_i$ are assigned to the decay products
of $\bar t$, while unbarred four-momenta $p_i$ are assigned to the decay
products of $t$.}
\label{fig:I-lbnlbn}
\end{figure}

With the momentum assignment of Fig.~\ref{fig:I-lbnlbn}, and assuming massless final-state leptons, the corresponding amplitude can be written as
\be
        \mathcal{M}= \frac{ig_{\alpha\beta}g_{\rho\sigma}}{\Delta_{t}(p_t)\Delta_{\bar{t}}(\bar{p}_t)\Delta_{W^+}(q)\Delta_{W^-}(\bar{q})}\bar{u}(p_n)V^\alpha_{L}v(p_\ell)\bar{u}(p_b)V^\beta_D(\slashed{p}_t+m_{t})V_{P}(\slashed{\bar{p}}_t-m_{t})V^\rho_{\bar{D}}v(\bar{p}_b)\bar{u}(\bar{p}_\ell)V^\sigma_Lv(\overline{p}_n), 
        \label{eq:matrix-element-full-process}
\ee
where $V_P$ denotes the Lorentz structure associated with the production subprocess.\footnote{In this notation we suppress the polarisation vectors, colour structure, spinors of the initial-state particles, and the propagators of possible intermediate virtual particles. Also note that for massless leptons, the longitudinal terms in the $W$-boson propagator
do not contribute after contraction with the leptonic currents; only the
terms proportional to $g_{\mu\nu}$ are therefore displayed.} The quantities $V_D^\mu$ and $V_{\bar D}^\mu$ denote the $Wtb$ decay vertices for the $t$ and $\bar t$ branches, respectively. The $W\ell\nu$ interaction is assumed to be SM-like and is denoted by $V_L^\mu$. The factors $\Delta_i$ are the denominators of the propagators of the unstable massive particles.

Working in the narrow-width approximation, the intermediate top and
antitop quarks are projected on shell,
$p_t^2=\bar p_t^{\,2}=m_t^2$. Their propagator numerators can therefore be
resolved into complete sets of helicity eigenstates,
\begin{align}
(\slashed p_t+m_t)
&=
\sum_{\lambda}
u(p_t,\lambda)\bar u(p_t,\lambda),
&
(\slashed{\bar p}_t-m_t)
&=
\sum_{\mu}
v(\bar p_t,\mu)\bar v(\bar p_t,\mu).
\end{align}
The short top-quark lifetime ensures that the spin information encoded at
production is transmitted to the decay products before hadronisation.

The full amplitude can then be written as
\begin{equation}
\mathcal M
=
\frac{1}{
\Delta_t(p_t)\Delta_{\bar t}(\bar p_t)
}
\sum_{\lambda,\mu}
\mathcal M^P(\lambda,\mu)\,
\mathcal M^{D,t}(\lambda)\,
\mathcal M^{D,\bar t}(\mu),
\label{eq:amplitude-factorisation}
\end{equation}
where the decay amplitudes include the corresponding $W$-boson
propagators.

An analogous decomposition can be performed for the hermitian-conjugate amplitude $\mathcal{M}^\dagger$, for which the helicity labels will be denoted by $\lambda'$ and $\mu'$.
It is convenient to express the squared matrix element for the process shown in Fig.~\ref{fig:I-lbnlbn} in terms of a spin production matrix $R^I_{\lambda\lambda',\mu\mu'}$, associated with $I(a_i)\to t\bar t$, and decay matrices $D^t_{\lambda\lambda'}$ and $D^{\bar t}_{\mu\mu'}$, associated with the decays $t\to b\,\ell^+\nu_\ell$ and $\bar t\to \bar b\,\ell^-\bar\nu_\ell$. They are defined as
\begin{align}
R^I_{\lambda\lambda',\mu\mu'} &=
\frac{1}{N_I}
\sum_{I\textnormal{-d.o.f.}}
\mathcal{M}^P(\lambda\mu)
\mathcal{M}^{P\dagger}(\lambda'\mu') ,
\\
D^t_{\lambda\lambda'} &=
\sum_{F\textnormal{-d.o.f.}}
\mathcal{M}^{D,t}(\lambda)
\mathcal{M}^{D,t\dagger}(\lambda') ,
\\
D^{\bar t}_{\mu\mu'} &=
\sum_{F\textnormal{-d.o.f.}}
\mathcal{M}^{D,\bar t}(\mu)
\mathcal{M}^{D,\bar t\dagger}(\mu') .
\end{align}
Here $N_I$ is the product of the initial-state spin, colour and any other
averaging factors, while the sums run over the relevant unobserved degrees of freedom of the initial and final states. The normalised production density matrix is defined as
\begin{equation}
\rho^I=\frac{R^I}{\Tr[R^I]} ,
\end{equation}
and describes the $t\bar t$ system as a bipartite two-qubit state. Analogously, the normalised decay density matrices are
\begin{equation}
\Gamma^{t}=\frac{D^{t}}{\Tr[D^{t}]} ,
\qquad
\Gamma^{\bar t}=\frac{D^{\bar t}}{\Tr[D^{\bar t}]} .
\label{eq:normalised-matrices}
\end{equation}
With these definitions, the squared matrix element can be written as
\begin{equation}
|\mathcal{M}|^2
=
\sum_{\lambda\lambda'\mu\mu'}
\frac{
R^{I}_{\lambda\lambda',\mu\mu'}
D^{t}_{\lambda\lambda'}
D^{\bar t}_{\mu\mu'}
}
{|\Delta_t|^2|\Delta_{\bar t}|^2}
=
\frac{
\Tr[R^I]\Tr[D^{t}]\Tr[D^{\bar{t} }]
}
{|\Delta_t|^2|\Delta_{\bar t}|^2}
\Tr\!\left[
\rho^I
[(\Gamma^{t})^T\otimes (\Gamma^{\bar t})^T]
\right] .
\label{eq:squared-M-density-matrices}
\end{equation}
Upon adopting the narrow-width approximation for the two top quarks, the differential cross section for
$I\to b\,\ell^+\nu_\ell\,\bar b\,\ell^-\bar\nu_\ell$
factorises into production and decay contributions. This allows the phase-space integrations associated with production and with the two decay branches to be performed independently, and in convenient reference frames~\cite{Boudjema:2009fz,Godbole:2006tq,Rahaman:2021fcz}. This factorised structure is the basis of the quantum-tomography procedure discussed in Sec.~\ref{sec:tomography}.

We also note that the traces of the production and decay matrices are related to the corresponding unpolarised squared amplitudes as
\begin{equation}
\Tr[R^I]=\overline{|\mathcal{M}^P|^2},
\qquad
\Tr[D^{t(\bar t)}]=(2s+1)\overline{|{\mathcal{M}}^{D,t(\bar t)}|^2} .
\end{equation}
For both $t$ and $\bar t$, $s=1/2$.\\


\section{Top decays}
\label{sec:decay}

In this section we derive the decay density matrices that enter the factorised expression of the full process. We first introduce the spinor projectors needed to retain the off-diagonal spin information of the intermediate top quarks. We then apply this formalism to the SM decay and to the SMEFT corrections to the $Wtb$ vertex induced by the operators $O_{\phi Q}^{3}, O_{\phi tb}, O_{bW}$, and $O_{tW}$:
\begin{align}
    O^{3}_{\phi Q}&=  \left( \phi^\dagger \sigma^I \overleftrightarrow{iD}_\mu \phi \right)
\left( \overline{Q}_{L} \gamma^\mu \sigma^I Q_{L} \right)\,,\label{eq:O3fQ} \\
^\ddagger O_{\phi tb} &=  i(\widetilde{\phi}^\dagger D_\mu \phi)(\overline{t}_{R} \gamma^\mu b_R)\,,\\
^\ddagger O_{bW}&=(\overline{Q}_L \sigma^{\mu\nu} b_R)\sigma^I \phi W^I_{\mu\nu\,},\\
^\ddagger O_{tW}&=(\overline{Q}_L \sigma^{\mu\nu} t_R)\sigma^I \widetilde{\phi}W^I_{\mu\nu}.\label{eq:Otw}
\end{align}
The operators $^\ddagger O_{\phi tb}$, $^\ddagger O_{bW}$, and $^\ddagger O_{tW}$ are non-Hermitian
and enter the SMEFT Lagrangian together with their Hermitian conjugates;
their Wilson coefficients can therefore be complex. By contrast,
$O_{\phi Q}^{3}$ is Hermitian and its Wilson coefficient is real. 
The conventions adopted here follow those of the companion
paper~\cite{Lamba:2026jfm}, where the operators affecting the dominant
$t\bar t$ production mechanisms are discussed. Here we focus on the
subset that modifies the $Wtb$ decay vertex. The resulting
production and decay density matrices determine how the spin information
encoded at the production stage is transferred to the angular distributions
of the final-state particles.

For an on-shell massive spin-$1/2$ particle with four-momentum
$p^\mu=(E,\vec p)$ and mass $m$, the helicity spin four-vector is
\begin{equation}
    s^\mu =
    \lambda
    \left(
    \frac{|\vec p|}{m},
    \frac{E}{m}\hat p
    \right),
    \qquad
    \hat p=\frac{\vec p}{|\vec p|},
    \qquad
    \lambda=\pm 1 .
    \label{eq:spin-vector}
\end{equation}
In the rest frame of the particle this becomes $s^\mu=(0,\lambda\hat p)$, so that the spatial components identify the spin direction. The usual diagonal helicity projectors determine the diagonal spin
components, but they do not retain the off-diagonal elements required for
the full decay density matrix. These off-diagonal components are essential in the production and decay of unstable particles, where spin-correlations between the two stages affect the angular distributions of the decay products.

To keep the full spin information we use the Bouchiat--Michel formulae~\cite{Bouchiat:1958yui,Haber:1994pe}. They are written in terms of three spin four-vectors $s^{a\mu}$, with $a=1,2,3$, satisfying
\begin{equation}
     p\cdot s^a=0,
     \qquad
     s^a\cdot s^b=-\delta^{ab},
     \qquad
     \sum_{a=1}^3 s^a_\mu s^a_\nu
     =
     -g_{\mu\nu}+\frac{p_\mu p_\nu}{m^2}.
\end{equation}
Thus $\{s^1,s^2,s^3,p/m\}$ form an orthonormal tetrad. For helicity amplitudes we choose $s^3$ along the direction of motion of the particle, while $s^1$ and $s^2$ span the transverse plane. In a frame where the spatial unit vectors $\{\hat \imath,\hat \jmath,\hat p\}$ form a right-handed orthonormal basis, one may take
\begin{equation}
    s^{1\mu}=(0,\hat \imath),
    \qquad
    s^{2\mu}=(0,\hat \jmath),
    \qquad
    s^{3\mu}=
    \left(
    \frac{|\vec p|}{m},
    \frac{E}{m}\hat p
    \right).
\end{equation}
The Bouchiat--Michel formulae then read
\begin{align}
u(p,\lambda)\bar u(p,\lambda')
&=
\frac{1}{2}
\left(
\delta_{\lambda'\lambda}
+
\gamma^5 \slashed{s}^a \sigma^a_{\lambda'\lambda}
\right)
(\slashed p+m),
\label{eq:bm-u}
\\
v(p,\lambda')\bar v(p,\lambda)
&=
\frac{1}{2}
\left(
\delta_{\lambda'\lambda}
+
\gamma^5 \slashed{s}^a \sigma^a_{\lambda'\lambda}
\right)
(\slashed p-m),
\label{eq:bm-v}
\end{align}
where $\sigma^a$ are the Pauli matrices and a sum over $a=1,2,3$ is understood. With this formalism, the decay matrix of a top quark can always be written as
\begin{equation}
D^{t({\rm BM})}_{\lambda\lambda'}
=
E_{\rm BM}\,\delta_{\lambda'\lambda}
+
\sigma^a_{\lambda'\lambda}\,s^a_\alpha F^\alpha_{\rm BM},
\label{eq:D_decay_t}
\end{equation}
where $E_{\rm BM}$ and $F^\alpha_{\rm BM}$ depend on the momenta and on the decay interaction. After normalisation, the transpose of the decay density matrix takes the standard form\footnote{We give the structure of the transpose of the decay density matrix, appearing explicitly in Eq.~\eqref{eq:squared-M-density-matrices}, where the indices are exchanged on the left-hand side compared to Eq.~\eqref{eq:D_decay_t}, allowing the compact matrix-form expression. }
\begin{equation}
(\Gamma^{t({\rm BM})})^T
=
\frac{1}{2}
\left[
\mathbb{I}
+
\pi_i\sigma^i
\right],
\qquad
\pi_i=
\frac{s^i_\alpha F^\alpha_{\rm BM}}{E_{\rm BM}} .
\label{eq:gamma-top-BM}
\end{equation}
The components $\pi_i$ form the spin-analyser vector associated with the chosen final-state variables. The antitop decay is treated analogously,
\begin{equation}
(\Gamma^{\bar t({\rm BM})})^T
=
\frac{1}{2}
\left[
\mathbb{I}
+
\bar\pi_i\sigma^i
\right],
\qquad
\bar\pi_i=
\frac{\bar s^i_\alpha \bar F^{\alpha}_{\rm BM}}{\bar E_{\rm BM}} .
\label{eq:gamma-antitop-BM}
\end{equation}

For the $t\bar t$ system we use a common spin basis, defined in the zero-momentum frame (ZMF) of the pair, see Ref.~\cite{Lamba:2026jfm}. The spin projections of the top and antitop are therefore referred to the same spatial axes, rather than to two independent helicity axes. With our conventions this amounts to
\begin{equation}
\Gamma^{t({\rm BM})}=\Gamma^t,
\qquad
\Gamma^{\bar t}=\sigma^1\Gamma^{\bar t({\rm BM})}\sigma^1 .
\end{equation}
It is useful to encode this convention through the sign vectors
\begin{equation}
    h=(1,1,1),
    \qquad
    \bar{h}=(1,-1,-1),
    \label{eq:h-vectors}
\end{equation}
so that the decay density matrices used in the $t\bar t$ spin basis are
\begin{align}
(\Gamma^t)^T
&=
\frac{1}{2}
\left[
\mathbb{I}
+
h_i\pi_i\sigma^i
\right],
&
(\Gamma^{\bar t})^T
&=
\frac{1}{2}
\left[
\mathbb{I}
+
\bar h_i\bar\pi_i\sigma^i
\right].
\label{eq:gamma-spin-basis}
\end{align}
Since $h_i=1$ for all components, we suppress these factors below, while
retaining the non-trivial antitop factors $\bar h_i$ explicitly.
\subsection{Standard Model}
\label{sec:top_decay_sm}

Using the momentum assignment of Fig.~\ref{fig:I-lbnlbn}, the SM decay matrix for the top quark can be computed directly from the Bouchiat--Michel formulae. For the decay
$t\to b\ell^+\nu_\ell$ one obtains~\footnote{For simplicity, in this work we take $V_{tb}=1$.}
\begin{equation}
D^{t}
=
\tilde A_D
\begin{pmatrix}
    1-\dfrac{m_t(p_\ell\cdot s^3)}{p_\ell\cdot p_t}
    &
    -\dfrac{m_t\,p_\ell\cdot(s^1+i s^2)}{p_\ell\cdot p_t}
    \\[3mm]
    -\dfrac{m_t\,p_\ell\cdot(s^1-i s^2)}{p_\ell\cdot p_t}
    &
    1+\dfrac{m_t(p_\ell\cdot s^3)}{p_\ell\cdot p_t}
\end{pmatrix},
\label{eq:decay-matrix-t-SM-before-frame}
\end{equation}
with
\begin{equation}
\tilde A_D
=2g^4 \left(
\frac{
p_\ell\cdot p_t  \, p_n\cdot p_b
}
{|\Delta_W|^2}\right).
\end{equation}
The transpose of the corresponding normalised density matrix can be written as
\begin{equation}
(\Gamma^t)^T
=
\frac{1}{2}
\left[
\mathbb{I}
+
f_{\rm SM}\,
p_\ell^\mu s^i_\mu \sigma^i
\right],
\qquad
f_{\rm SM}
=
-\frac{m_t}{p_t\cdot p_\ell}.
\label{eq:gamma-t-SM-covariant}
\end{equation}
In the rest frame of the decaying top quark this expression takes the familiar analyser form
\begin{equation}
(\Gamma^t)^T
=
\frac{1}{2}
\left[
\mathbb{I}
+
\alpha_\ell^{ \rm SM}\,
\hat p_\ell^{\,t{\rm -rest}}\cdot \hat s^i\,\sigma^i
\right],
\label{eq:gamma-t-SM}
\end{equation}
where $\hat p_\ell^{\,t{\rm -rest}}$ is the direction of the charged lepton in the top rest frame, and $\hat s^i$ are the spatial axes of the chosen spin basis boosted to that frame. At tree level in the SM the charged lepton is an optimal spin analyser,
\begin{equation}
\alpha_\ell^{ \rm SM}=1,
\end{equation}
in agreement with the standard results~\cite{Godbole:2006tq,Boudjema:2009fz,Aguilar-Saavedra:2022uye}. The antitop decay is obtained analogously. After expressing the antitop spin in the same basis used for the $t\bar t$ system, one finds
\begin{equation}
(\Gamma^{\bar t})^T
=
\frac{1}{2}
\left[
\mathbb{I}
+
\bar f_{\rm SM}\,
\bar h_i\,\bar p_\ell^\mu \bar s^i_\mu \sigma^i
\right]
=
\frac{1}{2}
\left[
\mathbb{I}
+
\bar\alpha_\ell^{\rm SM}\,
\hat{\bar p}_\ell^{\,\bar t{\rm -rest}}\cdot \hat s^i\,\sigma^i
\right],
\label{eq:gamma-tbar-SM}
\end{equation}
with
\begin{equation}
\bar f_{\rm SM}
=
\frac{m_t}{\bar p_t\cdot \bar p_\ell},
\qquad
\bar\alpha_\ell^{ \rm SM}=-1\,,
\end{equation}
at tree level in the SM. The opposite sign of $\bar\alpha_\ell$ follows from the charge conjugation structure of the top and antitop decay density matrices.
\subsection{Beyond the Standard Model} \label{sec:top_decay_bsm}

Employing the operators in Eqs.~\eqref{eq:O3fQ}--\eqref{eq:Otw}, the
dimension-six SMEFT $Wtb$ vertex relevant for our analysis can be written
as
\begin{equation}
V^\mu_D
=
\frac{ig}{\sqrt{2}}
\gamma^\mu
\left[
\left(1+\frac{v^2 C^{3}_{\phi Q}}{\Lambda^2}\right) P_L+ \frac{v^2 C^*_{\phi tb}}{2\Lambda^2} P_R
\right]
+
\frac{2v}{\Lambda^2}
\sigma^{\mu\alpha}
\left[
C_{tW}P_R+C^*_{bW}P_L
\right]q_\alpha \,,
\label{eq:vertex-BSM-decay-top}
\end{equation}
which comprises the SM left-handed interaction, SMEFT-induced
modifications of the left- and right-handed vector couplings, and
momentum-dependent dipole contributions ($q=p_t-p_b=p_\ell+p_n$ is the momentum carried by the outgoing
$W^+$ boson). The $\bar tW^-\bar b$ vertex is obtained from
Eq.~\eqref{eq:vertex-BSM-decay-top} by Hermitian conjugation, with all
Wilson coefficients complex conjugated and with the corresponding
charge-conjugate momentum convention.
The SM limit is recovered when all Wilson coefficients vanish. The coefficient $C^3_{\phi Q}$ is taken to be real, while the remaining coefficients can in general be complex. In the absence of absorptive phases, imaginary parts of the Wilson
coefficients, or CP-odd relative phases among them, can generate
CP-violating contributions to the decay density matrices. More generally,
the Levi--Civita structures are naive-$T$ odd; their interpretation as
genuine CP-odd effects follows from the comparison of the charge-conjugate
top and antitop decays.

With the effective vertex in Eq.~\eqref{eq:vertex-BSM-decay-top}, the transpose of the unnormalised decay matrices can be written in the form
\begin{align}
(D^t)^T
&=
T\,\mathbb{I}
+
s^i_\mu P^\mu \sigma^i ,
\label{eq:decay-matrix-t-BSM}
\\
(D^{\bar t})^T
&=
\bar T\,\mathbb{I}
+
\bar h_i\,\bar s^i_\mu \bar P^\mu \sigma^i .
\label{eq:decay-matrix-tbar-BSM}
\end{align}
In the computation of these two matrices, we retain all terms up to linear
order in the bottom-quark mass \(m_b\), while neglecting contributions of
\(\mathcal{O}(m_b^2)\) and higher. After imposing momentum conservation, the
vectors \(P^\mu\) and \(\bar P^\mu\) can be decomposed as\footnote{A possible term proportional to $p_t^\mu$ has been omitted because it
does not contribute to the decay matrix:
$s^i_\mu p_t^\mu=0$. The corresponding statement holds for the antitop
branch.}
\begin{align}
P^\mu
&=
f_\ell\,p_\ell^\mu
+
f_b\,p_b^\mu
+
f^{\rm CPV}_{\epsilon}\,
\epsilon^{\alpha\beta\gamma\mu}
p_{t\alpha}p_{\ell\beta}p_{b\gamma},
\label{eq:P-BSM-decay-t}
\\
\bar P^\mu
&=
\bar f_\ell\,\bar p_\ell^\mu
+
\bar f_b\,\bar p_b^\mu
+
\bar f^{\rm CPV}_{\epsilon}\,
\epsilon^{\alpha\beta\gamma\mu}
\bar p_{t\alpha}\bar p_{\ell\beta}\bar p_{b\gamma}.
\label{eq:P-BSM-decay-tbar}
\end{align}
The scalar functions $T$, $\bar T$, $f_i$, and $\bar f_i$ depend on the
decay kinematics and on the effective couplings. The terms proportional to
$f_\ell$ and $f_b$ arise from scalar products involving the top-spin
four-vector, whereas those proportional to
$f^{\rm CPV}_{\epsilon}$ and $\bar f^{\rm CPV}_{\epsilon}$ originate from
Levi-Civita contractions. These Levi-Civita terms give rise to the CP-odd
azimuthal structures discussed in Sec.~\ref{sec:discussion}.
The inclusion of SMEFT contributions in the \(Wtb\) decay vertex leads to a
richer decay density matrix structure than in the SM. Before performing any
partial integration over the decay phase space, the normalised decay density
matrix cannot, in general, be reduced to a form analogous to that of Eq.~\eqref{eq:gamma-t-SM}%
~\cite{Baumgart:2012ay}.

\begin{figure}[t]
    \centering
\includegraphics[width=0.60\linewidth]{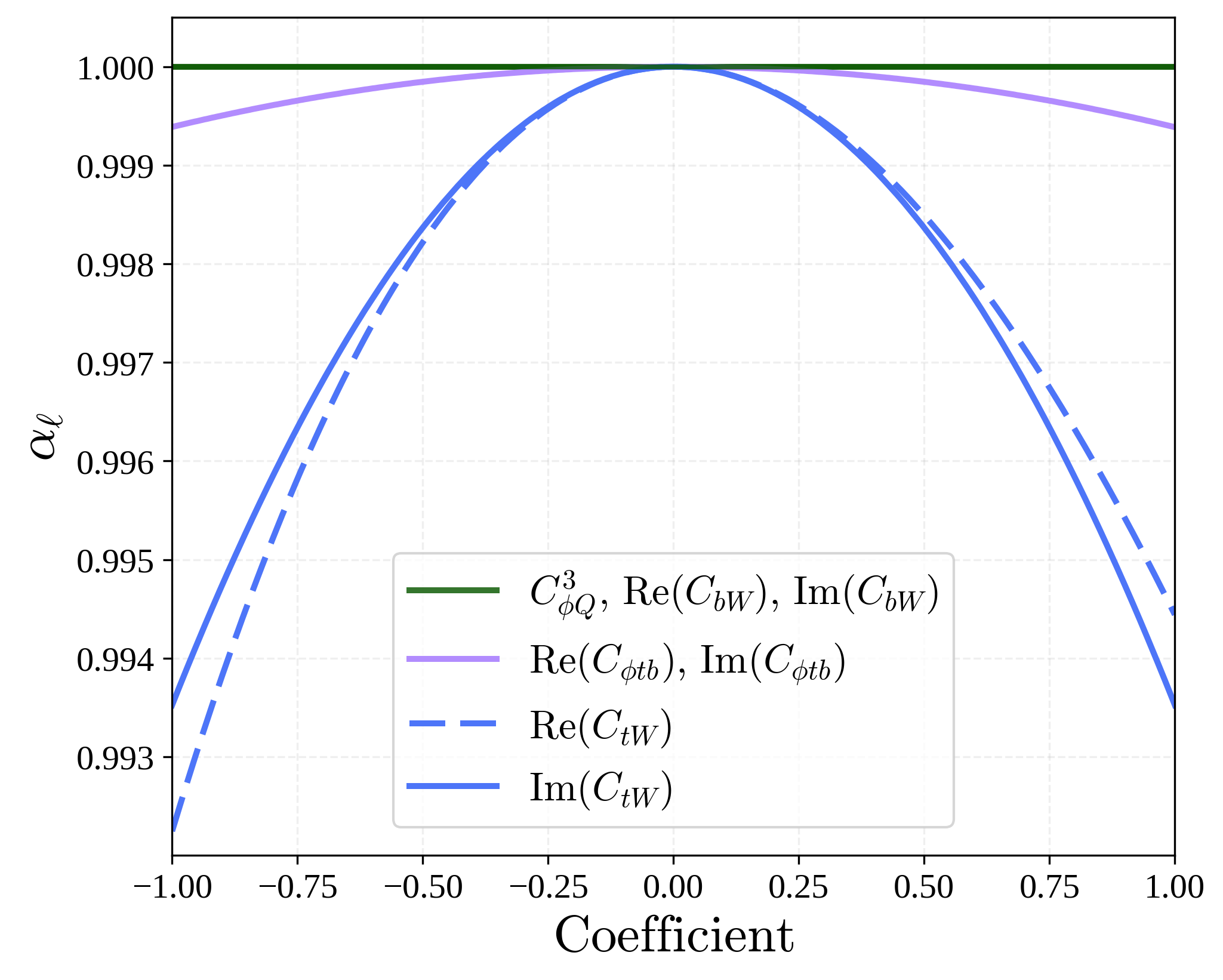}
    \caption{Dependence of the charged-lepton spin-analysing power
\(\alpha_\ell\) on the Wilson coefficients entering
Eq.~\eqref{eq:vertex-BSM-decay-top}, varied one at a time and taken to be
either purely real or purely imaginary, for
\(\Lambda=1~\mathrm{TeV}\). Within the displayed range, the predictions for
\(C_{\phi Q}^{3}\), \(\operatorname{Re}(C_{bW})\), and
\(\operatorname{Im}(C_{bW})\) coincide and are represented by a single green
curve. Likewise, the predictions for
\(\operatorname{Re}(C_{\phi tb})\) and
\(\operatorname{Im}(C_{\phi tb})\) are numerically indistinguishable and are
represented by a single purple curve.}
    \label{fig:sap}
\end{figure}

The modified interaction can also induce deviations
of the charged lepton spin-analysing power $\alpha_\ell$ from its SM value.
Indeed, the tree-level result $\alpha_\ell^{\rm SM}=1$ follows directly from
the purely left-handed $V-A$ structure of the charged-current interaction
and can be modified by the presence of additional Lorentz structures. The charged-lepton spin-analysing power modified by the \(Wtb\) vertex in
Eq.~\eqref{eq:vertex-BSM-decay-top} can be written as $\alpha_\ell = N(\alpha_\ell)/D(\alpha_\ell)$: the exact expressions for \(N(\alpha_\ell)\) and \(D(\alpha_\ell)\), including
the terms proportional to the bottom-quark mass \(m_b\), are given in
App.~\ref{app:alpha_analytical}. The same appendix also provides the expansion
of \(\alpha_\ell\) through \(\mathcal{O}(\Lambda^{-6})\), with terms of
\(\mathcal{O}(\Lambda^{-8})\) and higher neglected in Eq.~\eqref{eq:sap_expanded}. The effects obtained by
switching on the individual Wilson coefficients separately are shown in
Fig.~\ref{fig:sap}.

As shown explicitly in Eq.~\eqref{eq:sap_expanded}, the
\(\mathcal{O}(\Lambda^{-2})\) contributions cancel between the numerator
and denominator, so that the leading deviation from the SM prediction
arises at \(\mathcal{O}(\Lambda^{-4})\). The expansion of the normalised ratio also generates formal
$\mathcal O(\Lambda^{-6})$ terms containing products of three Wilson
coefficients. Nevertheless, all terms entering \(\alpha_\ell\) are CP-even
combinations. In particular, the imaginary parts of the Wilson coefficients
appear only in combinations containing an even number of CP-odd factors,
such as
\(\operatorname{Im}(C_i)\operatorname{Im}(C_j)\), \(\lvert C_i\rvert^2\),
\(\operatorname{Re}(C_k)\operatorname{Im}(C_i)\operatorname{Im}(C_j)\), or
\(\lvert C_i\rvert^2\operatorname{Re}(C_j)\).
Within the tree-level narrow-width treatment adopted here, and after
integration over the remaining decay variables defining the inclusive
spin-analysing power, the $\mathcal O(\Lambda^{-2})$ corrections cancel. Consequently, \(\alpha_\ell\) has no linear sensitivity to the Wilson coefficients that modify the \(Wtb\) vertex in Eq.~\eqref{eq:vertex-BSM-decay-top} and therefore does not directly probe CP violation in the decay.
In Fig.~\ref{fig:sap}, the Wilson coefficients are varied one at a time over
the interval \([-1,1]\), with \(\Lambda=1~\mathrm{TeV}\). For complex
coefficients, the real and imaginary parts are varied separately.

As can be seen from Eq.~\eqref{eq:sap_expanded}, the first contributions
involving \(C_{\phi Q}^{3}\) and \(C_{bW}\) arise at
\(\mathcal{O}(\Lambda^{-6})\). These terms, however, involve products of
Wilson coefficients associated with different operators and therefore
vanish in the one-coefficient-at-a-time scenario considered in the figure.
Consequently, \(C_{\phi Q}^{3}\), \(\operatorname{Re}(C_{bW})\), and
\(\operatorname{Im}(C_{bW})\) leave \(\alpha_\ell\) unchanged and are represented
by the same green curve. In particular, \(O_{\phi Q}^{3}\) only rescales
the SM left-handed \(Wtb\) coupling without modifying its \(V-A\) structure,
while the spin-analysing power is also unaffected by \(O_{bW}\) when this
operator is switched on individually.

The result shown in Fig.~\ref{fig:sap} is unchanged in the massless-bottom
limit. Indeed, the terms proportional to \(m_b\) in
Eq.~\eqref{eq:sap_expanded} are nonvanishing only when Wilson coefficients
from at least two different operators are simultaneously present, and hence
do not contribute in the single-coefficient scans considered here.

Within the displayed interval, the predictions obtained by varying
\(\operatorname{Re}(C_{\phi tb})\) and \(\operatorname{Im}(C_{\phi tb})\) are
numerically indistinguishable. The two behaviors begin to separate only for
substantially larger values, \(\lvert C_{\phi tb}\rvert\gtrsim 10\), well
outside the range shown. Among the operators considered, the strongest
variation of \(\alpha_\ell\) is induced by \(C_{tW}\).  The dependence on $\operatorname{Im}(C_{tW})$ is purely quadratic, as it is governed by the $\mathcal{O}(\Lambda^{-4})$ contribution. In contrast, for $\operatorname{Re}(C_{tW})$, both the $\mathcal{O}(\Lambda^{-4})$ and $\mathcal{O}(\Lambda^{-6})$ contributions are relevant in the one-coefficient-at-a-time scenario. The latter is numerically comparable and therefore causes the blue dashed curve to deviate slightly from an exact quadratic behavior. Although the \(C_{tW}\)
contribution can produce sizable departures from the SM value for
sufficiently large coefficient values, the variation remains below the
percent level throughout \(C_{tW}\in[-1,1]\). Since the experimentally
allowed range is even narrower, the corresponding phenomenologically
viable correction is expected to be smaller still. This behavior is
consistent with the absence of \(\mathcal{O}(\Lambda^{-2})\) corrections to
\(\alpha_\ell\), whose leading deviation from the SM value arises at
\(\mathcal{O}(\Lambda^{-4})\).
Finally, the tree-level relation $\bar{\alpha}_\ell=-\alpha_\ell$ continues to hold in the presence of the SMEFT contributions considered here.

The decay density matrix in Eq.~\eqref{eq:decay-matrix-t-BSM}, and
analogously that of the antitop quark, retains an explicit dependence on
the top-spin four-vector. This spin dependence determines the charged-lepton
spin-analysing power \(\alpha_\ell\) and provides sensitivity to both new CP-conserving and 
CP-violating structures in the \(Wtb\) vertex. The CP-violating contributions arise
from terms involving the Levi-Civita tensor. In a three-body decay,
momentum conservation renders the parent and final-state four-momenta
linearly dependent, so that any pseudoscalar constructed solely from these
momenta vanishes. A non-vanishing CP-odd structure must therefore also
involve the top spin four-vector, for example through
\begin{equation}
    \epsilon(p_t,p_\ell,p_b,s^i).
\end{equation}

Accessing these spin-dependent decay structures requires the spin
information of the decaying top quark to be retained in the production
process. This can occur in two distinct ways: through a non-zero polarisation
of an individually produced top quark, or through spin-correlations in
\(t\bar t\) production. We first illustrate the former case. The production
density matrix of a single top quark can be written as
\begin{equation}
    \rho_t
    =
    \frac{1}{2}
    \left[
        \mathbb{I}
        +
        \sum_j B_j\sigma^j
    \right],
\end{equation}
where \(B_j=\Tr[\rho_t\sigma_j]\) are the components of the top-quark
polarisation vector. If the top quark is produced together with additional
particles, \(\rho_t\) denotes the reduced density matrix obtained after
tracing over their degrees of freedom. Combining the production and decay
density matrices gives schematically
\begin{equation}
    d\sigma_t
    \sim
    \Tr\left[\rho_t(D^t)^T\right]
    =
    T+\sum_i B_i\,s^i_\mu P^\mu .
\end{equation}
Hence, in an inclusive single-top angular distribution, the spin-dependent
terms proportional to \(f_\ell\), \(f_b\), and
\(f_\epsilon^{\rm CPV}\) in Eq.~\eqref{eq:P-BSM-decay-t} contribute only
if the produced top quark has a non-zero polarisation component. This
condition is naturally realised in single-top production, where
CP-violating effects can therefore be probed using the decay of a polarised
top quark~\cite{deBeurs:2018pvs}.

The situation is different in \(t\bar t\) production. In the SM at the LHC,
the individual top- and antitop-quark polarisations are generated only by
electroweak corrections and are therefore small,
\(\mathcal{O}(10^{-3})\)--\(\mathcal{O}(10^{-2})\)~\cite{Bernreuther:2006vg}. Nevertheless, the top and antitop spins are
correlated. These correlations preserve access to the spin-dependent decay
structures even when the individual polarisation vectors vanish, because
the spin of one particle can be analysed relative to that of the other.
Consequently, the terms proportional to \(f_\ell\), \(f_b\), and
\(f_\epsilon^{\rm CPV}\) can contribute to the joint angular distributions
through their contraction with the \(t\bar t\) spin-correlation matrix.
This mechanism is derived explicitly in Sec.~\ref{sec:tomography}.


\section{Quantum tomography} \label{sec:tomography}

To access the information encoded in the composite state of the $t\bar t$ pair, one must determine the value of each independent coefficient appearing in the  Fano--Bloch decomposition \cite{Fano:1983zz}
\begin{equation}
\rho_{t\bar t} = \frac{1}{4} \left[ \mathbb{I} \otimes \mathbb{I} 
+  B_i\, \sigma^i \otimes \mathbb{I} 
+ \bar{B}_j\, \mathbb{I} \otimes \sigma^j 
+ C_{ij}\, \sigma^i \otimes \sigma^j \right].
\label{eq:rho2-qbit}
\end{equation}
Because of the extremely short top-quark lifetime, the spin state cannot be
measured directly and must be inferred from the decay products. Quantum-state
tomography reconstructs the polarisation and spin-correlation observables from
the differential distributions of final-state particles with respect to a
fixed orthonormal Cartesian basis. The reconstruction depends on the decay of
the two particles and is otherwise independent of the production mechanism.
It is therefore necessary to account for possible BSM contributions to the
\(Wtb\) vertex. We first review the SM-decay distributions, then extend the
analysis to the general decay structure, and finally identify observables
sensitive to CP violation in decay. CP violation in production instead enters
through \(B_i\), \(\bar B_i\), and \(C_{ij}\).

The general differential cross section for the full process $I\rightarrow t\bt\rightarrow b\,\ell^+\nu_\ell\,\bar{b} \ell^- \bar{\nu}_\ell$ is given by
\be
\qquad d\sigma= \frac{1}{\mathcal{F}_I}|\mathcal{M}|^2(2\pi)^4\delta^4(p_I-\sum_ip_i-\sum_i\bar{p}_i)d\Phi^td\Phi^{\bar{t}}\,,
\label{eq:dsigma-general-structure}
\ee
where $\mathcal{F}_I$ denotes the initial-state flux or normalisation factor and $d\Phi^i=\prod_f\frac{d^3\vec{p_f}}{(2\pi)^3 \, 2E_f}$ contains the integration over the final state momenta of the process $i\rightarrow\{f\}$. 

Using the narrow-width approximation for the intermediate top and antitop,
\begin{equation}
\frac{1}{|\Delta_t(p_t)|^2}
\longrightarrow
\frac{\pi}{m_t\Gamma_t}\,
\delta(p_t^2-m_t^2),
\end{equation}
and analogously for the antitop, the production and decay phase spaces
factorise while their spin-correlations remain encoded in the density-matrix
trace:
\be\begin{aligned}
d\sigma=&\biggl[\frac{(2\pi)^4}{\mathcal{F}_I}\delta^4(p_I-p_t-\bar{p}_t)d\Phi^I\biggl] \biggl[\frac{(2\pi)^4}{2m_{t}\Gamma_t}\delta^4(p_t-\sum_ip_i)d\Phi^t\biggr]\biggl[\frac{(2\pi)^4}{2m_{t}\Gamma_t}\delta^4(\bar{p}_t-\sum_i\bar{p}_i)d\Phi^{\bar{t}}\biggr] \\
&\Tr\biggl[R^I[(D^{ t})^T\otimes(D^{\bar{t}})^T]\biggr].
\label{eq:factorisation-cross-section}
\end{aligned}
\ee
The phase-space integrations for the production and decay parts are now
factorised. Because the factorised measures and contractions are Lorentz
invariant, the production and decay integrations may be evaluated in
independently chosen frames. We evaluate production in the \(t\bar t\) ZMF
and each decay in the rest frame of its parent top quark.

The production mechanism affects only the numerical values of the integrated
Fano--Bloch coefficients and does not alter the structure of the tomography
protocol. Using Eq.~\eqref{eq:normalised-matrices} to write
\[
R^I=\Tr[R^I]\,\rho^I,
\]
the production contribution can therefore be integrated through the factor 
\begin{equation} 
\Tr[R^I]\frac{(2\pi)^4}{\mathcal{F}_I}\delta^4(p_I-p_t-\bar{p}_t)d\Phi^I.
\label{eq:integration-production}
\end{equation}

We consider the \(1\to2\) decay and \(2\to2\) scattering mechanisms discussed
in Ref.~\cite{Lamba:2026jfm}. For \(S\to t\bar t\), the integration can be
performed in the rest frame of the decaying particle. Replacing the flux by
\(\mathcal F_S=2M_S\), where \(M_S\) is the mass of the decaying particle,
we obtain
\begin{equation}
\Tr[R^{S}]\frac{(2\pi)^4}{2M_S}\delta^4(p_S-p_t-\bar p_t)d\Phi^{S}
=\frac{k^2M_S\left(\beta^2\cos^2\delta+\sin^2\delta\right)\beta}{8\pi}
\equiv\Gamma_{S\to t\bar t}(\beta).
\label{eq:production-differential-stt}
\end{equation}
Here \(\beta=\beta(M_S)\), so the only kinematic parameter is the heavy-scalar mass.

For the $2\rightarrow2$ scattering $i_1i_2\rightarrow t\bar{t}$, the integration of Eq.~\eqref{eq:integration-production} can be performed in the ZMF of the $t\bar{t}$, where their kinematics will be completely defined by the velocity $\beta$ and the scattering angle $\theta$:
\begin{equation} 
\Tr[R^{i_1i_2}]\frac{(2\pi)^4}{\mathcal{F}_{i_1i_2}}\delta^4(p_{i_1}+p_{i_2} -p_t-\bar{p}_t)d\Phi^{i_1i_2}= \frac{\Tr[R^{i_1i_2}]}{32\pi s}\beta \ d\cos\theta\equiv\sigma^{i_1i_2}(\beta, \theta) \ d\cos\theta\,,
\label{eq:production-cross-section-2-2}
\end{equation}
with $\Tr[R^{i_1i_2}]= 4\tilde{A}^{i_1i_2}$ and the squared center-of-mass energy $s$ for the 2-particle system is  $s= 4m_t^2/(1-\beta^2)$. 
\subsection{SM decay} \label{sec:tomography-SM}

We perform the decay-phase-space integration in the rest frame of the decaying top quark. The Cartesian basis used in this frame is obtained by boosting the spin basis \(\hat s^i_\mu\) defined in the \(t\bar t\) ZMF. With our conventions, this gives
\begin{equation}
    \hat{s}^1=\hat{n}\,, \qquad
    \hat{s}^2=\hat{r}\,, \qquad
    \hat{s}^3=\hat{k}\, .
    \label{eq:hel-basis-in-trest}
\end{equation}
A convenient parametrisation is then obtained by choosing one of these axes, denoted by \(\hat a\), as the polar axis. The direction of one decay product, which will act as the spin analyser, is described by the polar and azimuthal angles measured with respect to \(\hat a\). A third angle specifies the relative orientation of one of the remaining decay products with respect to the analyser direction. The antitop decay is treated analogously in the \(\bar t\) rest frame, using the same Cartesian basis and allowing for an independent choice of polar axis, denoted by \(\hat b\).

The charged lepton is the natural choice of analyser. Indeed, from the SM decay matrix in Eq.~\eqref{eq:decay-matrix-t-SM-before-frame}, or equivalently from Eq.~\eqref{eq:gamma-t-SM-covariant}, the charged-lepton direction factorises from the remaining kinematic dependence, and its tree-level spin-analysing power is \(\alpha_\ell=1\). The same reasoning applies to the charged-lepton from the antitop decay. Choosing the two charged leptons as analysers, the three independent four-momenta in each decay can therefore be written as
\begin{equation}
    \begin{cases}
&p_t=(m_{t},  0)\\
&p_\ell= E_\ell (1, \hat{p_\ell}), \quad \hat{p_\ell}= \mathcal{R}(\theta^{(a)}_\ell,\phi^{(a)}_\ell)\hat{a} \\
    &p_b= E_b(1, \hat{p_b}), \quad  \hat{p_b}= \mathcal{R}(\theta^{(a)}_\ell,\phi^{(a)}_\ell)\mathcal{R}(\theta^{(a)}_{\ell b},\phi^{(a)}_{\ell b})\hat{a}\\
 \end{cases}
 \label{eq:momenta-top-decay}
\end{equation} 
\begin{equation}
    \begin{cases}
&\bp_t=(m_{t}, 0)\\
&\bp_{\ell}= \bar{E}_{\ell} (1, \hat{\bp}_{\ell}), \quad \hat{\bp}_{\ell}= \mathcal{R}(\btheta^{(b)}_{\ell},\bphi^{(b)}_{\ell})\hat{b} \\
    &\bp_{b}= \bar{E}_{b}(1, \hat{\bp}_{b}),\quad  \hat{\bp}_{b}= \mathcal{R}(\btheta^{(b)}_{\ell},\bphi^{(b)}_{\ell})\mathcal{R}(\btheta^{(b)}_{\ell b},\bphi^{(b)}_{\ell b})\hat{b}.\\
 \end{cases}
 \label{eq:momenta-antitop-decay}
\end{equation}

\begin{figure}[ht]
    \centering
\includegraphics[width=0.4\linewidth, trim=1cm 2cm 1cm 1.5cm, clip]{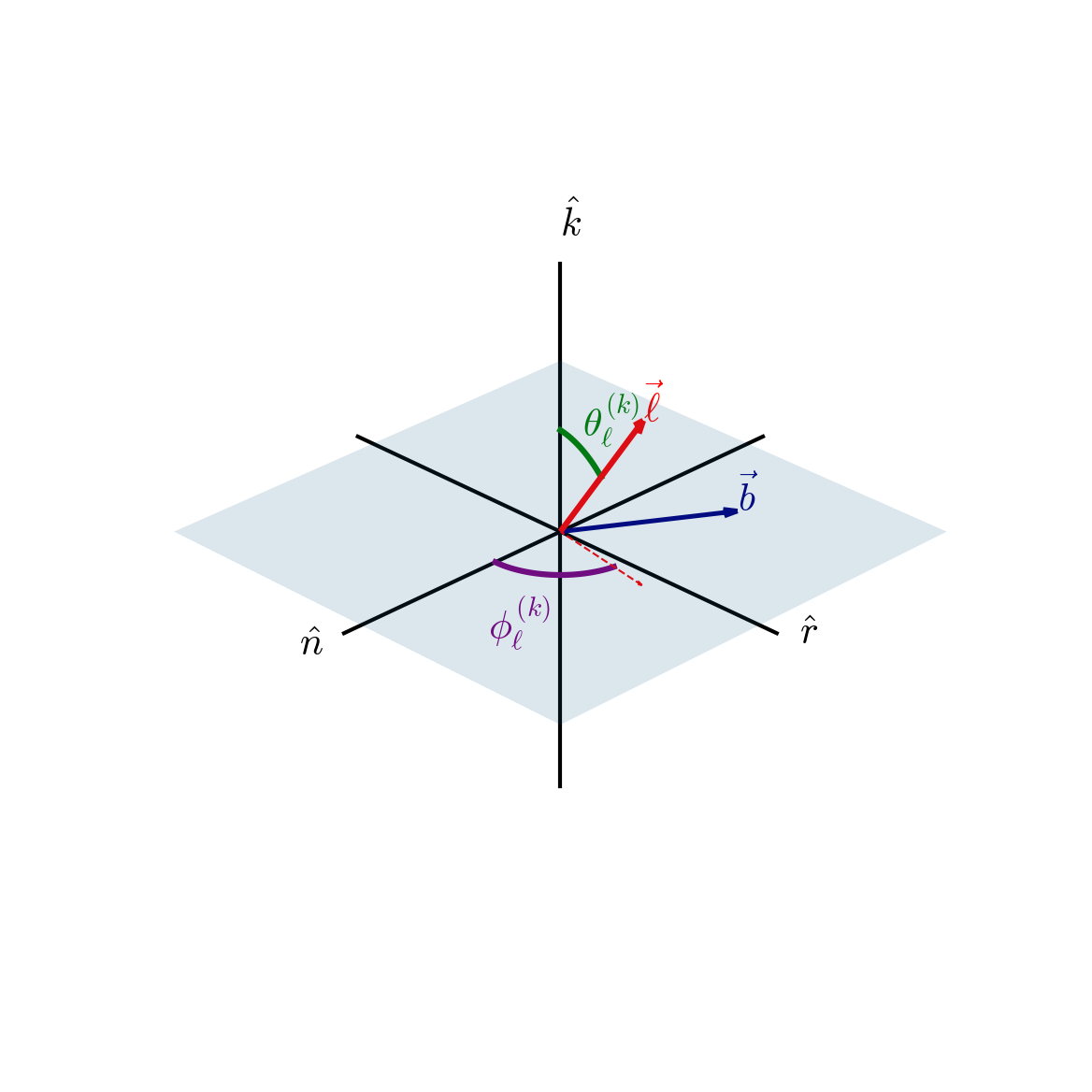}
\includegraphics[width=0.4\linewidth, trim=1cm 2cm 1cm 1.5cm, clip]{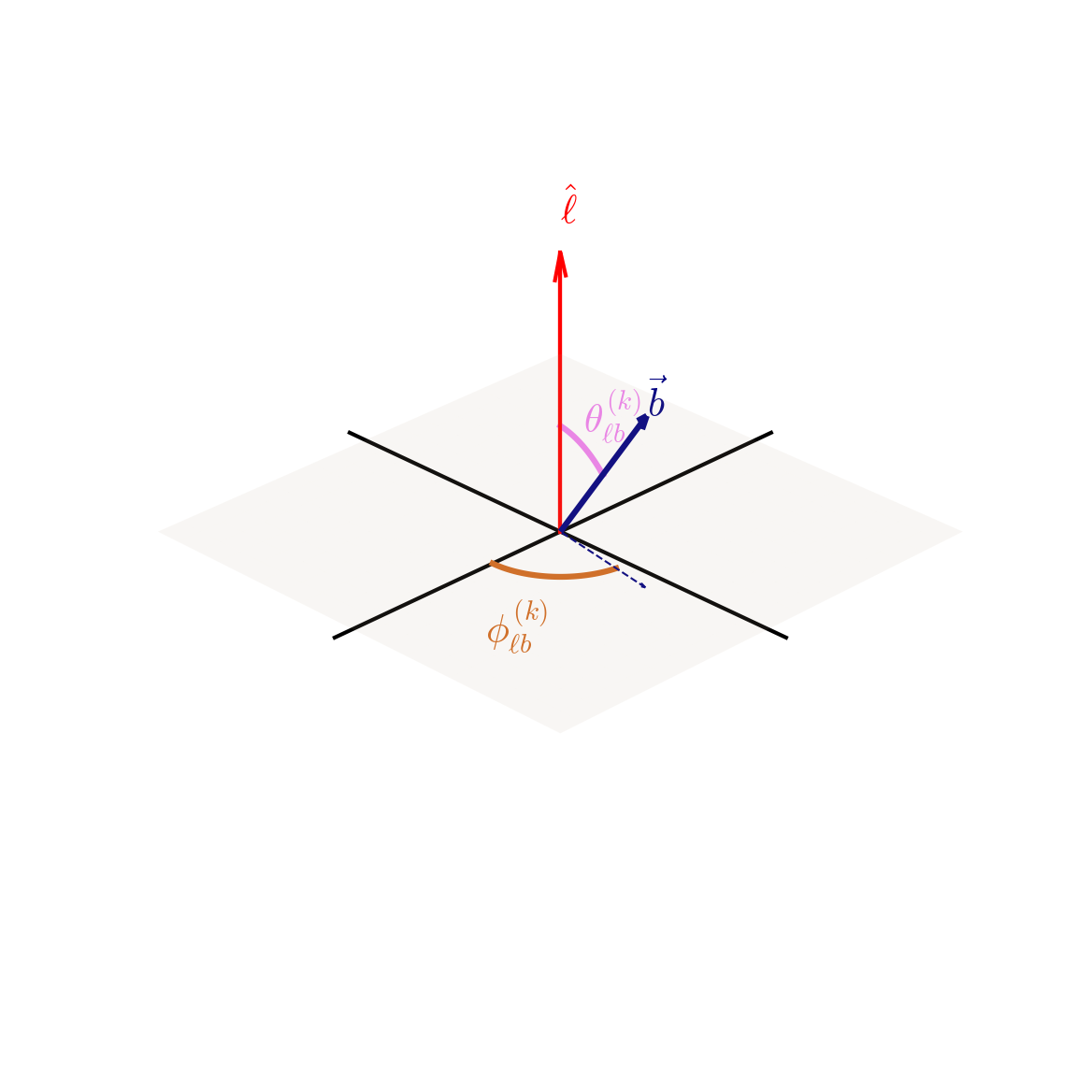}
    \caption{Example of the angular variables used for the \(t\) decay, choosing \(\hat a=\hat k\) as the polar axis. The left panel shows the definition of \(\theta_\ell^{(k)}\) and \(\phi_\ell^{(k)}\) in the helicity basis. In the right panel, the lepton direction is taken as the polar axis, and the relative angles \(\theta_{\ell b}^{(k)}\) and \(\phi_{\ell b}^{(k)}\) are defined. The angle \(\theta_{\ell b}^{(k)}\) is fixed by the decay kinematics. }
    \label{fig:angles-for-t-decay}
\end{figure}

We stress that the angles \(\phi_{\ell b}\) and \(\bar\phi_{\ell b}\) are defined as relative azimuthal angles between the bottom quark and the charged-lepton $\ell^+$, and between the antibottom quark and the charged-lepton $\ell^-$, respectively. Their definition therefore depends on the lepton directions in the helicity basis, specified by \(d\Omega_\ell^{(a)}\) and \(d\bar\Omega_\ell^{(b)}\), and consequently on the choice of polar axes \(\hat a\) and \(\hat b\). The full three-body phase space of the $t$ or $\bar t$  decay contains five independent
variables. Our parametrisation of these variables, together with the
definition of the invariant masses and the rotation matrices $\mathcal R$,
is given in App.~\ref{app:3-body-phase-space}. A graphical illustration of the angular variables is shown in Fig.~\ref{fig:angles-for-t-decay}.

Although the full three-body decay phase space contains five independent
variables, for the pure SM contribution to both the $t$ and $\bar t$ decays,
the decay amplitude depends only on the charged-lepton
direction in the corresponding helicity basis. The remaining phase-space variables can therefore be integrated out. Using,
for simplicity, the narrow-width approximation for the intermediate
$W$ boson, the following factor reduces to the simpler form
\be
\frac{\Tr[D^{t}]}{2m_{t}\Gamma_t}(2\pi)^4\delta^4(p_t-\sum_ip_i)d\Phi^t=\frac{\mathrm{Br}(t\rightarrow b\ell^+\nu_\ell)}{2\pi}d\cos\theta_\ell^{(a)} d \phi_\ell^{(a)} \,,
\ee
\be\frac{\Tr[D^{\bt}]}{2m_{t}\Gamma_t}(2\pi)^4\delta^4(\bar{p}_t-\sum_i\bar{p}_i)d\Phi^{\bar{t}}= \frac{\mathrm{Br}(\bar{t}\rightarrow \bar{b}\ell^-\bar{\nu}_\ell)}{2\pi}d\cos\btheta_\ell^{(b)} d\bar{\phi_\ell}^{(b)}\,.
\label{eq:SM_phasespace}
\ee
Using these reduced phase-space factors together with the normalised density matrices defined in Eq.~\eqref{eq:normalised-matrices}, the factorised cross section in Eq.~\eqref{eq:factorisation-cross-section} can be rewritten in terms of the production density matrix and the normalised top and antitop decay density matrices, as follows:
\be
\frac{4\pi^2}{\sigma(\beta,\theta)}\frac{d\sigma}{d\cos\theta\, d\cos\theta_\ell^{(a)}\, d \phi_\ell^{(a)}\,d\cos\btheta_\ell^{(b)} d\bar{\phi_\ell}^{(b)}}=\Tr\left[\rho^I[(\Gamma^{t})^T\otimes(\Gamma^{\bar{t}})^T]\right]. 
\label{eq:diff-angular-SM}
\ee
Here, $\sigma(\beta,\theta)$ denotes the cross section obtained by integrating over the complete phase spaces of the top and antitop decay products while keeping the production variables $\beta$ and $\theta$ fixed. Within the NWA, it factorises into the production differential cross section at fixed $\beta$ and the corresponding leptonic branching fractions,
\begin{equation}
    \sigma(\beta,\theta)
    \equiv
    \sigma^{I}(\beta,\theta)
    \mathrm{Br}(t\to b\ell^+\nu_\ell)
    \mathrm{Br}(\bar t\to \bar b\ell^-\bar\nu_\ell).
    \label{eq:sigmadecomposition}
\end{equation}
For production through the decay of a scalar resonance, the decay is isotropic in the scalar rest frame, so the production cross section is independent of $\theta$ and may be written simply as $\sigma(\beta)\equiv\Gamma_{S\to t\bar t}(\beta)$, and the dependence on $\cos\theta$ can be removed from Eq.~\eqref{eq:diff-angular-SM}. Substituting the general decomposition of production  density matrix in Eq.~\eqref{eq:rho2-qbit}, the SM decay density matrices given in Eqs.~\eqref{eq:gamma-t-SM} and \eqref{eq:gamma-tbar-SM}, and using the trace identities of the Pauli matrices, Eq.~\eqref{eq:diff-angular-SM} can be expressed as
\begin{align}
\notag\frac{16\pi^2}{\sigma(\beta,\theta)}\frac{d\sigma}{d\cos{\theta}\,d\Omega_\ell^{(a)}\,d\bar{\Omega}_\ell^{(b)}}&=1+\alpha_\ell^{ \rm SM}\sum_{i}(\hat{p}_\ell^{t-\text{rest}}\cdot \hat{s}^i)B_i(\beta,\theta)+ \bar{\alpha_\ell}^{\rm SM}\sum_j(\hat{\bp}_{\ell}^{\bt-\text{rest}}\cdot \hat{s}^j)\bar{B}_j(\beta,\theta)\\
&+\alpha_\ell^{\rm SM}\bar{\alpha_\ell}^{\rm SM}\sum_{ij}(\hat{p}_\ell^{t-\text{rest}}\cdot \hat{s}^i)(\hat{\bp}_{\ell}^{\bt-\text{rest}}\cdot \hat{s}^j)C_{ij}(\beta,\theta)\,.
\label{eq:diffdistribution}
\end{align}
Here, $i= \{\hat a,\hat \imath_1,\hat \imath_2\}$ denotes an orthonormal triad satisfying
$
\hat \imath_1\wedge \hat \imath_2=\hat a,
$
so that $\{\hat a,\hat \imath_1,\hat \imath_2\}$ forms a right-handed orthonormal basis. Similarly, $j=\{\hat b,\hat \jmath_1,\hat \jmath_2\}$ denotes the corresponding orthonormal basis for the antitop decay.
By integrating over the remaining angular variables, one obtains three classes of angular distributions from which the Fano--Bloch coefficients can be extracted:
\ba
\frac{2}{\sigma(\beta,\theta)}\frac{d\sigma}{d\cos{\theta}\,d\cos\theta_{\ell}^{(a)}}&=&1+\alpha_{\ell}^{\rm SM}B_a(\beta,\theta)\cos\theta_{\ell}^{(a)}\,,\label{eq:forB}\\
\frac{2}{\sigma(\beta,\theta)}\frac{d\sigma}{d\cos{\theta}\,d\cos\bar{\theta_{\ell}}^{(b)}}&=&1+\bar{\alpha_\ell}^{\rm SM}\bar{B}_b (\beta,\theta)\cos\bar{\theta_{\ell}}^{(b)}\,,\label{eq:forBbar}\\
\frac{4}{\sigma(\beta,\theta)}\frac{d\sigma}{d\cos{\theta}\,d\cos\theta_{\ell}^{(a)}\ d\cos\bar{\theta_{\ell}}^{(b)}}&=&1+\alpha_{\ell}^{\rm SM}B_a(\beta,\theta)\cos\theta_{\ell}^{(a)}+ \bar{\alpha_\ell}^{\rm SM}\bar{B}_b(\beta,\theta)\cos\bar{\theta_{\ell}}^{(b)}\nonumber\\
&&+ \alpha_\ell^{\rm SM}\bar{\alpha_\ell}^{\rm SM}C_{ab}(\beta,\theta)\cos\theta_{\ell}^{(a)}\cos\bar{\theta_{\ell}}^{(b)}\,.
\label{eq:diff-single-polar-angle-SM}
\ea
 The first two distributions isolate the individual polarisation components of the top and antitop quarks along the directions $\hat{a}$ and $\hat{b}$, respectively. Their linear dependence on the corresponding lepton polar angle allows the coefficients $B_a$ and $\bar{B}_b$ to be extracted directly. The third distribution contains both single-particle polarisation contributions and the spin-correlation contribution proportional to $C_{ab}$. To isolate the spin-correlation term in a one-dimensional distribution of decay angles, we introduce the variable $\xi^{(ab)}=\cos\theta_\ell^{(a)}\cos\bar{\theta_\ell}^{(b)} $. After integrating over the individual polar angles of leptons at fixed $\xi^{(ab)}$, the polarisation terms vanish, and the resulting distribution depends only on the spin-correlation coefficient:
\begin{equation}
\frac{2}{\sigma(\beta,\theta)}\frac{d\sigma}{d\cos{\theta}\,d\xi^{(ab)}}= (1+\alpha_{\ell}^{\rm SM}\bar{\alpha_{\ell}}^{\rm SM}C_{ab}(\beta,\theta)\xi^{(ab)})\log\left(\frac{1}{|\xi^{(ab)}|} \right)\,. 
\label{eq:diff-xi-SM}
\end{equation}
A similar construction can be used to probe the antisymmetric part of the spin-correlation matrix. Following Ref.~\cite{CMS:2019nrx}, we define $\xi^{(ab)}_{-}= \cos\theta_\ell^{(a)}\cos\bar{\theta_\ell}^{(b)}-\cos\theta_\ell^{(b)}\cos\bar{\theta_\ell}^{(a)}$. The corresponding one-dimensional distribution is
\begin{equation}
    \frac{2}{\sigma(\beta,\theta)}\frac{d\sigma}{d\cos{\theta}\,d\xi^{(ab)}_{-}}= \left(1+\alpha_{\ell}^{\rm SM}\bar{\alpha_{\ell}}^{\rm SM}C^A_{ab}(\beta,\theta)\xi^{(ab)}_{-}\right)\arccos|\xi^{(ab)}_{-}|\,.
    \label{eq:diff-xi_MINUS-SM}
\end{equation}
This distribution isolates
\(C^A_{ab}=(C_{ab}-C_{ba})/2\), the antisymmetric part of the
spin-correlation matrix. As shown in Ref.~\cite{Lamba:2026jfm}, a non-zero
\(C^A_{ab}\) is a CP-violation marker in the common spin basis when
CP-related kinematic configurations are compared. For \(C^A_{ab}=0\),
the distribution reduces to the reference shape proportional to
\(\arccos|\xi_-^{(ab)}|\); equivalently, its ratio to this reference
shape is flat.

It is worth emphasizing that the choice of spin analyser is not unique.
Although charged leptons provide the most convenient option, both
theoretically and experimentally, the \(b\) quark or, in principle, the
neutrino may also be used~\cite{Aguilar-Saavedra:2010ljg}. Different spin
analysers may likewise be chosen independently for the top and antitop
decays.

The particularly simple SM structure discussed above is, however, specific
to the choice of the charged lepton as spin analyser. If the \(b\) quark
or the neutrino is used instead, additional spin-dependent structures,
including terms proportional to \(p_b\cdot s\) and \(p_n\cdot s\), may
already arise in the SM. Nevertheless, no structures involving the
Levi-Civita tensor are generated.
\subsection{BSM decay}\label{subsect:BSM_decay}

The discussion below applies to the modifications of the \(Wtb\) vertex
introduced in Eq.~\eqref{eq:vertex-BSM-decay-top}. In the SM, when the
charged lepton is chosen as the spin analyser, the spin dependence of the
decay density matrix is entirely encoded in the Lorentz structure
\(p_\ell\cdot s\). This leads to a simple factorisation between the decay
kinematics and its angular dependence, so that the decay density matrix
depends only on the two angles specifying the charged-lepton direction,
rather than on all five independent variables of the three-body phase
space.

This simplification no longer holds once the \(Wtb\) vertex is modified by
the SMEFT operators considered in Eq.~\eqref{eq:vertex-BSM-decay-top}. These contributions generate additional spin-dependent Lorentz structures,
such as \(p_b\cdot s\), as well as terms involving the Levi-Civita tensor,
e.g., \(\epsilon_{p_t p_\ell p_b s}\), as shown in
Eqs.~\eqref{eq:P-BSM-decay-t} and
\eqref{eq:P-BSM-decay-tbar}. Consequently, the angular dependence can no
longer be factorised from the remaining kinematic variables, and the decay
density matrix generally depends on all five independent phase-space
parameters, as made explicit in
Eq.~\eqref{eq:phasespace_t}.

In the following, we focus on the case in which the \(Wtb\) vertex receives
the SMEFT contributions specified in
Eq.~\eqref{eq:vertex-BSM-decay-top}. To make explicit how the
production Fano--Bloch coefficients combine with the spin-independent and
spin-dependent components of the top and antitop decay density matrices, we
write
\be
\Tr[\rho^I[(D^{t})^T\otimes (D^{\bar{t}})^T]]= \Bigl[T\bar{T}+\bar{T}\sum_iB_i(s^i_{\mu}P^\mu)+T\sum_j \bar{B}_j (\bh_j\bs^j_{\mu}\bar{P}^\mu)+ \sum_{ij}C_{ij}(s^i_{\mu}P^\mu)(\bh_j\bs^j_{\nu}\bar{P}^\nu)\Bigr]. 
\ee
Here, $T$ and $\bar T$ denote the spin-independent decay contributions, whereas $s_\mu P^\mu$  and $\bar s_\nu \bar{P}^\nu$ encode the corresponding spin-dependent structures given in Eqs.~\eqref{eq:P-BSM-decay-t}-\eqref{eq:P-BSM-decay-tbar}. In the presence of anomalous $Wtb$ interactions, these quantities generally depend on the full set of variables parameterizing the respective decay phase spaces. 
Using this decomposition, the normalised differential cross section for the process $I\rightarrow t\bt\rightarrow b\,\ell^+\nu_\ell\,\bar{b} \ell^- \bar{\nu}_\ell$ can be constructed directly. To obtain the angular distribution relevant for the tomographic analysis, we integrate independently over the invariant-mass variables of the factorised top and antitop decay phase spaces. The resulting normalised angular distribution can be written as follows:
\begin{align}
\frac{64\pi^4}{\sigma(\beta,\theta)}
\frac{d\sigma}{
d\cos\theta\,
d\Omega_{\ell}^{(a)}\,d\phi_{\ell b}^{(a)}\,
d\bar\Omega_{\ell}^{(b)}\,d\bar\phi_{\ell b}^{(b)}
}
={}&
1
+\sum_i B_i(\beta,\theta)\,\mathcal{A}_i
+\sum_j \bar{B}_j(\beta,\theta)\,\bar{\mathcal{A}}_j
+\sum_{ij}C_{ij}(\beta,\theta)\,
\mathcal{A}_i\bar{\mathcal{A}}_j\,,
\label{eq:angular-distribution-BSM}
\end{align}
where the invariant-mass-integrated decay analysing functions are defined by
\begin{equation}
\mathcal{A}_i
\equiv
\frac{
\displaystyle \int dq^2\,dm_{23}^2\,
s^i_\mu P^\mu
}{
\displaystyle \int dq^2\,dm_{23}^2\,T
},
\qquad
\bar{\mathcal{A}}_j
\equiv
\frac{
\displaystyle \int d\bar q^{\,2}\,d\bar m_{23}^{\,2}\,
\bar h_j\,\bar s^j_\mu \bar P^\mu
}{
\displaystyle \int d\bar q^{\,2}\,d\bar m_{23}^{\,2}\,\bar T
}.
\label{eq:integrated-decay-analysers}
\end{equation}
For the distributions considered below, we integrate over the production
angle and use the corresponding production-angle-integrated Fano--Bloch
coefficients,
\begin{equation}
X_i(\beta)
=
\frac{
\displaystyle\int d\cos\theta\,
\widetilde{X}_i(\beta,\theta)
}{
\displaystyle\int d\cos\theta\,
\widetilde{A}(\beta,\theta)
},
\qquad
X_i\in\{B_i,\bar B_i\},
\qquad
C_{ij}(\beta)
=
\frac{
\displaystyle\int d\cos\theta\,
\widetilde{C}_{ij}(\beta,\theta)
}{
\displaystyle\int d\cos\theta\,
\widetilde{A}(\beta,\theta)
}.
\end{equation}
Here, \(\widetilde A\), \(\widetilde B_i\), \(\widetilde {\bar{B}}_j\), and
\(\widetilde C_{ij}\) are the coefficients entering the Fano--Bloch decomposition
of the unnormalised production density matrix, defined in
Ref.~\cite{Lamba:2026jfm}. We denote by $\sigma(\beta,\theta)\equiv d\sigma(\beta)/d\cos\theta$
the production cross section differential in \(\cos\theta\), while the
corresponding production-angle-integrated cross section is 
\begin{equation}
\sigma(\beta)
=
\int d\cos\theta\,
\sigma(\beta,\theta).
\end{equation}
The distributions considered below are normalised with respect to
\(\sigma(\beta)\), but can equivalently be retained at fixed
production angle by using \(B_i(\beta,\theta)\), \(\bar{B}_j(\beta,\theta)\) and
\(C_{ij}(\beta,\theta)\), replacing the normalisation
\(\sigma(\beta)\) by
\(\sigma(\beta,\theta)\), and keeping the
differential \(d\cos\theta\) on the left-hand side.
In the following, we suppress the explicit \(\beta\) dependence and denote
the production-angle-integrated quantities
\(\sigma(\beta)\), \(B_i(\beta)\), \(\bar{B}_j(\beta)\), and \(C_{ij}(\beta)\)
simply by \(\sigma\), \(B_i\), \(\bar{B}_j\), and \(C_{ij}\), respectively.

After integrating over the production angle, the distribution in Eq.~\eqref{eq:angular-distribution-BSM} depends on six angular variables, three associated with each of the top and antitop decay phase spaces. Lower-dimensional distributions can then be obtained by integrating over selected angular variables or suitable combinations thereof.

Before constructing the joint \(t\bar t\) angular distributions, we introduce
the one-angle top-decay functions obtained by integrating \(\mathcal A_i\)
over two of the three angular variables. Since the spin-independent function
\(T\) entering the normalisation of \(\mathcal A_i\) is independent of the
decay angles, these integrations can be performed directly on
\(\mathcal A_i\). The resulting functions are
\begin{align}
\mathcal A_i^{[\phi_{\ell b}^{(a)}]}
&\equiv
\displaystyle
\int d\cos\theta_\ell^{(a)}\,d\phi_\ell^{(a)}\,
\mathcal A_i
=
\delta_{ia}
\left[
v_c\cos\phi_{\ell b}^{(a)}
+w_s^{\rm CPV}\sin\phi_{\ell b}^{(a)}
\right],\label{eq:partially-integrated-top-functions_lb}
\\[2mm]
\mathcal A_i^{[\phi_\ell^{(a)}]}
&\equiv
\displaystyle
\int d\cos\theta_\ell^{(a)}\,d\phi_{\ell b}^{(a)}\,
\mathcal A_i
=
\pi^2\alpha_\ell
\left[
\delta_{ii_1}\cos\phi_\ell^{(a)}
+\delta_{ii_2}\sin\phi_\ell^{(a)}
\right],
\\[2mm]
\mathcal A_i^{[\cos\theta_\ell^{(a)}]}
&\equiv
\displaystyle
\int d\phi_\ell^{(a)}\,d\phi_{\ell b}^{(a)}\,
\mathcal A_i
=
4\pi^2\alpha_\ell\,
\delta_{ia}\cos\theta_\ell^{(a)},
\qquad
\hat a\wedge \hat{\imath}_1=\hat{\imath}_2 .
\label{eq:partially-integrated-top-functions}
\end{align}
The coefficients \(v_c\), \(w_s^{\rm CPV}\), and \(\alpha_\ell\) encode the
dependence of the top-decay angular distributions on the modified \(Wtb\)
vertex of Eq.~\eqref{eq:vertex-BSM-decay-top}. All three are functions of the
corresponding Wilson coefficients, see Appendix~\ref{app:analytic-expressions}. The charged-lepton spin-analysing power
\(\alpha_\ell\) was introduced previously, whereas \(v_c\) and
\(w_s^{\rm CPV}\) parametrize the cosine and sine modulations of
\(\phi_{\ell b}^{(a)}\), respectively. 

The coefficient \(w_s^{\rm CPV}\) originates from the Levi-Civita
structures in the top decay density matrix, in particular from contractions
of the form
\begin{equation}
\epsilon(p_t,p_\ell,p_b,s^i).
\end{equation}
It is therefore sensitive to the CP-violating contributions generated by
the imaginary parts of the relevant Wilson coefficient combinations. By contrast, \(v_c\) originates from spin-dependent structures proportional
to \(p_b\cdot s^i\) and depends on CP-even combinations of Wilson
coefficients. At \(\mathcal{O}(\Lambda^{-2})\), these contributions are
proportional to \(\operatorname{Re}(C_{\phi tb})\) and
\(\operatorname{Re}(C_{tW})\), while \(\alpha_\ell\) receives contributions
from both \(p_\ell\cdot s^i\) and \(p_b\cdot s^i\). In the SM limit,
\begin{equation}
v_c^{\rm SM}=0,
\qquad
w_s^{{\rm CPV},\,{\rm SM}}=0,
\qquad
\alpha_\ell^{\rm SM}=1.
\end{equation}
Similarly, for the antitop decay, we define the one-angle functions obtained by integrating  \(\bar{\mathcal A}_j\)
over two of the three antitop decay angles:
\begin{align}
\bar{\mathcal A}_j^{[\bar\phi_{\ell b}^{(b)}]}
&\equiv
\int d\cos\bar\theta_\ell^{(b)}\,d\bar\phi_\ell^{(b)}\,
\bar{\mathcal A}_j
=
\delta_{jb}
\left[
\bar v_c\,\cos{\bar\phi_{\ell b}^{(b)}}
+\bar w_s^{\rm CPV}\,\sin{\bar\phi_{\ell b}^{(b)}}
\right], \label{eq:partially-integrated-antitop-functions_lb}
\\
\bar{\mathcal A}_j^{[\bar\phi_\ell^{(b)}]}
&\equiv\int d\cos\bar\theta_\ell^{(b)}\,d\bar\phi_{\ell b}^{(b)}\,
\bar{\mathcal A}_j
=
\pi^2\bar\alpha_\ell
\left[
\delta_{j j_1}\,\cos{\bar\phi_\ell^{(b)}}
+\delta_{j j_2}\,\sin{\bar\phi_\ell^{(b)}}
\right],
\\
\bar{\mathcal A}_j^{[\cos\bar\theta_\ell^{(b)}]}
&\equiv\int d\bar\phi_\ell^{(b)}\,d\bar\phi_{\ell b}^{(b)}\,
\bar{\mathcal A}_j
=
4\pi^2\bar\alpha_\ell\,
\delta_{jb}\,\cos{\bar\theta_\ell^{(b)}},
\qquad
\hat b\wedge\hat \jmath_1=\hat \jmath_2 .
\label{eq:partially-integrated-antitop-functions}
\end{align}
The corresponding antitop decay coefficients are related to the top decay coefficients by
\begin{equation}
    \bar{v}_c = -v_c,\qquad\bar{w}^{ \rm CPV}_s = -w^{ \rm CPV}_s,\qquad \bar\alpha_\ell= -\alpha_\ell.
    \label{eq:rel_tandtbar_decaycoeff}
\end{equation}
These relations follow from the charge-conjugate structure of the top and
antitop decay density matrices and from the adopted spin-basis and angular
conventions. The analytic expressions for \(v_c\), \(w_s^{\rm CPV}\), and
\(\alpha_\ell\) are collected in
Appendices~\ref{app:alpha_analytical} and~\ref{App:vcwcpv}.
\begin{figure}[t]
    \centering
    \includegraphics[width=0.48\linewidth]{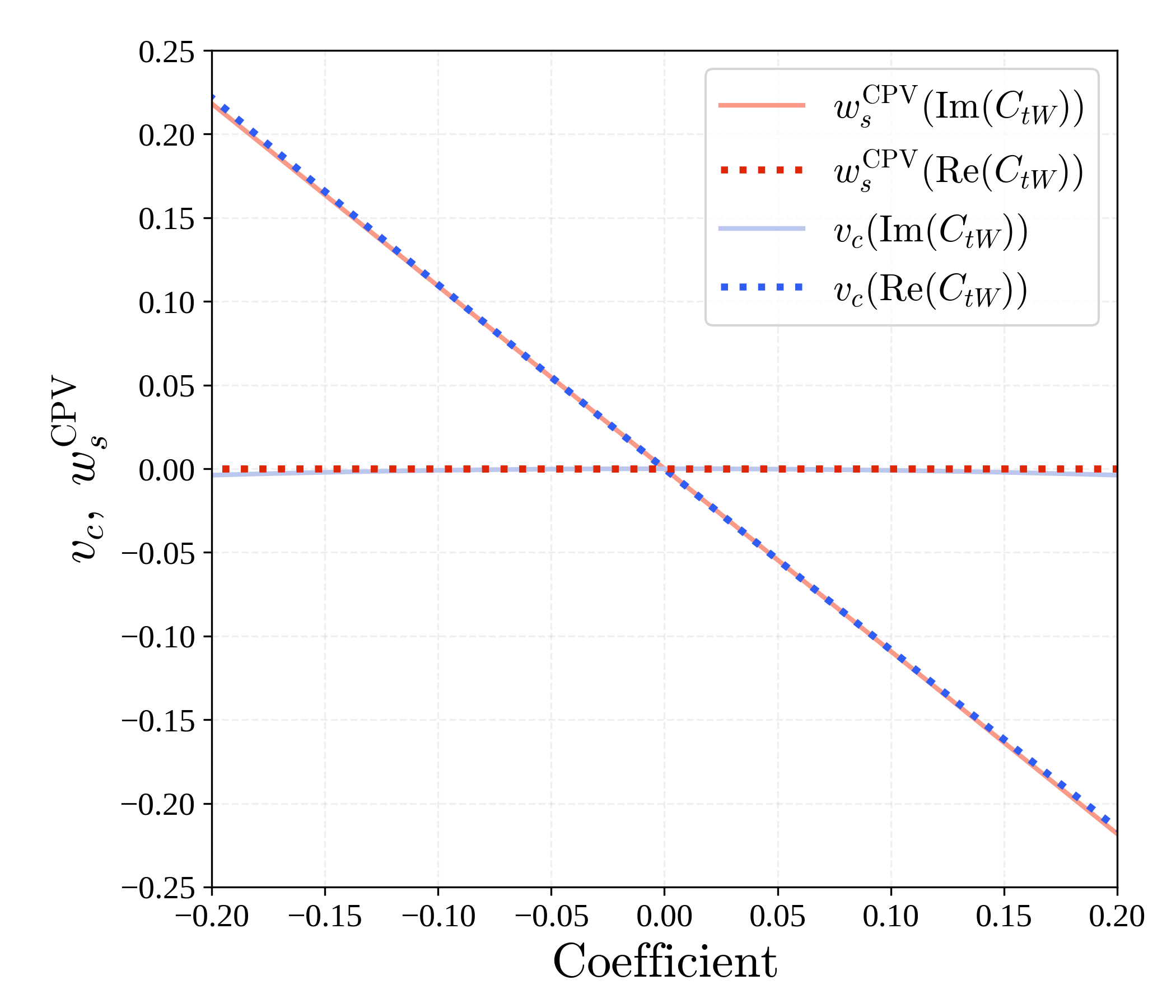}
    \includegraphics[width=0.48\linewidth]{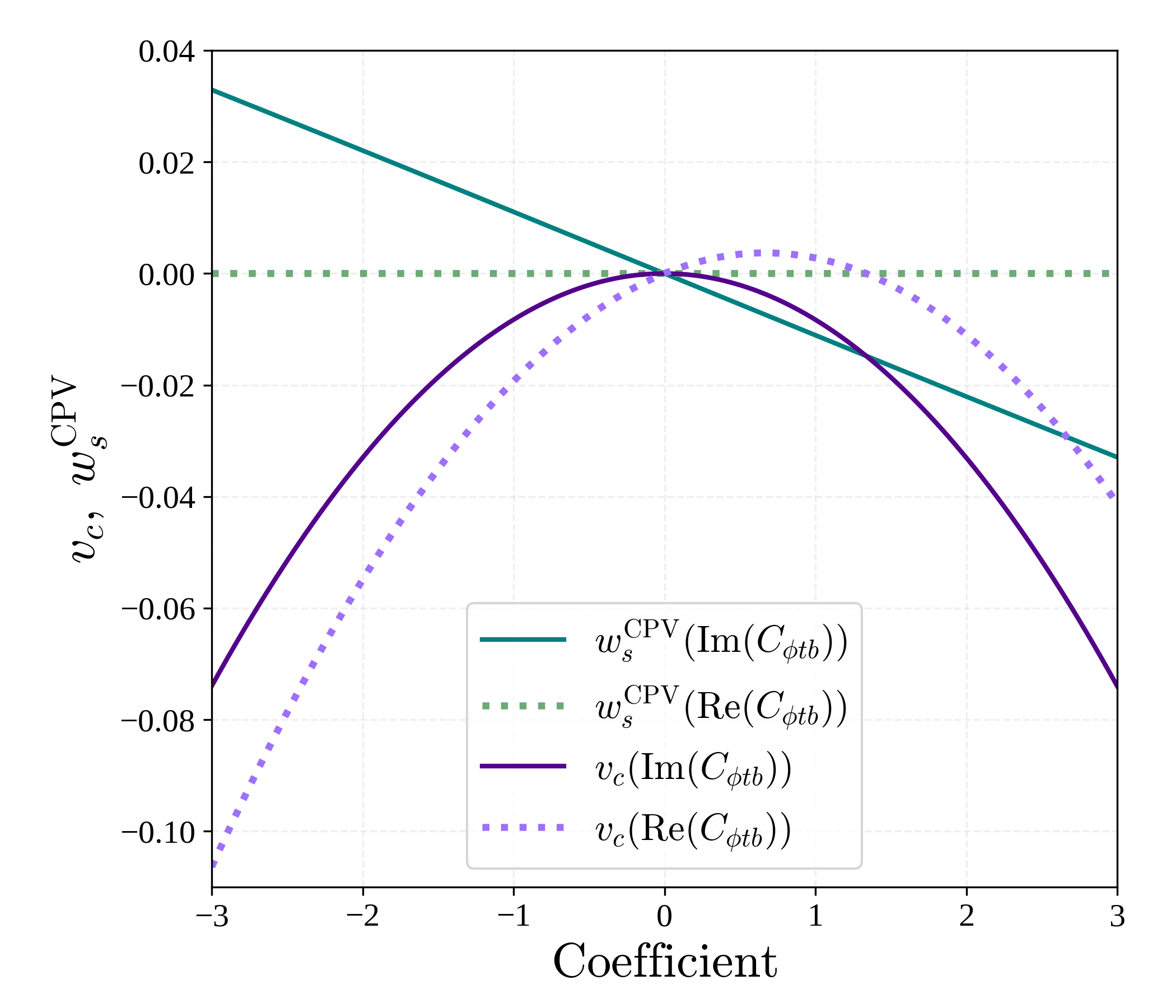}
    \caption{Dependence of \(v_c\) and \(w_s^{\rm CPV}\) on the Wilson coefficients
\(C_{tW}\) and \(C_{\phi tb}\), with only one coefficient taken to be
non-zero at a time and chosen to be either purely real or purely imaginary.
The ranges shown are selected according to the experimental bounds
summarised in Tab.~\ref{tab:EFT-op-limits}. The left and right panels correspond to
\(C_{tW}\) and \(C_{\phi tb}\), respectively.
The value of $\Lambda$ has been set to 1 TeV.}
    \label{fig:vc-and-ws}
\end{figure}

We then proceed to visualise how these decay coefficients depend on the Wilson coefficients. We first consider the case in which only one Wilson coefficient is non-zero
at a time and is taken to be either purely real or purely imaginary, as shown in Fig.~\ref{fig:vc-and-ws}. The ranges used in
the scans are guided by the experimental bounds summarised in
Tab.~\ref{tab:EFT-op-limits}, but are not intended to reproduce the exact
multidimensional allowed region. Instead, the quoted limits are used as
reference scales, and the corresponding coefficients are varied
symmetrically over positive and negative values.
\begin{table}[t]
\centering
\begin{tabular}{|c|c|c|}
\hline  
  Quantity & Limits & Reference\\
\hline
\hline
  $\operatorname{Re}(C_{tW})$& [0.02, 0.34]  &\cite{ATLAS:2025adk} 95\% C.L. 
  \\
   $\operatorname{Im}(C_{tW})$& [-0.08, 0.12]  &\cite{ATLAS:2025adk} 95\% C.L. 
   \\
 \hline
 \hline
  $\operatorname{Re}(C_{bW})$& [-0.5, 0.7]  &\cite{ATLAS:2025adk} 95\% C.L. 
  \\
  $\operatorname{Im}(C_{bW})$& [-0.6, 0.6]  &\cite{ATLAS:2025adk} 95\% C.L. 
  \\
 \hline
 \hline
 $\operatorname{Re}(C_{\phi tb})$&  [-3.14, 3.18] &\cite{CMS:2023xyc} $2\sigma$ C.L. 
 \\
  $\operatorname{Im}(C_{\phi tb})$& [-2.6, 2.6]  & Reinterpreted from \cite{ATLAS:2025adk} 
  \\
 \hline
 \hline

    $C^{3}_{\phi Q}$& [-0.75,0.41]&\cite{ATLAS:2025adk} 95\% C.L. \\ 
\hline
\end{tabular}
\caption{Bounds on the Wilson coefficients for $\Lambda=1~\mathrm{TeV}$, 
taken from Refs.~\cite{ATLAS:2025adk,CMS:2023xyc}. 
Ref.~\cite{ATLAS:2025adk} assumes $C_{\phi tb}$ to be real and positive 
and assigns the complex phase to the relative phase between 
$C_{\phi tb}$ and $C_{bW}$. The range shown for 
$\operatorname{Im}(C_{\phi tb})$ is obtained by reinterpreting the quoted 
bound as a constraint on $r=\lvert C_{\phi tb}\rvert$, writing 
$C_{\phi tb}=r e^{i\eta}$, and varying the allowed relative phase $\eta$.}
\label{tab:EFT-op-limits}
\end{table}

For \(C_{tW}\), the dominant dependence of \(v_c\) on
\(\operatorname{Re}(C_{tW})\) and of \(w_s^{\rm CPV}\) on
\(\operatorname{Im}(C_{tW})\) arises already at the interference order and
is therefore approximately linear over the range shown. By contrast, the
dependence of \(v_c\) on \(\operatorname{Im}(C_{tW})\) originates from terms
quadratic in \(C_{tW}\). These contributions are strongly suppressed within
the experimentally allowed interval considered here, although their
magnitude increases for larger absolute values of the Wilson coefficient.
Moreover, \(w_s^{\rm CPV}\) vanishes when only
\(\operatorname{Re}(C_{tW})\) is non-zero. Contributions involving the real
part can arise only through products with the imaginary part of another
Wilson coefficient, which are absent in our one-operator-at-a-time
analysis.
 
The dependence on \(C_{\phi tb}\) exhibits a similar CP nature.
The coefficient \(w_s^{\rm CPV}\) is generated by
\(\operatorname{Im}(C_{\phi tb})\), whereas it vanishes identically for a
purely real \(C_{\phi tb}\) in the one-operator-at-a-time setup. By contrast, \(v_c\) receives contributions from both the real and
imaginary parts of \(C_{\phi tb}\). Its dependence on
\(\operatorname{Im}(C_{\phi tb})\) is symmetric under
\(\operatorname{Im}(C_{\phi tb})\to-\operatorname{Im}(C_{\phi tb})\),
reflecting the quadratic origin of this contribution. For
\(\operatorname{Re}(C_{\phi tb})\), the interplay between interference and
quadratic terms leads to an asymmetric dependence. The substantially weaker
dependence of 
\(w_s^{\rm CPV}\) on \(\operatorname{Im}(C_{\phi tb})\),
compared with its dependence on \(\operatorname{Im}(C_{tW})\), follows from
the different chiral structures of the two contributions. At interference
order, the term proportional to 
\(\operatorname{Im}(C_{\phi tb})\) is
suppressed by the bottom-quark mass, whereas the contribution proportional
to \(\operatorname{Im}(C_{tW})\) is not. Consequently, for Wilson
coefficients of comparable magnitude, \(w_s^{\rm CPV}\) is parametrically
more sensitive to \(\operatorname{Im}(C_{tW})\).

The coefficients \(C_{bW}\) and \(C_{\phi Q}^{3}\) are not shown because,
when either is switched on individually, the numerators of both \(v_c\)
and \(w_s^{\rm CPV}\) remain zero. Although these coefficients modify the
corresponding normalisation factors, \(C_{bW}\) contributes to the
numerators only through products with \(C_{\phi tb}\) or \(C_{tW}\), while
\(C_{\phi Q}^{3}\) contributes only through products with other Wilson
coefficients. Consequently, \(v_c=w_s^{\rm CPV}=0\) in either
one-coefficient scenario. For \(C_{\phi Q}^{3}\), this conclusion is independent of the truncation
in \(m_b\). For \(C_{bW}\), contributions to \(v_c\) can arise at
quadratic order when terms beyond linear order in \(m_b\) are retained;
these terms are outside the approximation adopted here.
\begin{figure}[t]
    \centering
    \includegraphics[width=0.325\linewidth]{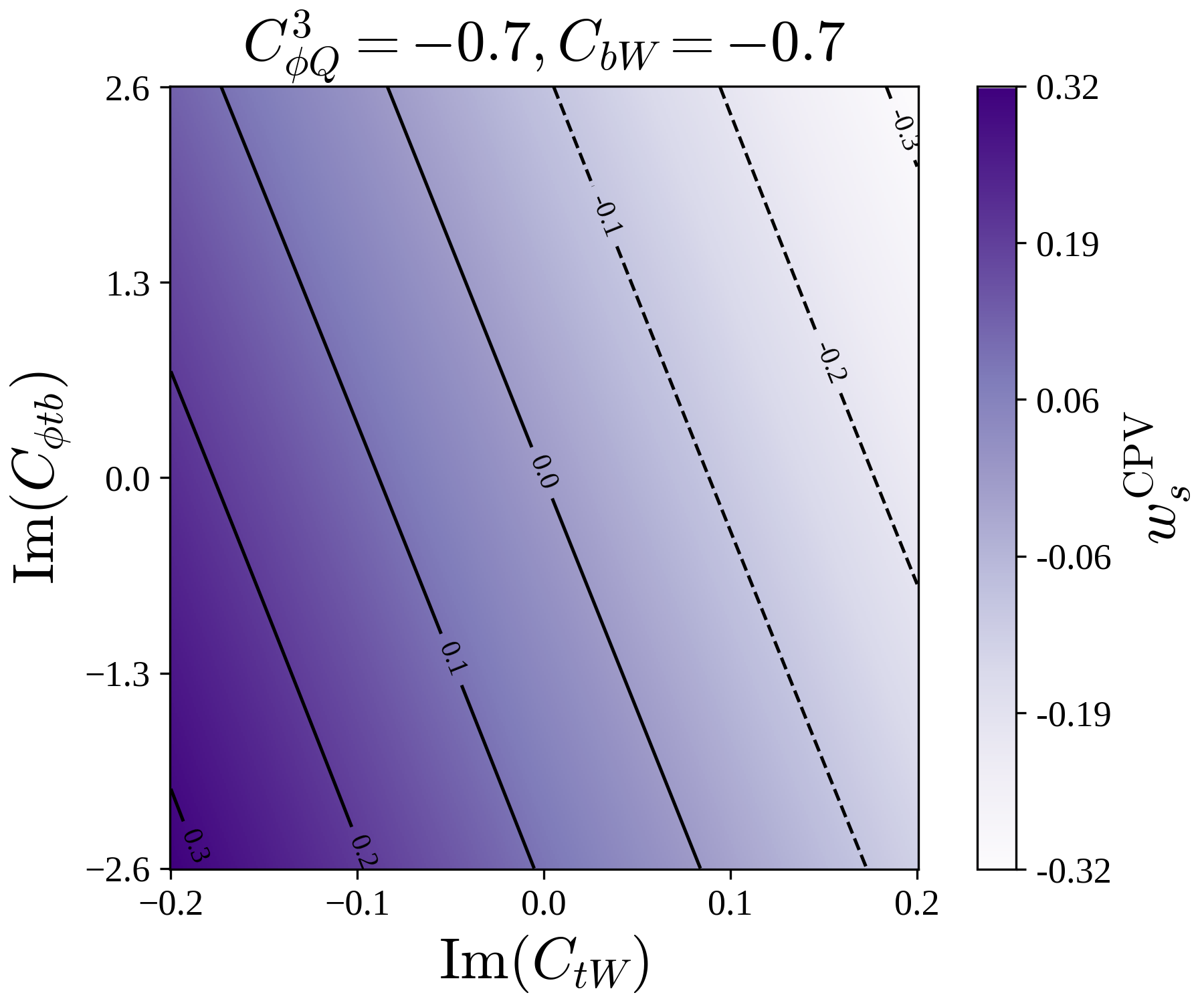}
    \includegraphics[width=0.325\linewidth]{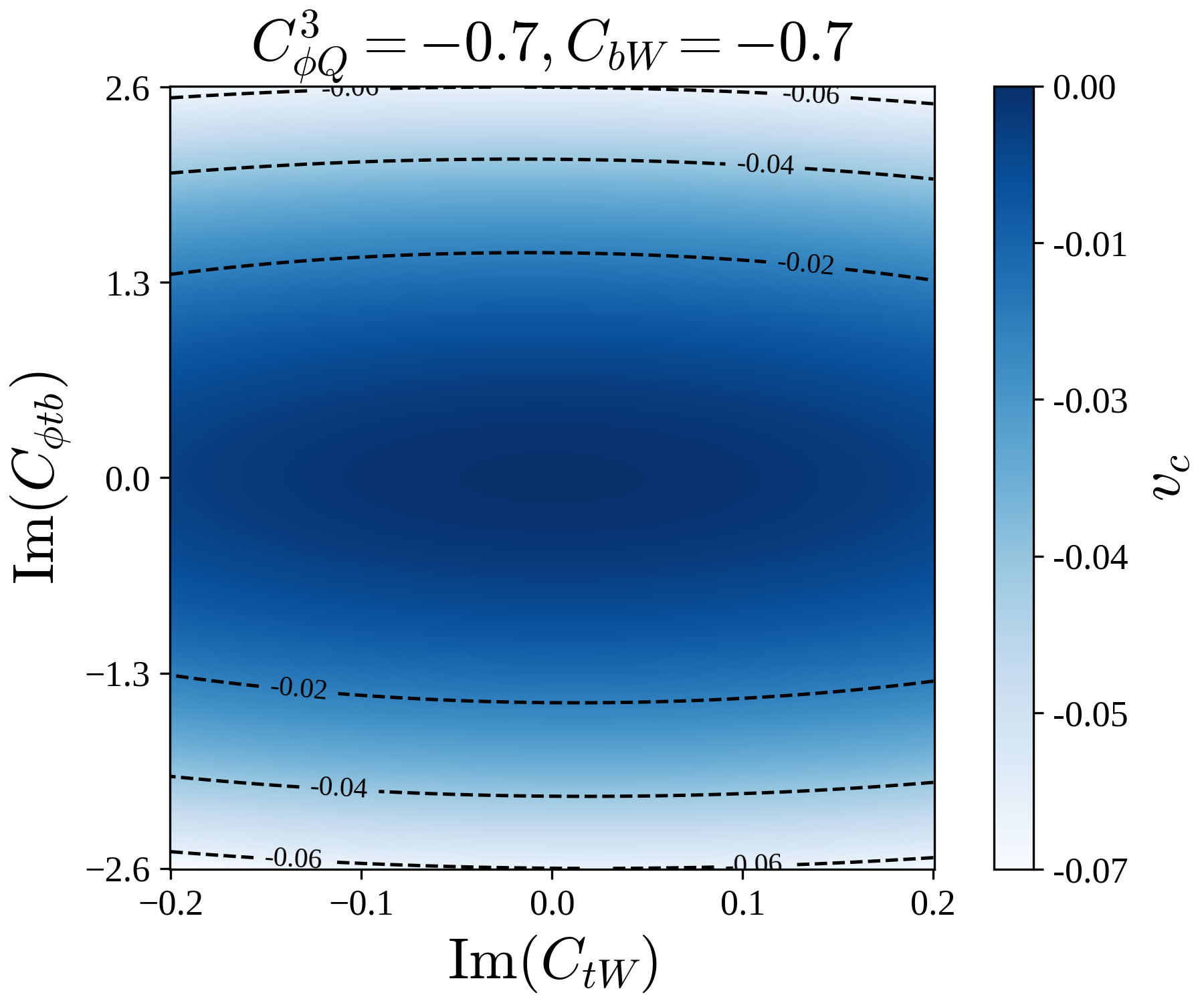} 
     \includegraphics[width=0.325\linewidth]{ 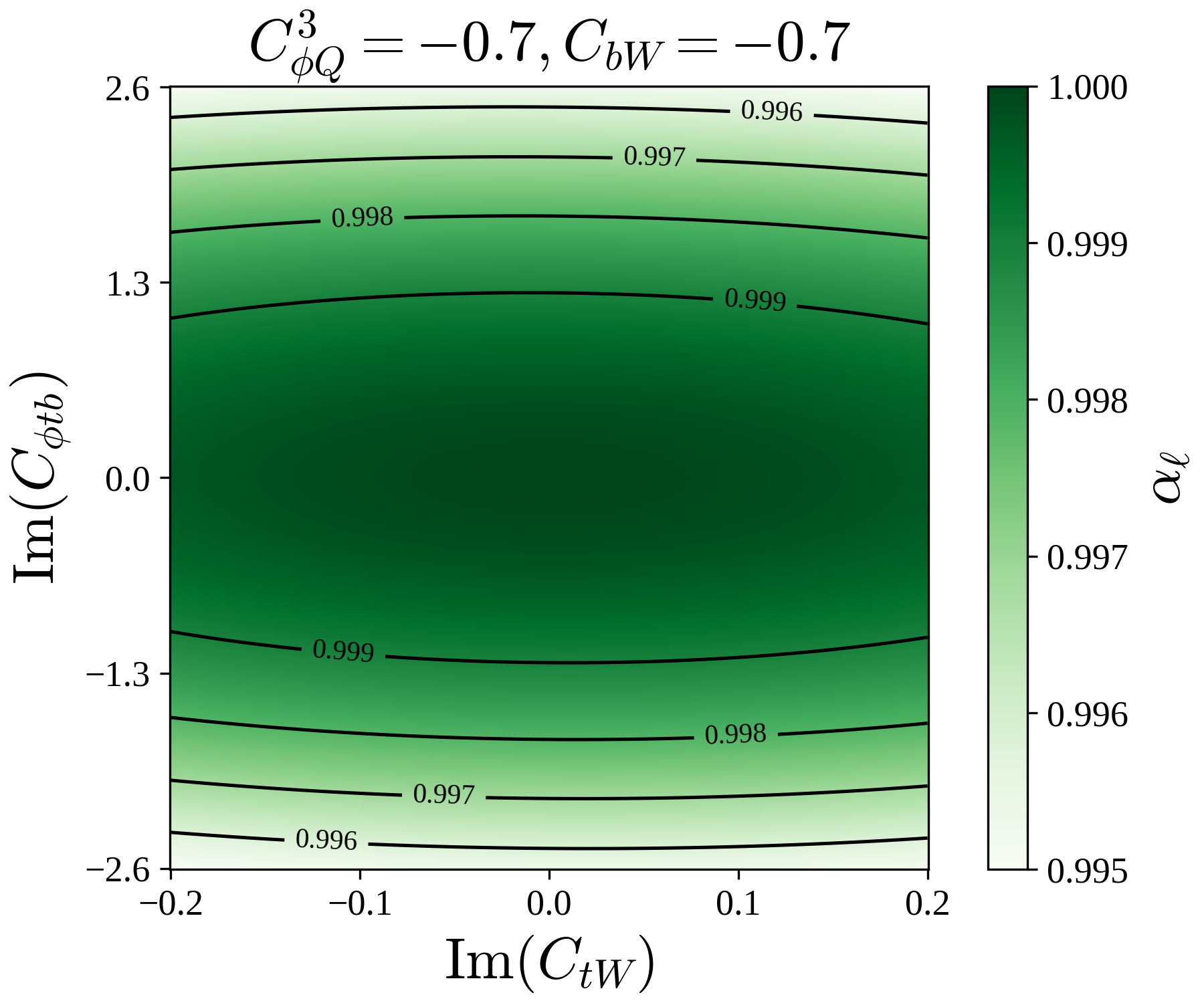}
\caption{Dependence of \(w_s^{\rm CPV}\), \(v_c\), and \(\alpha_\ell\) on
\(\operatorname{Im}(C_{tW})\) and \(\operatorname{Im}(C_{\phi tb})\), with
\(\operatorname{Re}(C_{tW})=\operatorname{Re}(C_{\phi tb})=0\), and all
remaining Wilson coefficients fixed to the benchmark values indicated above
the corresponding panels. The benchmark points are chosen to illustrate the
largest CP-violating effects within the reference ranges considered.}
\label{fig:v_w_sap_2D_Wmax}
\end{figure}

We next consider benchmark configurations in which all the relevant Wilson
coefficients are simultaneously non-zero. In each scan, the coefficients
not shown on the axes are fixed to the benchmark values indicated above the
corresponding panels. These values are chosen within reference intervals
motivated by the experimental constraints summarised in
Tab.~\ref{tab:EFT-op-limits}, but are not intended to represent a statistically
correlated allowed point.

In Fig.~\ref{fig:v_w_sap_2D_Wmax}, the fixed benchmark is chosen to
illustrate the largest CP-violating effects within the reference ranges
considered, while \(\operatorname{Im}(C_{tW})\) and
\(\operatorname{Im}(C_{\phi tb})\) are varied simultaneously over intervals
symmetric about zero. Within the scanned region, \(w_s^{\rm CPV}\) reaches absolute values of
order \(0.3\), while \(\lvert v_c\rvert\) remains below approximately
\(0.07\). The charged-lepton spin-analysing power is much less sensitive,
with \(\lvert\alpha_\ell-1\rvert\lesssim 5\times10^{-3}\).
\begin{figure}[t]
    \centering
    \includegraphics[width=0.42\linewidth]{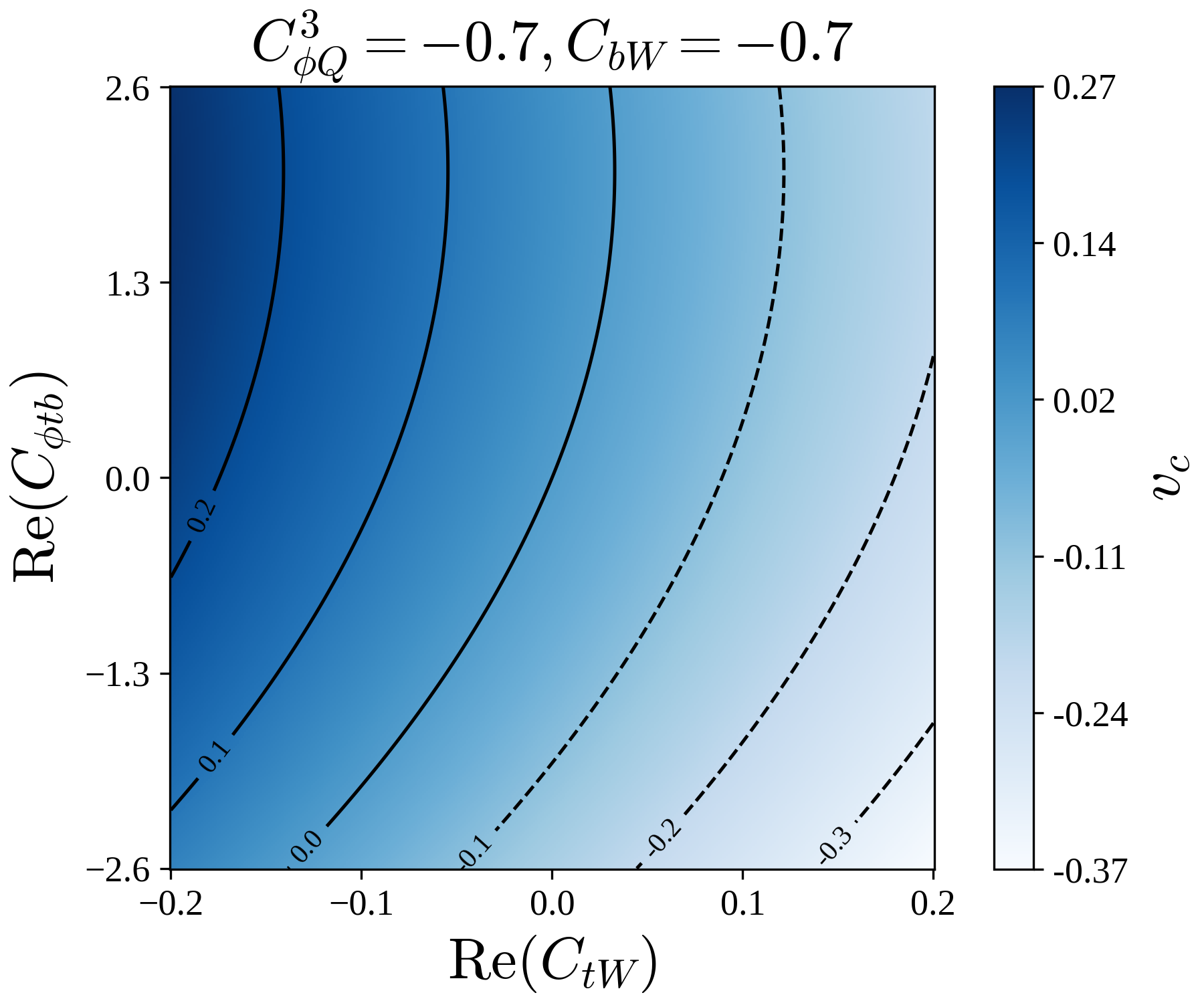} 
     \includegraphics[width=0.42\linewidth]{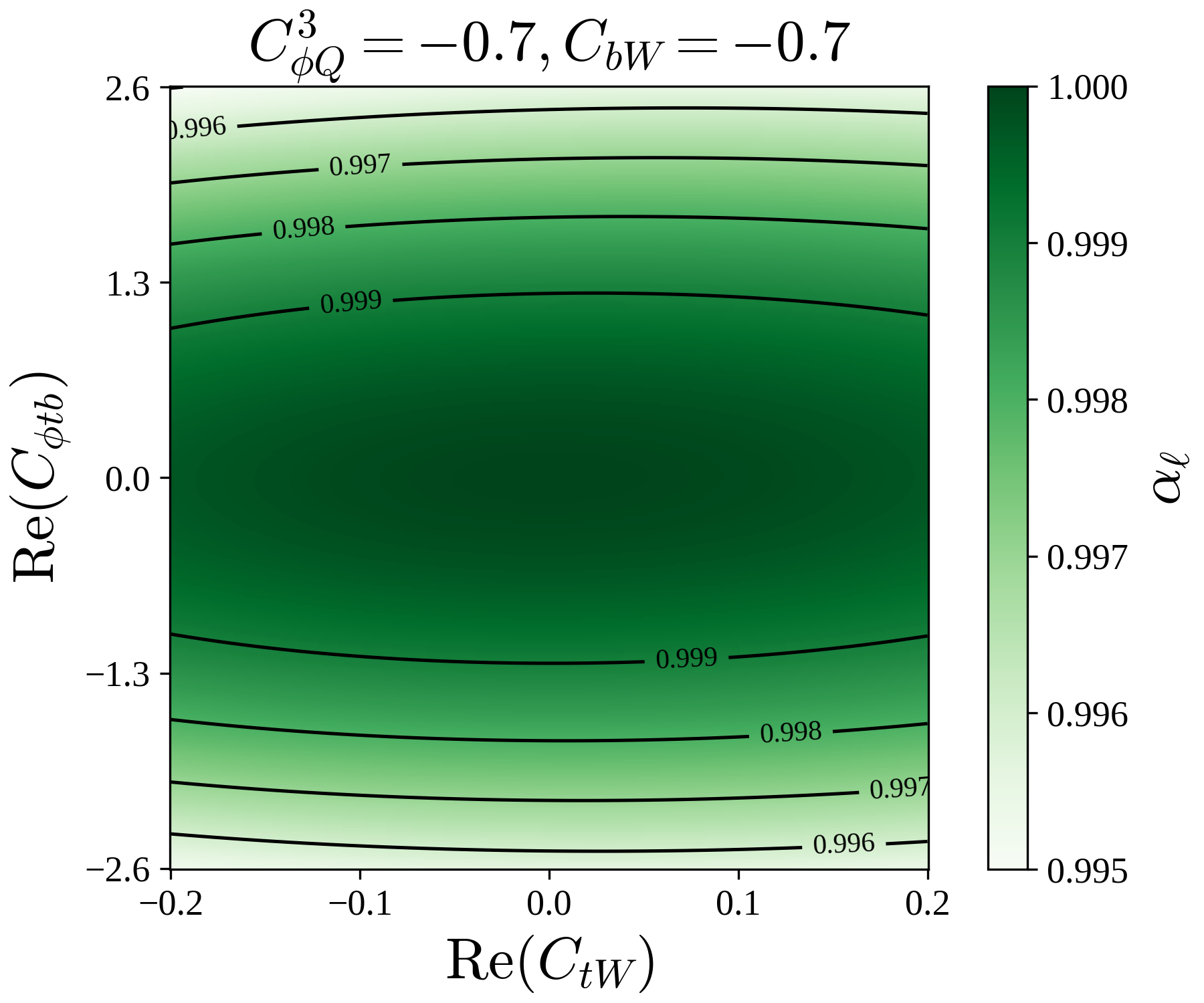}
\caption{Dependence of \(v_c\) and \(\alpha_\ell\) on
\(\operatorname{Re}(C_{tW})\) and \(\operatorname{Re}(C_{\phi tb})\), with
\(\operatorname{Im}(C_{tW})=\operatorname{Im}(C_{\phi tb})=0\) and all
remaining Wilson coefficients fixed to the benchmark values indicated above
the corresponding panels. The benchmark points are chosen to illustrate the
largest CP-even effects within the reference ranges considered. In this
configuration, \(w_s^{\rm CPV}=0\).}

    \label{fig:v_w_sap_2D_Vmax}
\end{figure}
In Fig.~\ref{fig:v_w_sap_2D_Vmax}, the remaining coefficients are fixed to
a benchmark chosen to illustrate the largest CP-even effects within the
reference ranges considered, while \(\operatorname{Re}(C_{tW})\) and
\(\operatorname{Re}(C_{\phi tb})\) are varied simultaneously. Over the scanned region, \(v_c\) reaches absolute values of order \(0.4\),
whereas the deviation of the charged-lepton spin-analysing power remains
below approximately \(0.5\%\). The coefficient
\(w_s^{\rm CPV}\) vanishes in this CP-conserving configuration. 

Taken together, these benchmark scans show that \(v_c\) and
\(w_s^{\rm CPV}\) can reach absolute values of order a few tenths when
multiple Wilson coefficients are simultaneously non-zero, while the
deviation of \(\alpha_\ell\) from its SM value remains below the percent
level. Variations of comparable
magnitude can, however, already be generated by switching on \(C_{tW}\)
alone, as shown in Fig.~\ref{fig:vc-and-ws}.

The angular distributions used to extract the polarisation and
spin-correlation coefficients retain the same functional form as in the SM,
Eqs.~\eqref{eq:forB}--\eqref{eq:diff-xi_MINUS-SM}.
The only modification is the replacement of the SM charged-lepton
spin-analysing powers by their values in the presence of modified
\(Wtb\) interactions,
 \begin{equation}
    \alpha _{\ell} = \alpha_{\ell}^{\rm SM}+{\delta\alpha_\ell}, \qquad \bar \alpha _{\ell} = \bar{\alpha_\ell}^{\rm SM}+{\delta\bar{\alpha_\ell}}
    \label{eq:alpha_l_SMplusdelta}
 \end{equation}
where the analytic expressions for
\({\delta\alpha_\ell}\) and \({\delta\bar{\alpha_\ell}}\) are given in
App.~\ref{app:analytic-expressions}. Consequently, these distributions can
still be used to extract \(B_a\), \(\bar{B}_b\), \(C_{ab}\), after accounting for the
modified decay analysing powers.

In addition, further two-angle distributions involve
the decay coefficients \(v_c\) and \(w_s^{\rm CPV}\). They are obtained from
the one-angle functions defined in
Eqs.~\eqref{eq:partially-integrated-top-functions_lb}--\eqref{eq:partially-integrated-top-functions}
and
\eqref{eq:partially-integrated-antitop-functions_lb}--\eqref{eq:partially-integrated-antitop-functions}
by retaining one angular variable from each decay phase space and integrating
over all remaining variables.
Where applicable, the resulting distributions occur in pairs corresponding
to the two possible assignments of the retained angles to the top and
antitop decay branches. The two members are related by interchange of the
two spin subsystems, which in our notation amounts to
\[
\text{barred}\leftrightarrow\text{unbarred},
\qquad
a\leftrightarrow b,
\qquad
\mathbf{C}\leftrightarrow \mathbf{C}^{T}.
\]
We label the two members of each pair by \(n\) and \(\bar n\), with
\(n=1,2,3,\ldots\), and display both explicitly below.

To keep the expressions compact, we use the shorthand
\(c_x\equiv\cos x\) and \(s_x\equiv\sin x\) for any angular variable \(x\).
We also recall that
\((\hat a,\hat{\imath}_1,\hat{\imath}_2)\) and
\((\hat b,\hat{\jmath}_1,\hat{\jmath}_2)\) are right-handed orthonormal
triads. The joint angular distributions can then be organised into two
groups according to whether they contain explicit dependence on
\(w_s^{\rm CPV}\).
The first group collects distributions with an explicit dependence on CPV terms:
\begin{align}
 1)\  \frac{4\pi}{\sigma}\frac{d\sigma}{d\phi_{\ell b}^{(a)}dc_{\btheta_{\ell}^{(b)}}}&=1+ B_a\Bigl[\frac{v_c}{4\pi} c_{\phi^{(a)}_{\ell b}}+  \frac{w_s^{  \rm  CPV}}{4\pi}s_{\phi^{(a)}_{\ell b}}\Bigr]+ \bar{\alpha}_\ell\bar{B}_bc_{\bar{\theta}_{l}^{(b)}}+C_{ab}\bar{\alpha}_\ell c_{\btheta_\ell^{(b)}}\Bigl[\frac{v_c}{4\pi} c_{\phi^{(a)}_{\ell b}}+  \frac{w_s^{  \rm  CPV}}{4\pi}s_{\phi^{(a)}_{\ell b}}\Bigr]  \label{eq:dist1} \\[10pt] 
 \bar{1})\ \frac{4\pi}{\sigma}\frac{d\sigma}{dc_{\theta_{\ell}^{(a)}}d\bphi_{\ell b}^{(b)}}&= 1+ \alpha_{\ell}B_ac_{\theta_{\ell}^{(a)}}  + \bar{B}_b\Bigl[\frac{\bar{v}_c}{4\pi} c_{\bphi^{(b)}_{\ell b}}+  \frac{\bar{w}_s^{\rm  CPV}}{4\pi}s_{\bphi^{(b)}_{\ell b}}\Bigr]+C_{ab}\alpha_{\ell}c_{\theta_{\ell}^{(a)}}\Bigl[\frac{\bar{v}_c}{4\pi} c_{\bphi^{(b)}_{\ell b}}+  \frac{\bar{w}_s^{\rm  CPV}}{4\pi}s_{\bphi^{(b)}_{\ell b}}\Bigr] 
\label{eq:dist1s} \\[10pt]
2)\ \frac{4\pi^2}{\sigma}\frac{d\sigma}{d\phi_{\ell b}^{(a)}d\bphi_{\ell}^{(b)}}  &= 1+ B_a\Bigl[\frac{v_c}{4\pi} c_{\phi^{(a)}_{\ell b}}+  \frac{w_s^{  \rm  CPV}}{4\pi}s_{\phi^{(a)}_{\ell b}}\Bigr] 
+\bar{\alpha}_\ell\frac{\pi}{4}\Bigl[\bar{B}_{ j_1}c_{\bphi_{\ell}^{(b)}}+ \bar{B}_{ j_2}s_{\bphi_{\ell}^{(b)}}\Bigr]\nonumber\\[10pt]
&+\bar{\alpha}_\ell\frac{\pi}{4}\Bigl[\frac{v_c}{4\pi} c_{\phi^{(a)}_{\ell b}}+  \frac{w_s^{  \rm  CPV}}{4\pi}s_{\phi^{(a)}_{\ell b}}\Bigr]\Bigl[
   C_{a j_1}c_{\bphi_{\ell}^{(b)}}
 + C_{a j_2}s_{\bphi_{\ell}^{(b)}}
\Bigr]\label{eq:dist2}\\[10pt]
\bar{2})\ \frac{4\pi^2}{\sigma}\frac{d\sigma}{d\phi_{\ell}^{(a)}d\bphi_{\ell b}^{(b)}}  &=1+  \alpha_\ell\frac{\pi}{4} \Bigl[B_{i_1}c_{\phi_{\ell}^{(a)}}+ B_{i_2}s_{\phi_{\ell}^{(a)}}\Bigr] +  \bar{B}_b\Bigl[\frac{\bar{v}_c}{4\pi} c_{\bphi^{(b)}_{\ell b}}+  \frac{\bar{w}_s^{\rm  CPV}}{4\pi}s_{\bphi^{(b)}_{\ell b}}\Bigr] \nonumber\\[10pt]
 &+\alpha_\ell\frac{\pi}{4}\Bigl[\frac{\bar{v}_c}{4\pi} c_{\bphi^{(b)}_{\ell b}}+  \frac{\bar{w}_s^{\rm  CPV}}{4\pi}s_{\bphi^{(b)}_{\ell b}}\Bigr]\Bigl[C_{i_1b}c_{\phi_{\ell}^{(a)}}+ C_{i_2b}s_{\phi_{\ell}^{(a)}}\Bigr]
\label{eq:dist2s} \\[10pt]
3)\ \frac{4\pi^2}{\sigma}\frac{d\sigma}{d\phi_{\ell b}^{(a)}d\bphi_{\ell b}^{(b)}}  &= 1+ B_a\Bigl[\frac{v_c}{4\pi} c_{\phi^{(a)}_{\ell b}}+  \frac{w_s^{  \rm  CPV}}{4\pi}s_{\phi^{(a)}_{\ell b}}\Bigr]  + \bar{B}_b\Bigl[\frac{\bar{v}_c}{4\pi} c_{\bphi^{(b)}_{\ell b}}+  \frac{\bar{w}_s^{  \rm  CPV}}{4\pi}s_{\bphi^{(b)}_{\ell b}}\Bigr] \nonumber\\[10pt]
&+C_{ab}\Bigl[\frac{v_c}{4\pi} c_{\phi^{(a)}_{\ell b}}+  \frac{w_s^{  \rm  CPV}}{4\pi}s_{\phi^{(a)}_{\ell b}}\Bigr]\Bigl[\frac{\bar{v}_c}{4\pi} c_{\bphi^{(b)}_{\ell b}}+  \frac{\bar{w}_s^{\rm  CPV}}{4\pi}s_{\bphi^{(b)}_{\ell b}}\Bigr]\,.\label{eq:dist3}
\end{align}
The second group has no direct dependence on CP-violating decay terms, apart
from possible contributions through \(\alpha_\ell\):
\begin{align}
4)\ \frac{4\pi}{\sigma
   }\frac{d\sigma}{dc_{\theta_{\ell}^{(a)}}d\bphi_{\ell}^{(b)}}  &= 1+\alpha_{\ell}B_ac_{\theta_{\ell}^{(a)}}+ \bar{\alpha}_\ell\frac{\pi}{4}\Bigl[\bar{B}_{j_1}c_{\bphi_{\ell}^{(b)}}+ \bar{B}_{j_2}s_{\bphi_{\ell}^{(b)}}\Bigr]+\alpha_{\ell}\bar{\alpha}_\ell c_{\theta_{\ell}^{(a)}} \frac{\pi}{4}\Bigl[C_{aj_1}c_{\bphi_{\ell}^{(b)}}+ C_{aj_2}s_{\bphi_{\ell}^{(b)}}\Bigr] \label{eq:dist4}  \\[10pt]
\bar{4})\ \frac{4\pi}{\sigma}\frac{d\sigma}{d\phi_{\ell}^{(a)}dc_{\btheta_{\ell}^{(b)}}} &=1+ \alpha_\ell\frac{\pi}{4} \Bigl[B_{i_1}c_{\phi_{\ell}^{(a)}}+ B_{i_2}s_{\phi_{\ell}^{(a)}}\Bigr]+\bar{\alpha}_\ell\bar{B}_bc_{\bar{\theta_{\ell}}^{(b)}}+\bar{\alpha}_\ell\alpha_\ell c_{\bar{\theta_{\ell}}^{(b)}} \frac{\pi}{4}\Bigl[C_{i_1b}c_{\phi_{\ell}^{(a)}}+ C_{i_2b}s_{\phi_{\ell}^{(a)}}\Bigr] \label{eq:dist4s} \\[10pt]
5)\ \frac{4\pi^2}{\sigma}\frac{d\sigma}{d\phi_{\ell}^{(a)}d\bphi_{\ell}^{(b)}} &=  1+ \alpha_\ell\frac{\pi}{4} \Bigl[B_{i_1}c_{\phi_{\ell}^{(a)}}+ B_{i_2}s_{\phi_{\ell}^{(a)}}\Bigr]+\bar{\alpha}_\ell\frac{\pi}{4}\Bigl[\bar{B}_{j_1}c_{\bphi_{\ell}^{(b)}}+ \bar{B}_{j_2}s_{\bphi_{\ell}^{(b)}}\Bigr]\nonumber  \\[10pt]
     &+ \alpha_\ell\bar{\alpha}_\ell\frac{\pi^2}{16}\Bigl[C_{i_1j_1}c_{\phi_{\ell}^{(a)}}c_{\bphi_{\ell}^{(b)}}+ C_{i_2j_1}s_{\phi_{\ell}^{(a)}}c_{\bphi_{\ell}^{(b)}}+ C_{i_1j_2}c_{\phi_{\ell}^{(a)}}s_{\bphi_{\ell}^{(b)}}+ C_{i_2j_2}s_{\phi_{\ell}^{(a)}}s_{\bphi_{\ell}^{(b)}}\Bigr]\, . \label{eq:dist5}
\end{align}
The distributions in this group retain their SM functional form. Modified
decay interactions enter only through the spin-analysing powers, whereas the
production dependence remains encoded in \(B_i\), \(\bar B_j\), and
\(C_{ij}\).

The distributions in Eqs.~\eqref{eq:dist1}--\eqref{eq:dist2s} already
contain non-trivial angular dependences in the SM, although new 
\(Wtb\) interactions introduce additional structures through
\(v_c\) and \(w_s^{\rm CPV}\). By contrast, the distribution in Eq.~\eqref{eq:dist3} becomes non-trivial
only in the presence of new decay contributions and therefore provides a
particularly clean null test for anomalous \(Wtb\) interactions within
the framework considered here. However, for production processes in which
the relevant polarisation components vanish, the new contribution to
Eq.~\eqref{eq:dist3} enters only through products of decay coefficients and
is therefore quadratic and potentially suppressed. In such cases, the
distributions in Eqs.~\eqref{eq:dist1}--\eqref{eq:dist2s} provide
complementary sensitivity. The phenomenologically most relevant
distributions will be discussed in greater detail in the next section.

Further integration over one of the retained angular variables yields the
corresponding one-dimensional distributions. In addition to those already
given in Eqs.~\eqref{eq:forB} and \eqref{eq:forBbar}, the remaining
one-dimensional distributions are
\begin{align}
   &  6) \ \frac{2\pi}{\sigma}\frac{d\sigma}{d\phi_{\ell}^{(a)}} =  1+ \alpha_\ell\frac{\pi}{4} \Bigl[B_{i_1}\cos\phi_{\ell}^{(a)}+ B_{i_2}\sin\phi_{\ell}^{(a)}\Bigr]\,,\label{eq:dist6}\\
    & \bar{6}) \ \frac{2\pi}{\sigma}\frac{d\sigma}{d \bar\phi_{\ell}^{(b)}} =  1+ \bar \alpha_\ell\frac{\pi}{4} \Bigl[\bar{B}_{j_1}\cos\bar\phi_{\ell}^{(b)}+ \bar{B}_{j_2}\sin\bar\phi_{\ell}^{(b)}\Bigr]\,, \label{eq:dist6s}\\
     & 7) \  \frac{2\pi}{\sigma}\frac{d\sigma}{d\phi_{\ell b}^{(a)}}  = 1+ B_a\Bigl[\frac{v_c}{4\pi} \cos\phi^{(a)}_{\ell b}+  \frac{w_s^{ \rm CPV}}{4\pi}\sin\phi^{(a)}_{\ell b}\Bigr]\,,\label{eq:dist7}\\
   &  \bar{7}) \  \frac{2\pi}{\sigma}\frac{d\sigma}{d\bar\phi_{\ell b}^{(b)}}  = 1+ \bar{B}_b\Bigl[\frac{\bar v_c}{4\pi} \cos\bar\phi^{(b)}_{\ell b}+  \frac{\bar w_s^{ \rm CPV}}{4\pi}\sin\bar\phi^{(b)}_{\ell b}\Bigr] \,,\label{eq:dist7s}
\end{align}
which are non-trivial only if the corresponding 
polarisations, $\mathbf{B}$ or $\mathbf{\bar{B}}$, 
are non-zero. 
The distributions in Eqs.~\eqref{eq:dist6} and
\eqref{eq:dist6s} retain a SM-like angular structure coming from the transverse-polarisation components of the top and
antitop quark, respectively. By contrast, the distributions in Eqs.~\eqref{eq:dist7}
and \eqref{eq:dist7s} are flat in the SM but acquire angular dependence in the presence of new physics decay contributions. 
Depending on the CP properties of the underlying interaction, these
contributions generate cosine or sine modulations, corresponding to
CP-conserving and CP-violating effects, respectively. Related
distributions have been considered in the literature as probes of CP
 violation in the decay of polarised top quarks \cite{Baumgart:2012ay,Bernreuther:2010ny,Bernreuther:2015yna}. If the top or
antitop quark is unpolarised along the relevant direction, the
corresponding distributions remain flat.

This completes our discussion of the angular distributions. In the next
section, we investigate their phenomenological sensitivity to anomalous
\(Wtb\) interactions.


\section{Discussion}
\label{sec:discussion}

The angular distributions derived in Sec.~\ref{sec:tomography} provide
complementary information on the possible origin of CP-odd effects. Some of
these distributions retain the same functional form as in the SM decay
scenario and therefore remain directly applicable to the tomographic
reconstruction of the production density matrix. The impact of new physics
on this class of distributions was discussed in the previous section. By
contrast, other distributions develop additional azimuthal structures when
the \(Wtb\) decay vertex is modified, particularly in the presence of
CP-violating phases. In this section, we investigate these distributions in
detail, beginning with the two-angle and single-angle distributions and
subsequently introducing several informative one-dimensional projections.

To illustrate the resulting modulations, we present projected distributions
for \(pp\to t\bar t\) production at the LHC with
\(\sqrt{s}=13~\mathrm{TeV}\), as well as for \(e^+e^-\to t\bar t\)
production at a center-of-mass energy of \(365~\mathrm{GeV}\), as proposed
at the FCC-ee \cite{FCC:2025lpp,FCC:2025uan}. For the LHC setup, we consider two phase-space regions, motivated by the experimental binning adopted in Ref.~\cite{CMS:2024zkc}. These regions are defined in terms of the invariant mass of the top-quark pair, \(m_{t\bar t}\), and the absolute value of the cosine of the partonic scattering angle, \(\lvert\cos\theta\rvert\):
\begin{align}
     &\mathrm{bin}\  1: 300~ \mathrm{ GeV}<m_{t\bt}< 400~ \mathrm{ GeV}, \ 0.4<|\cos\theta|< 0.7 \label{eq:bin1} \\
     &\mathrm{bin}\  2: m_{t\bt}> 800~ \mathrm{ GeV},\  |\cos\theta|< 0.4\,. \label{eq:bin2}
\end{align} 
For each bin, we use the Fano--Bloch coefficients extracted from Monte Carlo
simulations, performed with
{\sc MadGraph5\_aMC@NLO}~\cite{Alwall:2014hca}, recently extended with a framework dedicated to the computation of spin density matrices~\cite{Durupt:2025wuk}. Further details of the simulation setup are
provided in the companion paper \cite{Lamba:2026jfm}. The simulations are
performed at leading order in QCD, where the individual top and antitop polarisations vanish, while non-trivial spin-correlations remain present.

The threshold and high-\(m_{t\bar t}\) regions are selected on the basis of
their experimental relevance. The threshold region corresponds to the
phase-space domain in which the first observation of quantum entanglement
in top-quark pairs was reported by the ATLAS Collaboration
\cite{ATLAS:2023fsd}. The high-mass region defined in
Eq.~\eqref{eq:bin2} corresponds instead to the region in which the CMS
Collaboration obtained its most significant entanglement signal
\cite{CMS:2024zkc}. In each region, we choose the pair of spin-quantisation
axes \(a\) and \(b\), together with the \(\lvert\cos\theta\rvert\) interval
in the threshold case, so as to maximise the magnitude of the
spin-correlation coefficient \(\lvert C_{ab}\rvert\) entering the
distributions considered below, following the results reported in
Ref.~\cite{CMS:2024zkc}.

For \(t\bar t\) production in \(e^+e^-\) collisions, we use the theoretical
predictions presented in Ref.~\cite{Lamba:2026jfm} and consider
distributions inclusive in the production angle \(\theta\). In contrast to
the hadron-collider case, top and antitop quarks produced in \(e^+e^-\)
annihilation are individually polarised already at Born level
\cite{Arens:1992wh}. This makes the lepton-collider setup particularly well
suited for studying single-angle distributions. The numerical values of the
polarisation and spin-correlation coefficients used in the figures are
collected in App.~\ref{app:fano_num}, respectively in Tabs.~\ref{tab:values_distrib_bins_1}-\ref{tab:values_distrib_bins_2} for $pp\rightarrow t\bt$ and in Tab.~\ref{tab:spin-pol-corr-ee-fig} for $e^+e^-\rightarrow t\bt$.

For the Wilson coefficients \(C_i\) entering the \(Wtb\) vertex in
Eq.~\eqref{eq:vertex-BSM-decay-top}, we consider two benchmark scenarios.
The first, denoted by ``CPV'', is chosen to maximise
\(w_s^{\rm CPV}\), under the assumption that each Wilson coefficient is
either purely real or purely imaginary. The second, denoted by ``CPC'',
satisfies \(w_s^{\rm CPV}=0\) and is chosen to maximise \(v_c\) under
the same assumption. Finally, the SM configuration corresponds to
\(C_i=0\) for all coefficients: 
\begin{align}
    \{C_i\}= \mathrm{CPV}:\ &\  C^{3}_{\phi Q}= -0.7,\ 
        C_{\phi tb}= -2.6\, i,\  C_{bW}= -0.7, \ C_{tW}= -0.2\, i  \label{eq:wc_set_CPV}\\
   \{C_i\}= \mathrm{CPC}:\ &\ C^{3}_{\phi Q}= -0.7,\  C_{\phi tb}= 2.0,\  C_{bW}= -0.7,\ C_{tW}=  -0.2 \label{eq:wc_set_CPC}\\
     \{C_i\}= \mathrm{SM}:\ &\ C^{3}_{\phi Q}= 0,\  C_{\phi tb}=0,\  C_{bW}= 0,\ C_{tW}=  0. \label{eq:wc_set_SM}
\end{align}
For
\(\Lambda=1~\mathrm{TeV}\), the corresponding values of
\(w_s^{\mathrm{CPV}}\), \(v_c\), and \(\alpha_\ell\) in the SM, CPV, and
CPC scenarios are reported in Tab.~\ref{tab:benchmark_decay_parameters}.

The numerical distributions presented below are intended to illustrate
the characteristic angular structures and the relative sizes of the
different contributions, rather than to provide projections of
experimental sensitivity. A realistic assessment would require a
dedicated analysis including event yields, detector acceptance,
reconstruction efficiencies, backgrounds, angular resolution, and
statistical and systematic uncertainties. In particular, modulations at
the \(10^{-4}\) level are unlikely to be experimentally accessible, while
even percent-level effects may be challenging and must be assessed in a
collider- and observable-specific analysis.
\subsection{Double differential distributions sensitive to CP violation in top or antitop decays} \label{sec:decay-sensitive-distributions}

Among the two-dimensional distributions derived in the previous section,
only the five distributions collected in
Eqs.~\eqref{eq:dist1}--\eqref{eq:dist3} are directly sensitive to
CP-violating effects induced by the modified decay vertex in
Eq.~\eqref{eq:vertex-BSM-decay-top}. These effects enter through terms
proportional to \(\sin\phi_{\ell b}^{(a)}\) or
\(\sin\bar{\phi}_{\ell b}^{(b)}\), and therefore vanish when the
corresponding azimuthal angle is integrated over. The same distributions
are also sensitive to CP-even new physics contributions to the decay
vertex, which instead generate terms proportional to
\(\cos\phi_{\ell b}^{(a)}\) or
\(\cos\bar{\phi}_{\ell b}^{(b)}\).
In the SM, these distributions are independent of
\(\phi_{\ell b}^{(a)}\) and \(\bar{\phi}_{\ell b}^{(b)}\), so that neither
sine nor cosine modulations are present. Consequently, within the framework considered here, any non-trivial
dependence on either azimuthal angle signals the presence of
new physics contributions induced by the operators modifying the
\(Wtb\) decay vertex in Eq.~\eqref{eq:vertex-BSM-decay-top}. Moreover, the angular structure provides
information on the CP nature of these interactions: the cosine modulation
is controlled by the CP-even coefficient \(v_c\), whereas the sine
modulation is controlled by the CP-odd coefficient
\(w_s^{\rm CPV}\).

More generally, a configuration with
\(w_s^{\rm CPV}\neq 0\) is generically accompanied by
\(v_c\neq 0\). Indeed, the CP-violating couplings contribute to
\(w_s^{\rm CPV}\) through interference terms that are linear in their
imaginary parts, while their quadratic contributions are CP-even and enter
\(v_c\). Therefore, even when the anomalous \(Wtb\) interaction contains
only CP-violating couplings, it generally induces both sine and cosine
modulations. A configuration with
\(w_s^{\rm CPV}\neq 0\) but \(v_c=0\) can arise only through a specific
cancellation among the different contributions to \(v_c\). Conversely,
in the absence of CP-violating phases in the \(Wtb\) vertex,
\(w_s^{\rm CPV}=0\), and the anomalous azimuthal dependence is entirely
described by the cosine modulation. 

For a generic dependence on \(\phi_{\ell b}^{(a)}\) of the form
\begin{equation}
    a_1
    +a_2 \cos\phi_{\ell b}^{(a)}
    +a_3 \sin\phi_{\ell b}^{(a)},
\end{equation}
the reflection
\begin{equation}
    \phi_{\ell b}^{(a)}
    \longrightarrow
    2\pi-\phi_{\ell b}^{(a)}
\end{equation}
leaves the constant and cosine terms invariant while reversing the sign
of the sine term. The absence of the CP-odd contribution therefore implies
an exact symmetry of the distribution about the
\(\phi_{\ell b}^{(a)}=\pi\) axis. The same conclusion applies to
distributions depending on \(\bar{\phi}_{\ell b}^{(b)}\). Conversely, a
breaking of this reflection symmetry directly probes the sine modulation
and hence CP violation in the decay vertex. Barring accidental
cancellations in \(w_s^{\rm CPV}\), this symmetry therefore provides a
useful marker for the absence of CP-violating phases in the modified
\(Wtb\) interaction. The asymmetry becomes more pronounced when both
\(\lvert w_s^{\rm CPV}\rvert\) and the magnitude of the relevant
production polarisation or spin-correlation coefficient are large.
Accordingly, the sensitivity of a given distribution to CP violation
depends not only on the decay vertex, but also on the polarisation and
spin-correlation structure of the production process.

We now examine the five distributions ~\eqref{eq:dist1}--\eqref{eq:dist3} in detail, highlighting the
different terms through which the modified decay vertex reshapes their
angular dependence. In the following expressions, the new CP-even
structures are shown in brown, while the CP-violating structures are shown
in blue. Although it will not always be displayed explicitly, the charged-lepton
spin-analysing powers are themselves modified by the anomalous \(Wtb\)
interaction. Their structure may be written as
\begin{equation}
    \alpha_{\ell}
    =
    \alpha_{\ell}^{\rm SM}
    +\textcolor{Bittersweet}{\delta\alpha_{\ell}},
    \qquad
    \bar{\alpha}_{\ell}
    =
    \bar{\alpha}_{\ell}^{\rm SM}
    +\textcolor{Bittersweet}{\delta\bar{\alpha}_{\ell}}.
    \label{eq:sap_BSM}
\end{equation}
For brevity, this decomposition will be left implicit in the expressions
below.

For the distributions that already possess a non-trivial dependence on the
remaining angular variables in the SM, we also introduce SM-subtracted
combinations. Removing the SM contribution makes the structures induced by
the modified decay vertex more transparent and enhances the visibility of
their characteristic angular modulations.
\subsubsection{Charged-lepton polar angle and the lepton-bottom azimuthal angle distributions} \label{sect:dist1_discussion}

The first CP-sensitive pair consists of distributions 
$1)$ and~$\bar{1})$ defined in
Eqs.~\eqref{eq:dist1} and~\eqref{eq:dist1s}, respectively. We discuss distribution $1)$ explicitly; the charge-conjugate discussion follows by exchanging the top and antitop branches. For compactness, and to introduce a short notation for the distributions
shown in the figures, we define
\begin{equation}
    \mathcal{W}_{i}^{ab}(x,y)
    \equiv
    \frac{\Delta x\,\Delta y}{\sigma}
    \frac{d^{2}\sigma}{dx\,dy},
    \qquad
    \mathcal{W}_{i}^{a}(x)
    \equiv
    \frac{\Delta x}{\sigma}
    \frac{d\sigma}{dx},
\end{equation}
for the two-angle and one-angle distributions, respectively. Here,
\(x\) and \(y\) denote the angular variables, \(\Delta x\) and
\(\Delta y\) their full ranges, and \(a,b\in\{n,r,k\}\) specify the
chosen polar axes. The subscript \(i\) labels the distributions introduced above. With this convention, a flat
distribution is equal to unity. With this notation, distribution 1) reads
\begin{align}
    \mathcal{W}_{1}^{ab}(\phi_{\ell b}^{(a)},\cos\btheta_{\ell}^{(b)}) &= 1+ B_a\Bigl[\textcolor{Bittersweet}{\frac{v_c}{4\pi} \cos\phi^{(a)}_{\ell b}}+  \textcolor{teal}{\frac{w_s^{ \rm CPV}}{4\pi}\sin\phi^{(a)}_{\ell b}}\Bigr]  + \bar{\alpha_\ell}\bar{B}_b\cos\bar{\theta_\ell}^{(b)}\nonumber\\
    &+\bar{\alpha_\ell}\cos\btheta_\ell^{(b)}C_{ab}\Bigl[\textcolor{Bittersweet}{\frac{v_c}{4\pi}  \cos\phi^{(a)}_{\ell b}}+ \textcolor{teal}{ \frac{w_s^{ \rm  CPV}}{4\pi}\sin\phi^{(a)}_{\ell b}}\Bigr]\,.
    \label{eq:dist1_col}
\end{align}
Modifications of the antitop decay vertex enter this distribution only
through the spin-analysing power \(\bar{\alpha}_{\ell}\) and do not generate
an additional dependence on \(\phi_{\ell b}^{(a)}\). By contrast, the
CP-odd sine modulation associated with the top decay can arise from the
interference between the SM and dimension-six decay amplitudes and
therefore appears already at \(\mathcal{O}(\Lambda^{-2})\). Importantly,
this sensitivity does not require non-zero single-spin polarisations.
Even when \(B_a=\bar{B}_b=0\), the contribution proportional to
\begin{equation}
C_{ab}\cos\bar{\theta}_{\ell}^{(b)}
\,\textcolor{teal}{w_s^{\rm CPV}\sin\phi_{\ell b}^{(a)}}
\end{equation}
survives provided that the spin-correlation coefficient \(C_{ab}\) is
non-zero. Distribution~$1)$ is therefore sensitive to CP violation in the
top-quark decay vertex already at \(\mathcal{O}(\Lambda^{-2})\), including
for production processes in which the top and antitop quarks are
individually unpolarised but remain spin-correlated.

\begin{figure}[t]
    \centering
    \includegraphics[width=0.445\linewidth]{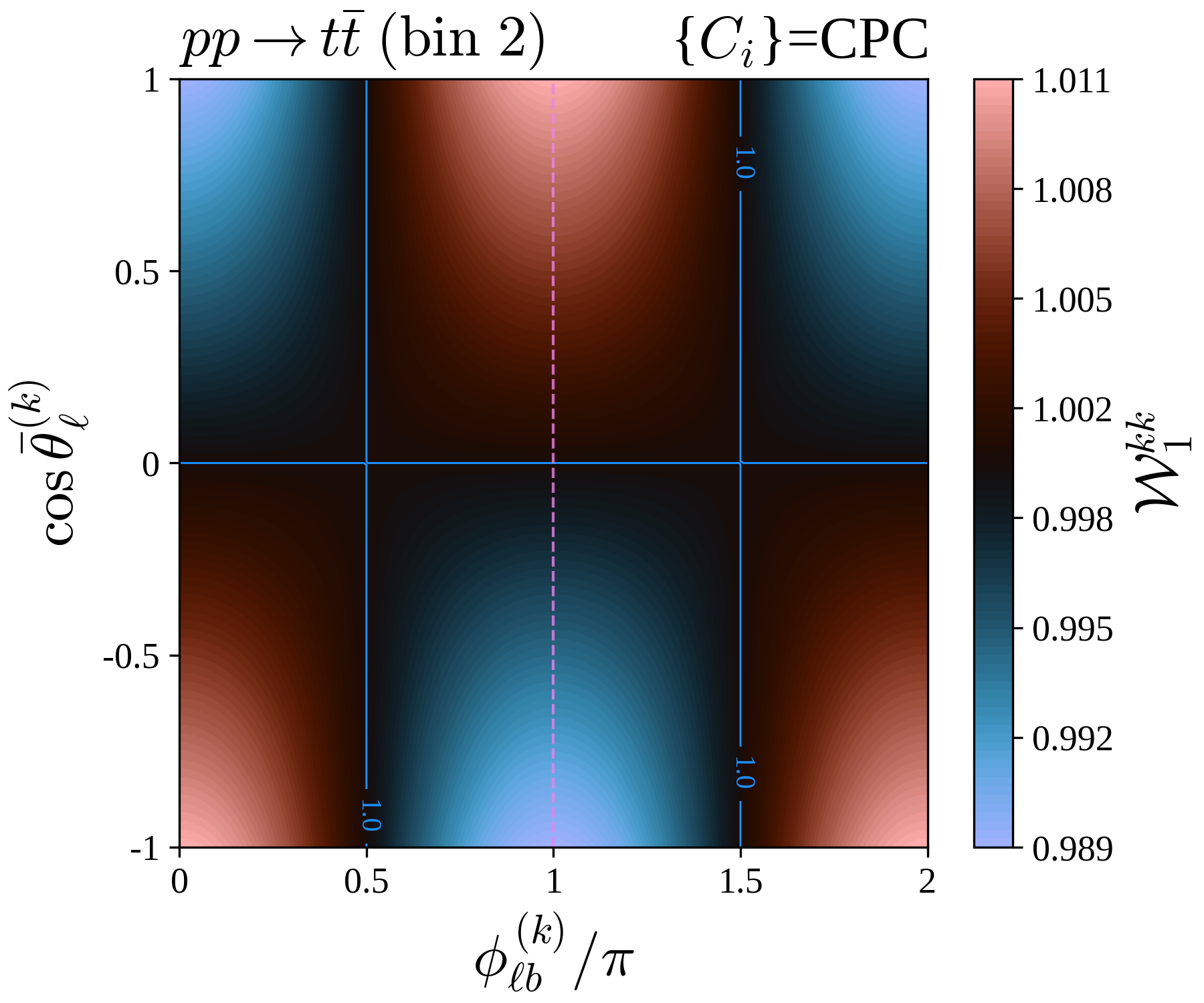}
    \includegraphics[width=0.445\linewidth]{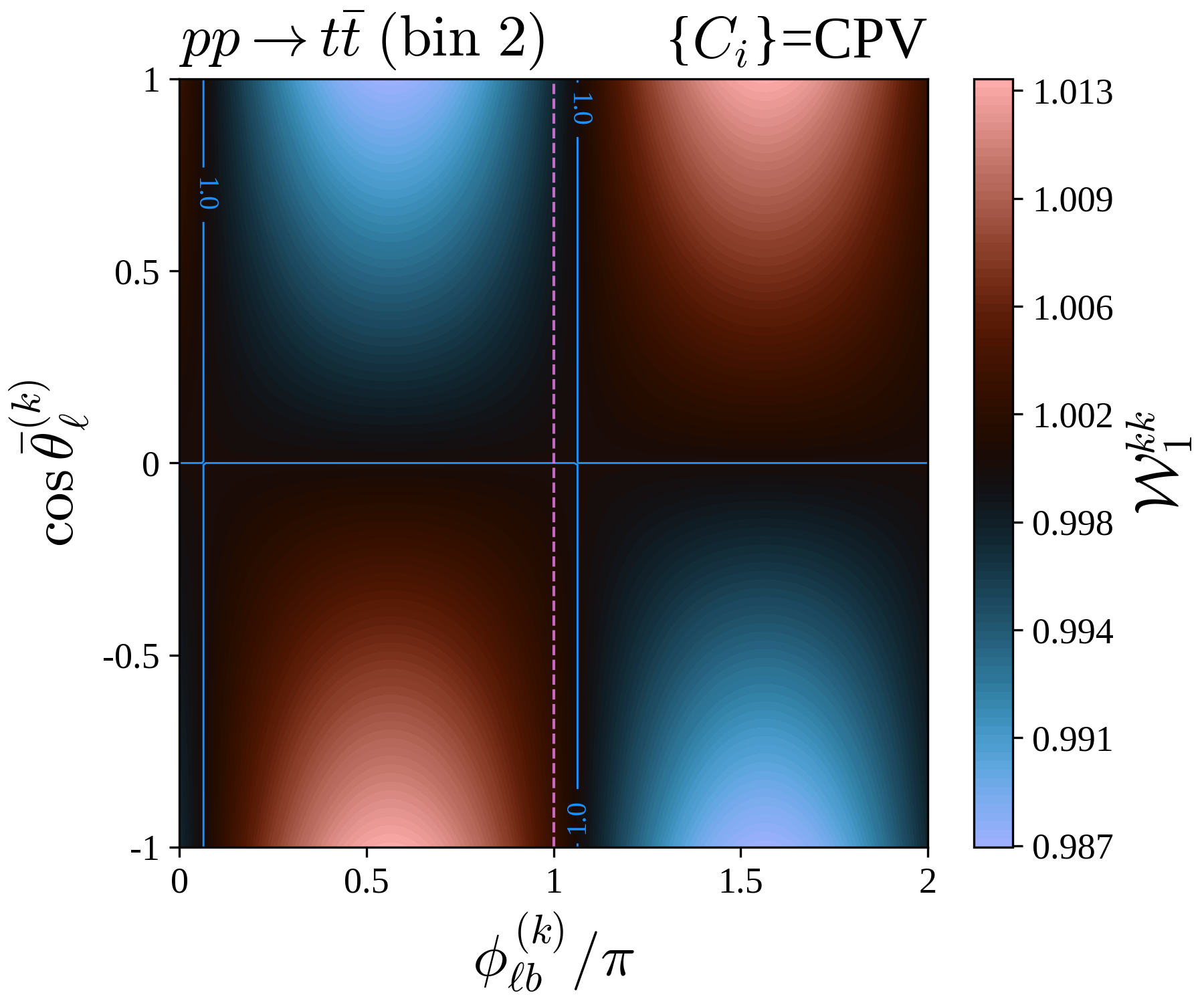}\\
\includegraphics[width=0.445\linewidth]{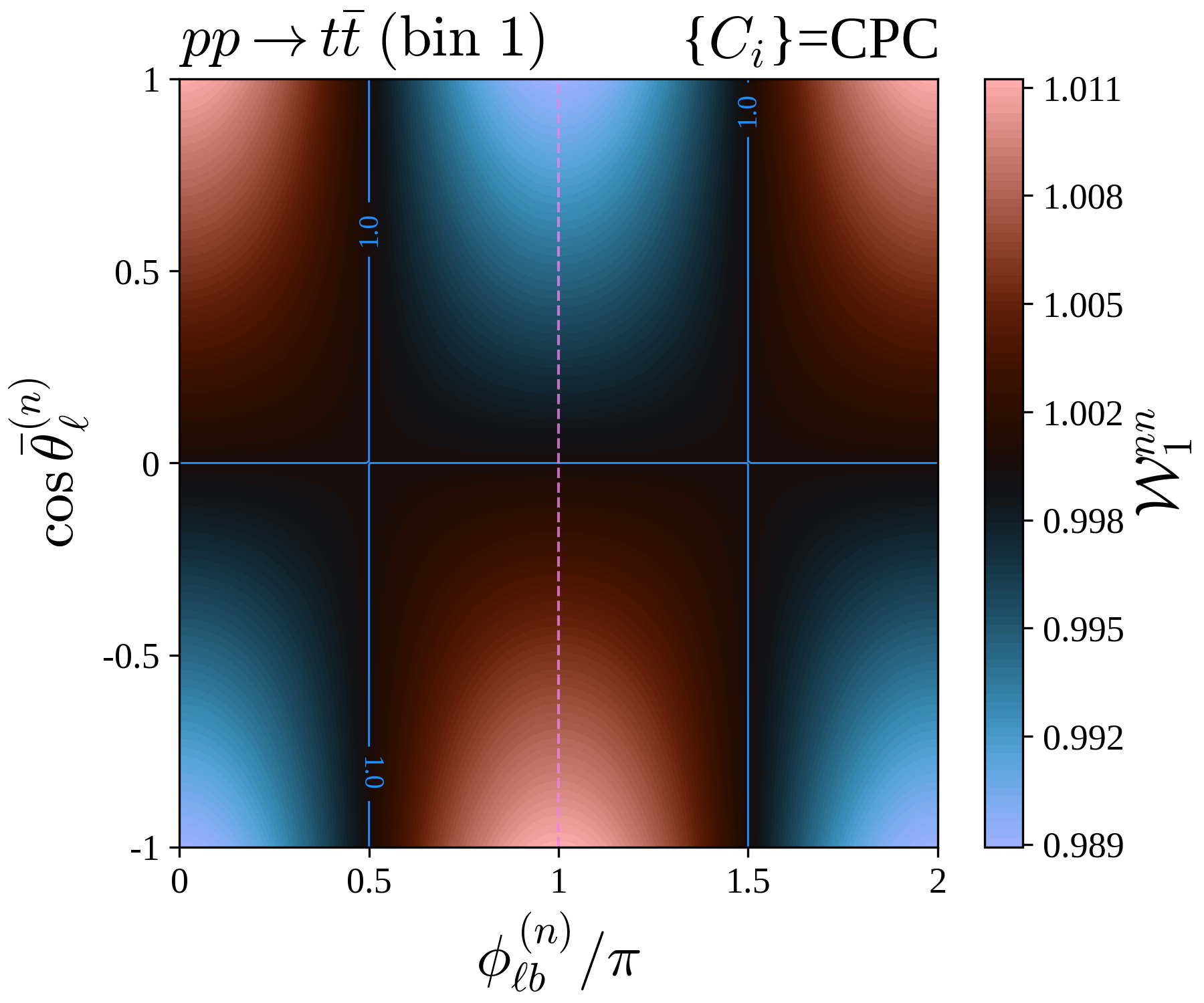}
    \includegraphics[width=0.445\linewidth]{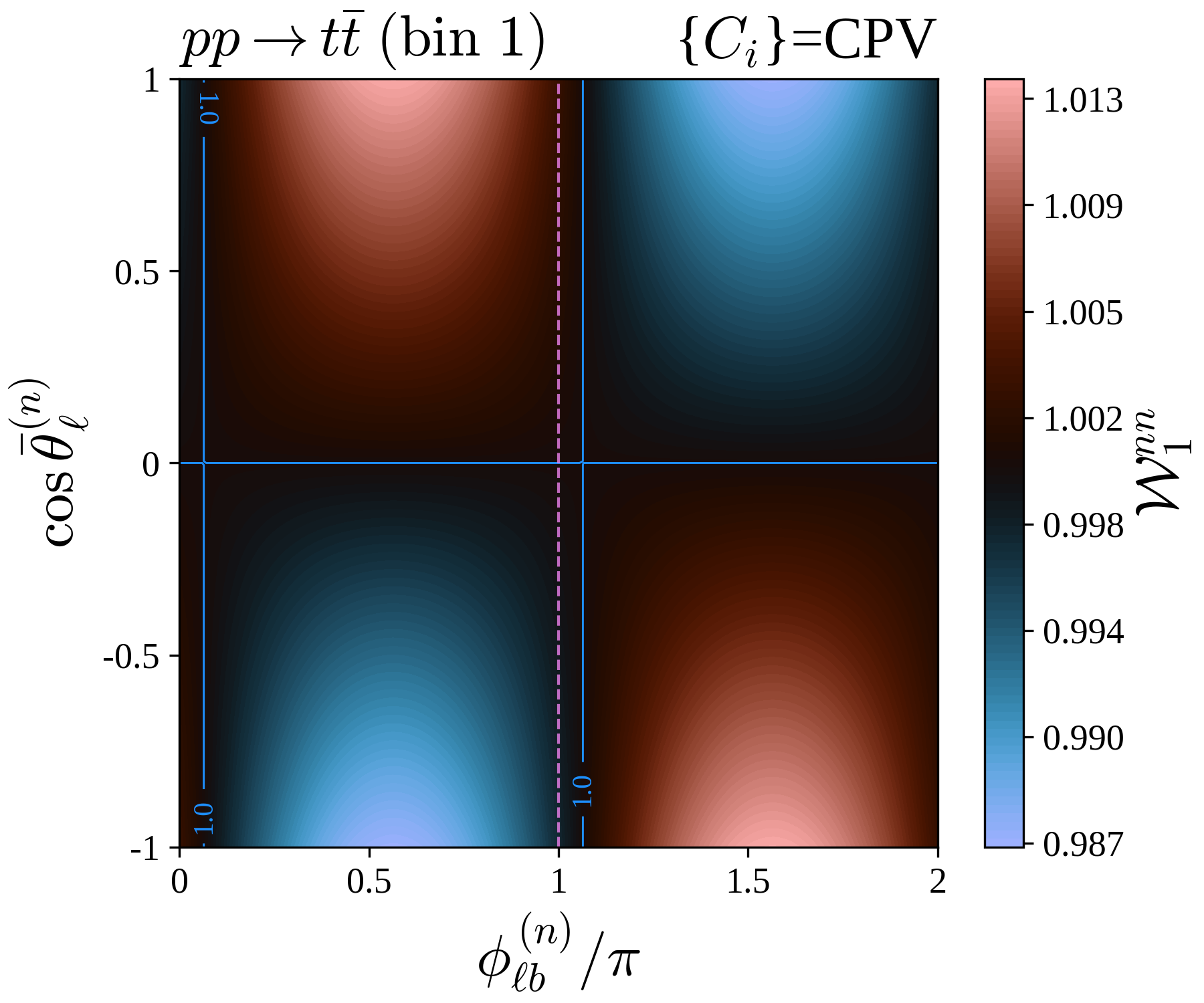}
\caption{
Normalised two-angle distribution
\(\mathcal{W}_{1}^{ab}
(\phi_{\ell b}^{(a)},\cos\bar{\theta}_{\ell}^{(b)})\),
defined in Eq.~\eqref{eq:dist1_col}, for \(pp\to t\bar t\) production at
the LHC, using the polarisation and spin-correlation coefficients extracted
from the simulations. The top row corresponds to \((a,b)=(k,k)\) in
bin~2, while the bottom row corresponds to \((a,b)=(n,n)\) in bin~1.
The left and right columns show the CPC and CPV benchmark scenarios,
respectively, defined in Eqs.~\eqref{eq:wc_set_CPC}
and~\eqref{eq:wc_set_CPV}. The dashed vertical line marks
\(\phi_{\ell b}^{(a)}=\pi\), while the solid blue contour indicates
\(\mathcal{W}_{1}^{ab}=1\).}
\label{fig:dist1_pp}
\end{figure}

\begin{figure}[t]
    \centering
     \includegraphics[width=0.325\linewidth]{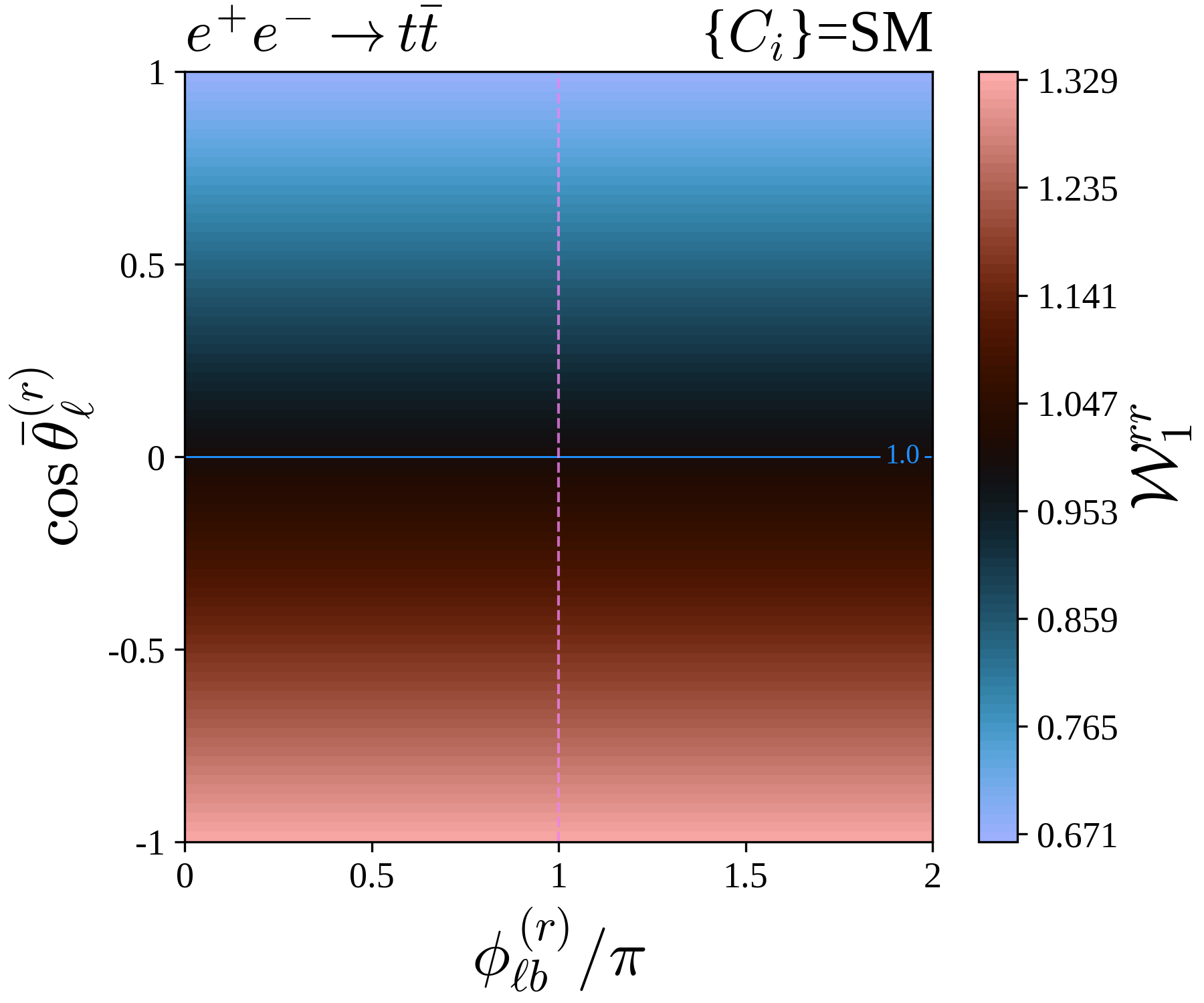}
    \includegraphics[width=0.325\linewidth]{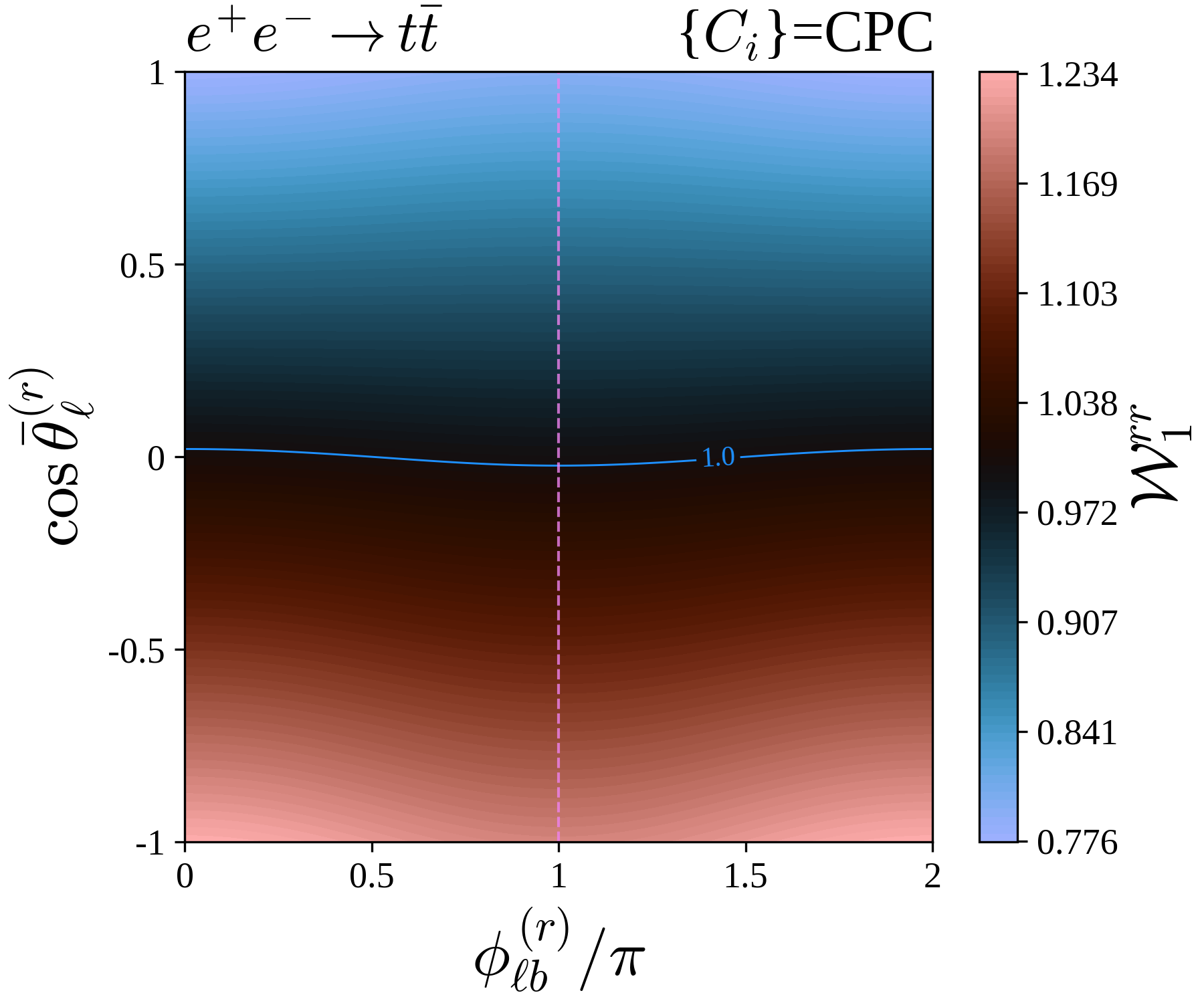} 
    \includegraphics[width=0.325\linewidth]{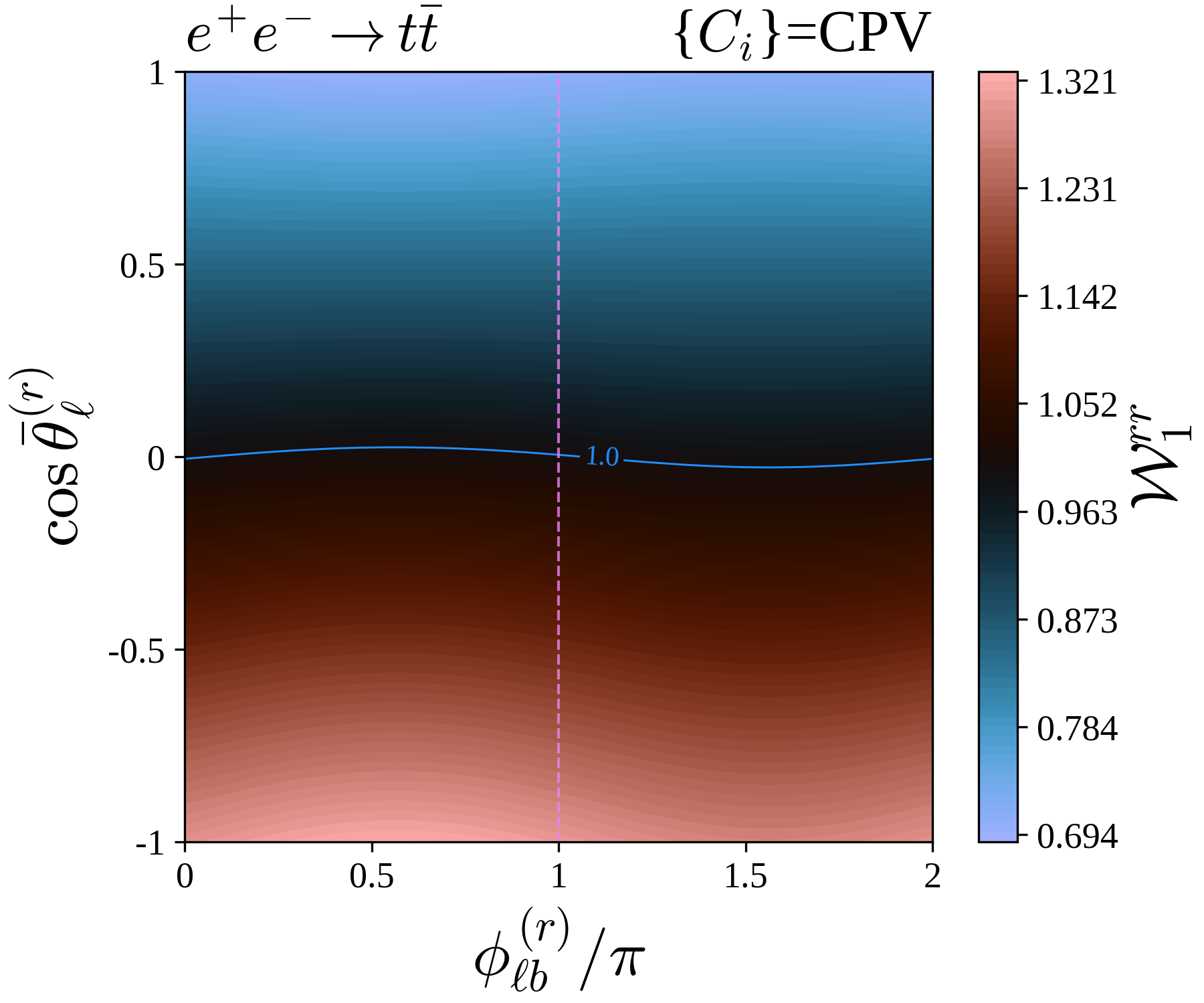}
   
    \caption{Normalised two-angle distribution \(\mathcal{W}_{1}^{rr}
(\phi_{\ell b}^{(r)},\cos\bar{\theta}_{\ell}^{(r)})\), defined in Eq.~\eqref{eq:dist1_col},
for \(e^+e^-\to t\bar t\) production at
\(\sqrt{s}=365~\mathrm{GeV}\). The curves correspond to the SM, CPC, and CPV benchmark coefficients defined
in Eqs.~\eqref{eq:wc_set_SM}, \eqref{eq:wc_set_CPC},
and~\eqref{eq:wc_set_CPV}, respectively. The dashed vertical line marks
\(\phi_{\ell b}^{(r)}=\pi\), while the solid blue contour indicates
\(\mathcal{W}_{1}^{rr}=1\). }
\label{fig:dist1_ee_rr}    
\end{figure}

In Fig.~\ref{fig:dist1_pp}, we present distribution~1,
Eq.~\eqref{eq:dist1_col}, for \(pp\to t\bar t\) production. We choose
\((a,b)=(n,n)\) in bin~1, defined in Eq.~\eqref{eq:bin1}, and
\((a,b)=(k,k)\) in bin~2, defined in Eq.~\eqref{eq:bin2}. These choices
are motivated by the experimentally determined spin-correlation pattern~\cite{CMS:2024zkc}:
among the combinations relevant to this distribution, \(C_{nn}\) in the
threshold region and \(C_{kk}\) in the high-\(m_{t\bar t}\) region have
the largest magnitudes. For \(e^+e^-\to t\bar t\) production, shown in
Fig.~\ref{fig:dist1_ee_rr}, we instead consider
\((a,b)=(r,r)\), following the polarisation and spin-correlation pattern
obtained in our companion analysis~\cite{Lamba:2026jfm}.
For the \(pp\) case, Fig.~\ref{fig:dist1_pp} shows the predictions for
the CPC and CPV benchmark configurations. The SM result is not displayed,
since at leading order in QCD the top and antitop quarks are individually
unpolarised, $B_a=\bar{B}_b=0,$ and the SM decay vertex does not generate any dependence on
\(\phi_{\ell b}^{(a)}\). Consequently,
\begin{equation}
    \mathcal{W}_{1}^{ab}
    \left(
        \phi_{\ell b}^{(a)},
        \cos\bar{\theta}_{\ell}^{(b)}
    \right)
    =1
    \qquad
    \mathrm{for}
    \ 
    pp\to t\bar t\ \mathrm{at\  LO \ QCD},\ 
    \{C_i\} \, \mathrm{given\ by\ the \ } \mathrm{SM}.
\end{equation}
The CP characterisation of the modified decay interaction can be identified from
the reflection properties of the distribution with respect to
\(\phi_{\ell b}^{(a)}=\pi\). The CPC benchmark contains only the cosine
modulation and is therefore symmetric under
\(\phi_{\ell b}^{(a)}\to2\pi-\phi_{\ell b}^{(a)}\), whereas the additional
sine modulation present in the CPV benchmark breaks this symmetry. This
separation can be made explicit by defining the symmetric and
antisymmetric components
\begin{align}
    \mathcal{W}_{1,S}^{ab}(\phi_{\ell b}^{(a)},c)
    &\equiv
    \frac{1}{2}
    \left[
        \mathcal{W}_{1}^{ab}(\phi_{\ell b}^{(a)},c)
        +
        \mathcal{W}_{1}^{ab}(2\pi-\phi_{\ell b}^{(a)},c)
    \right],\label{eq:symm}
    \\
    \mathcal{W}_{1,A}^{ab}(\phi_{\ell b}^{(a)},c)
    &\equiv
    \frac{1}{2}
    \left[
        \mathcal{W}_{1}^{ab}(\phi_{\ell b}^{(a)},c)
        -
        \mathcal{W}_{1}^{ab}(2\pi-\phi_{\ell b}^{(a)},c)
    \right],
    \label{eq:antisymm}
\end{align}
where \(c\equiv\cos\bar{\theta}_{\ell}^{(b)}\). For
\(pp\to t\bar t\) production at LO, these combinations satisfy
\begin{align}
    \mathcal{W}_{1,S}^{ab}(\phi_{\ell b}^{(a)},c)-1
    &\propto
    C_{ab}\,c\,\textcolor{Bittersweet}{v_c\cos\phi_{\ell b}^{(a)}},
    \\
    \mathcal{W}_{1,A}^{ab}(\phi_{\ell b}^{(a)},c)
    &\propto
    C_{ab}\,c\,\textcolor{teal}{w_s^{\rm CPV}\sin\phi_{\ell b}^{(a)}}.
\end{align}
The symmetric component therefore isolates the CP-even contribution,
whereas the antisymmetric component isolates the CP-odd one and vanishes
for the CPC benchmark.

All panels exhibit a nodal line at
\(\cos\bar{\theta}_{\ell}^{(b)}=0\), where
\(\mathcal{W}_{1}^{ab}=1\), while the deviations in the two hemispheres
have opposite signs because of the linear dependence on
\(\cos\bar{\theta}_{\ell}^{(b)}\). For the benchmark values considered
here, the maximum absolute departure from the flat SM prediction is approximately
\(1\%\). The distributions obtained in the two phase-space bins show
deviations of comparable magnitude but opposite sign. This sign reversal
follows from the chosen bins and spin axes, for which the corresponding
spin-correlation coefficients satisfy
\(C^{\rm bin\, 1}_{nn}\simeq-C_{kk}^{\rm bin\, 2}\). The correlation contribution changes
sign while retaining approximately the same magnitude.

Figure~\ref{fig:dist1_ee_rr} shows distribution~1 for
\(e^+e^-\to t\bar t\) production at \(\sqrt{s}=365~\mathrm{GeV}\), using
the axis choice \((a,b)=(r,r)\). In contrast to \(pp\to t\bar t\) at LO,
the non-vanishing single-spin polarisations generate a non-trivial SM
dependence on \(\cos\bar{\theta}_{\ell}^{(r)}\). The SM prediction is
independent of \(\phi_{\ell b}^{(r)}\) and is therefore linear in the polar
variable,
\begin{equation}
    \mathcal{W}_{1,\mathrm{SM}}^{rr}
    =
    1+
    \bar{\alpha}_{\ell}^{\mathrm{SM}}\bar{B}_r
    \cos\bar{\theta}_{\ell}^{(r)} .
\end{equation}

In the SMEFT benchmark scenarios considered here, the Wilson coefficients
modify both the production and decay stages. Consequently, the production
coefficients \(B_r\), \(\bar B_r\), and \(C_{rr}\) can differ among the SM,
CPC, and CPV benchmarks. At fixed production density matrix, the explicit
dependence on the anomalous \(Wtb\) decay vertices enters in two distinct
ways. First, the modification of \(\bar{\alpha}_{\ell}\) changes the slope
along the \(\cos\bar{\theta}_{\ell}^{(r)}\) direction. Second, the terms
proportional to \(v_c\) and \(w_s^{\rm CPV}\) introduce an additional
dependence on \(\phi_{\ell b}^{(r)}\), weighted by
\[
B_r+\bar{\alpha}_{\ell}C_{rr}
\cos\bar{\theta}_{\ell}^{(r)}.
\]
The CPC benchmark produces a mild cosine modulation that preserves the
reflection symmetry about \(\phi_{\ell b}^{(r)}=\pi\), whereas the sine
contribution present in the CPV benchmark breaks this symmetry. The
distributions shown in Fig.~\ref{fig:dist1_ee_rr} include both the modifications associated with each benchmark in production and decay.

The azimuthal modulations are less visually pronounced than the dominant
polar-angle dependence because they are suppressed by the factors
\(v_c/(4\pi)\) and \(w_s^{\rm CPV}/(4\pi)\), while the leading variation
along \(\cos\bar{\theta}_{\ell}^{(r)}\) is controlled by
\(\bar{\alpha}_{\ell}\). Nevertheless, the deformation of the unit
contour makes the azimuthal structure and its reflection properties visible.

The symmetric--antisymmetric decomposition in Eqs.~\eqref{eq:symm}--\eqref{eq:antisymm} remains applicable in the
presence of non-vanishing single-spin polarisations. In this case, the
azimuthal modulations are weighted by both the polarisation \(B_a\) and
the spin-correlation coefficient \(C_{ab}\), whereas the term proportional
to \(\bar{B}_b\) contributes only to the
\(\phi_{\ell b}^{(a)}\)-independent part of the distribution. The
antisymmetric component therefore continues to isolate the contribution
proportional to \(w_s^{\rm CPV}\), also for
\(e^+e^-\to t\bar t\) production.

Since the dominant polar-angle dependence can partially obscure the much
smaller azimuthal modulation, it is useful to introduce the subtracted
distribution
\begin{align}
    \mathcal{W}_{8}^{ab}
    \left(
        \phi_{\ell b}^{(a)},
        \cos\bar{\theta}_{\ell}^{(b)}
    \right)
    &\equiv
    \mathcal{W}_{1}^{ab}
    \left(
        \phi_{\ell b}^{(a)},
        \cos\bar{\theta}_{\ell}^{(b)}
    \right)
    -
    \mathcal{W}_{1}^{b}
    \left(
        \cos\bar{\theta}_{\ell}^{(b)}
    \right)
    \nonumber\\
    &=
    \frac{4\pi}{\sigma}
    \frac{d^{2}\sigma}
    {d\phi_{\ell b}^{(a)}
     \,d\cos\bar{\theta}_{\ell}^{(b)}}
    -
    \frac{2}{\sigma}
    \frac{d\sigma}
    {d\cos\bar{\theta}_{\ell}^{(b)}}
    \nonumber\\
    &=
    \left[
        B_a
        +
        \bar{\alpha}_{\ell}C_{ab}
        \cos\bar{\theta}_{\ell}^{(b)}
    \right]
    \left[
        \textcolor{Bittersweet}{
        \frac{v_c}{4\pi}
        \cos\phi_{\ell b}^{(a)}}
        +
        \textcolor{teal}{
        \frac{w_s^{\rm CPV}}{4\pi}
        \sin\phi_{\ell b}^{(a)}}
    \right].
    \label{eq:distSUB_1}
\end{align}
This is not a subtraction of a fixed SM template. Rather, it removes the
full \(\phi_{\ell b}^{(a)}\)-independent marginal distribution evaluated in
the same parameter scenario. It therefore subtracts both the constant term
and the polar-angle dependence, including possible anomalous contributions
entering through \(\bar\alpha_\ell\), and retains only the azimuthal
modulations proportional to \(v_c\) and \(w_s^{\rm CPV}\).
Consequently, \(\mathcal W_8^{ab}\) vanishes for an SM decay vertex,
independently of the production polarisations and spin-correlations.
A non-zero value provides a null test for anomalous azimuthal structures in
the \(Wtb\) decay vertex.
\begin{figure}[t]
    \centering
    \includegraphics[width=0.445\linewidth]{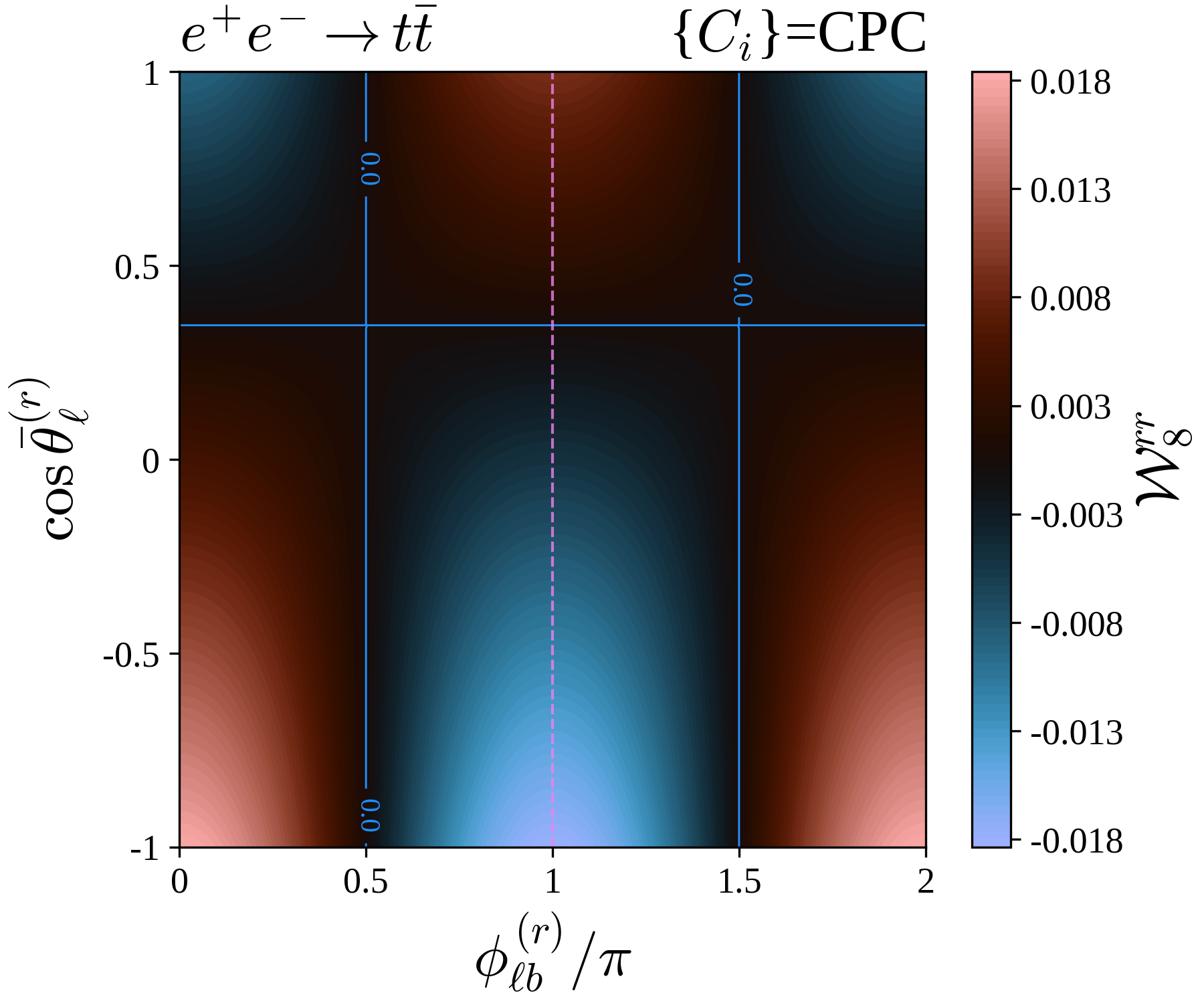}
    \includegraphics[width=0.445\linewidth]{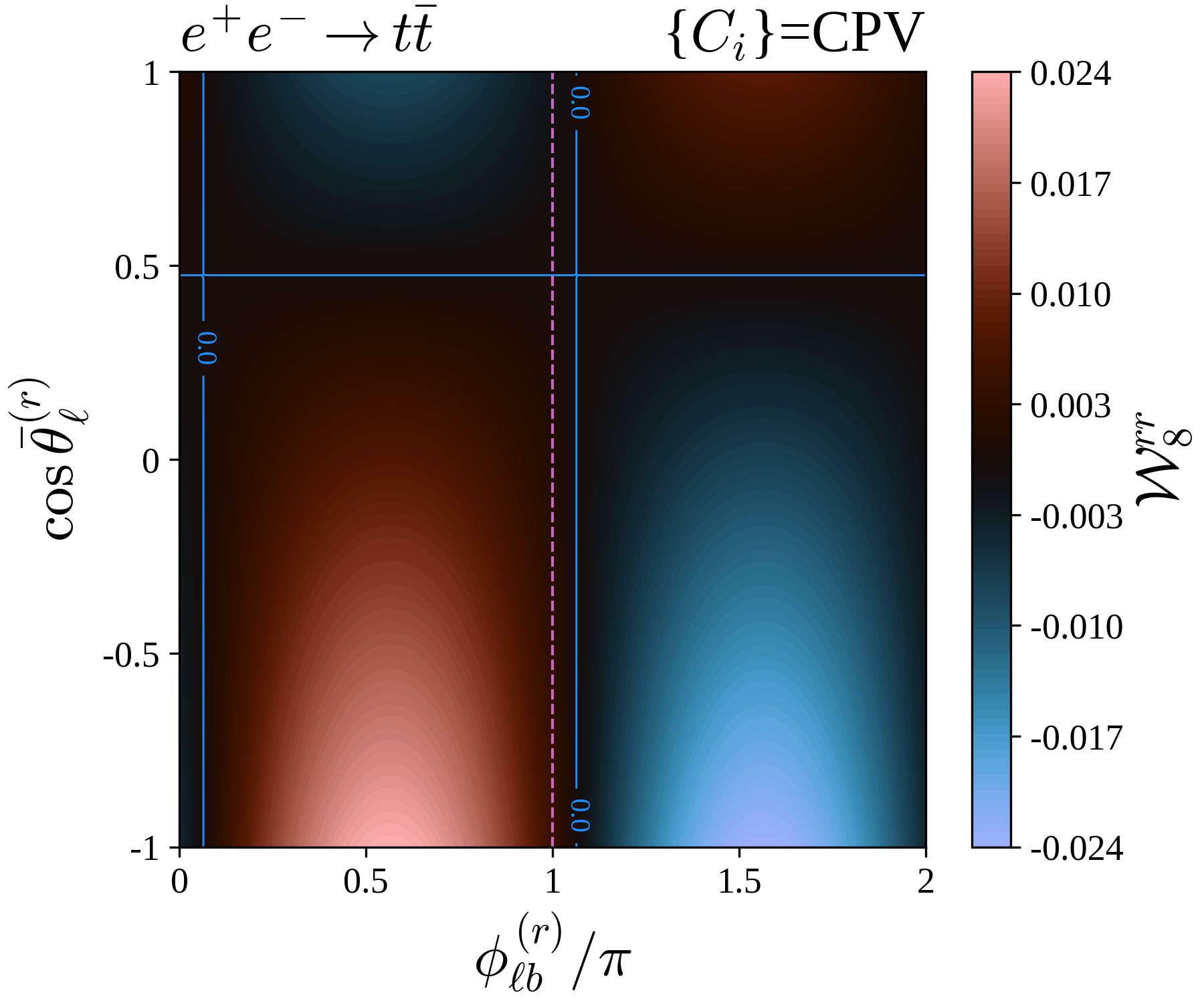}
\caption{Subtracted distribution
\(\mathcal{W}_{8}^{rr}
(\phi_{\ell b}^{(r)},\cos\bar{\theta}_{\ell}^{(r)})\),
defined in Eq.~\eqref{eq:distSUB_1}, for
\(e^+e^-\to t\bar t\) production at
\(\sqrt{s}=365~\mathrm{GeV}\). The left and right panels correspond to
the CPC and CPV benchmark scenarios, respectively. The dashed vertical
line marks \(\phi_{\ell b}^{(r)}=\pi\), while the solid blue contour
indicates \(\mathcal{W}_{8}^{rr}=0\).}
\label{fig:distSUBTRACTION_1}
\end{figure}
Figure~\ref{fig:distSUBTRACTION_1} shows
\(\mathcal{W}_{8}^{rr}\) for \(e^+e^-\to t\bar t\) production in the CPC
and CPV benchmark scenarios. Although the absolute deviations from zero
are small, reaching the \(2\%\) to \(3\%\) level for the benchmarks considered here,
the subtraction makes the new-physics-induced azimuthal structures clearly
visible. The CPC prediction remains symmetric with respect to
\(\phi_{\ell b}^{(r)}=\pi\), whereas this reflection symmetry is broken
by the sine modulation in the CPV scenario. Experimentally, a non-zero
value significantly exceeding the statistical and systematic
uncertainties would provide evidence for a modification of the \(Wtb\)
decay vertex. The analogous distribution can be defined by exchanging
the roles of the top and antitop quarks.

The CP structure can be further separated through the symmetric and
antisymmetric components under
\(\phi_{\ell b}^{(a)}\to2\pi-\phi_{\ell b}^{(a)}\):
\begin{align}
    \mathcal{W}_{8,S}^{ab}(\phi_{\ell b}^{(a)},c)
    &=
    \left(
        B_a+\bar{\alpha}_{\ell}C_{ab}c
    \right)
    \textcolor{Bittersweet}{\frac{v_c}{4\pi}\cos\phi_{\ell b}^{(a)}},
    \\
    \mathcal{W}_{8,A}^{ab}(\phi_{\ell b}^{(a)},c)
    &=
    \left(
        B_a+\bar{\alpha}_{\ell}C_{ab}c
    \right)
    \textcolor{teal}{\frac{w_s^{\rm CPV}}{4\pi}\sin\phi_{\ell b}^{(a)}},
\end{align}
where \(c\equiv\cos\bar{\theta}_{\ell}^{(b)}\). The symmetric component
therefore isolates the CP-even modulation, whereas the antisymmetric
component directly probes the CP-odd contribution.
\subsubsection{Charged-lepton azimuthal angle and the lepton-bottom azimuthal angle distributions}
\label{sect:dist2_discussion}

Distributions~$2)$ and~$\bar{2})$, defined in Eqs.~\eqref{eq:dist2}
and~\eqref{eq:dist2s}, respectively, constitute the second pair of
two-angle distributions sensitive to CP violation in the decay vertex.
Distribution~2 takes the form
\begin{align}
    \mathcal{W}_{2}^{ab}
    \left(
        \phi_{\ell b}^{(a)},
        \bar{\phi}_{\ell}^{(b)}
    \right)
    &=
     1+ B_a\Bigl[\textcolor{Bittersweet}{\frac{v_c}{4\pi} \cos\phi^{(a)}_{\ell b}}+  \textcolor{teal}{\frac{w_s^{ \rm CPV}}{4\pi}\sin\phi^{(a)}_{\ell b}}\Bigr] 
+\bar{\alpha_\ell}\frac{\pi}{4}\Bigl[\bar{B}_{ j_1}\cos\bphi_{\ell}^{(b)}+ \bar{B}_{ j_2}\sin\bphi_{\ell}^{(b)}\Bigr]\nonumber\\
&+\bar{\alpha_\ell}\frac{\pi}{4}\Bigl[\textcolor{Bittersweet}{\frac{v_c}{4\pi} \cos\phi^{(a)}_{\ell b}}+  \textcolor{teal}{\frac{w_s^{\rm CPV}}{4\pi}\sin\phi^{(a)}_{\ell b}}\Bigr]\Bigl[
   C_{a j_1}\cos\bphi_{\ell}^{(b)}
 + C_{a j_2}\sin\bphi_{\ell}^{(b)}
\Bigr].
\label{eq:dist2_col}
\end{align}
The corresponding expression for distribution~$\bar{2})$ is obtained by
exchanging the roles of the top and antitop quarks.

As for distribution~$1)$, the CP-odd sine modulation can arise from the
interference between the SM and dimension-six decay amplitudes and therefore
appears already at \(\mathcal{O}(\Lambda^{-2})\). The distinctive feature
of distribution~2 is that the opposite decay branch is analysed through the
charged-lepton azimuthal angle \(\bar{\phi}_{\ell}^{(b)}\). Consequently,
even when \(B_a=\bar B_{j_1}=\bar B_{j_2}=0\), the contribution proportional to
\begin{equation}
    w_s^{\rm CPV}\sin\phi_{\ell b}^{(a)}
    \left[
        C_{a j_1}\cos\bar{\phi}_{\ell}^{(b)}
        +
        C_{a j_2}\sin\bar{\phi}_{\ell}^{(b)}
    \right]
\end{equation}
survives in the presence of non-vanishing spin-correlations.
\begin{figure}[ht]
    \centering
    \includegraphics[width=0.445\linewidth]{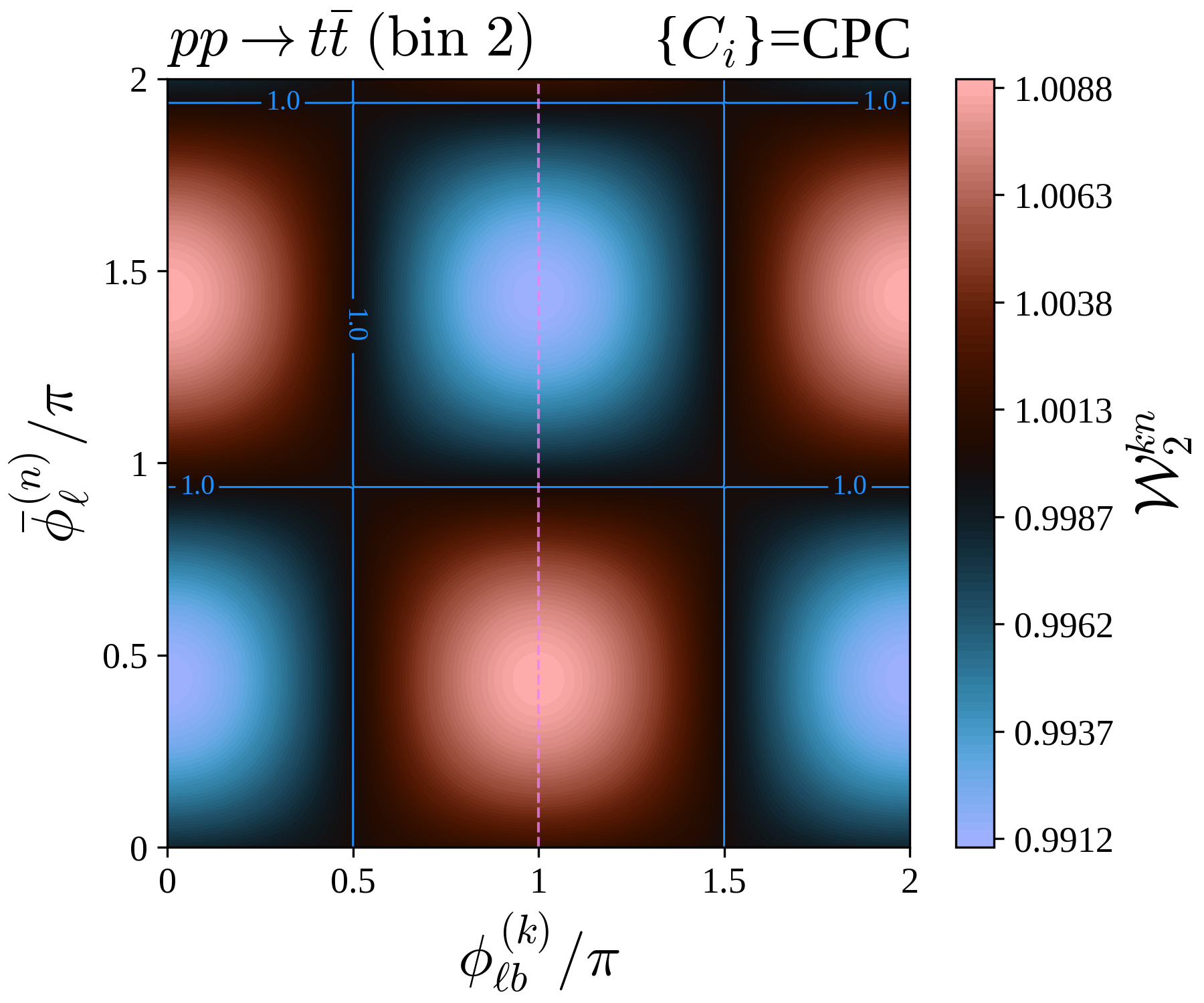}
    \includegraphics[width=0.445\linewidth]{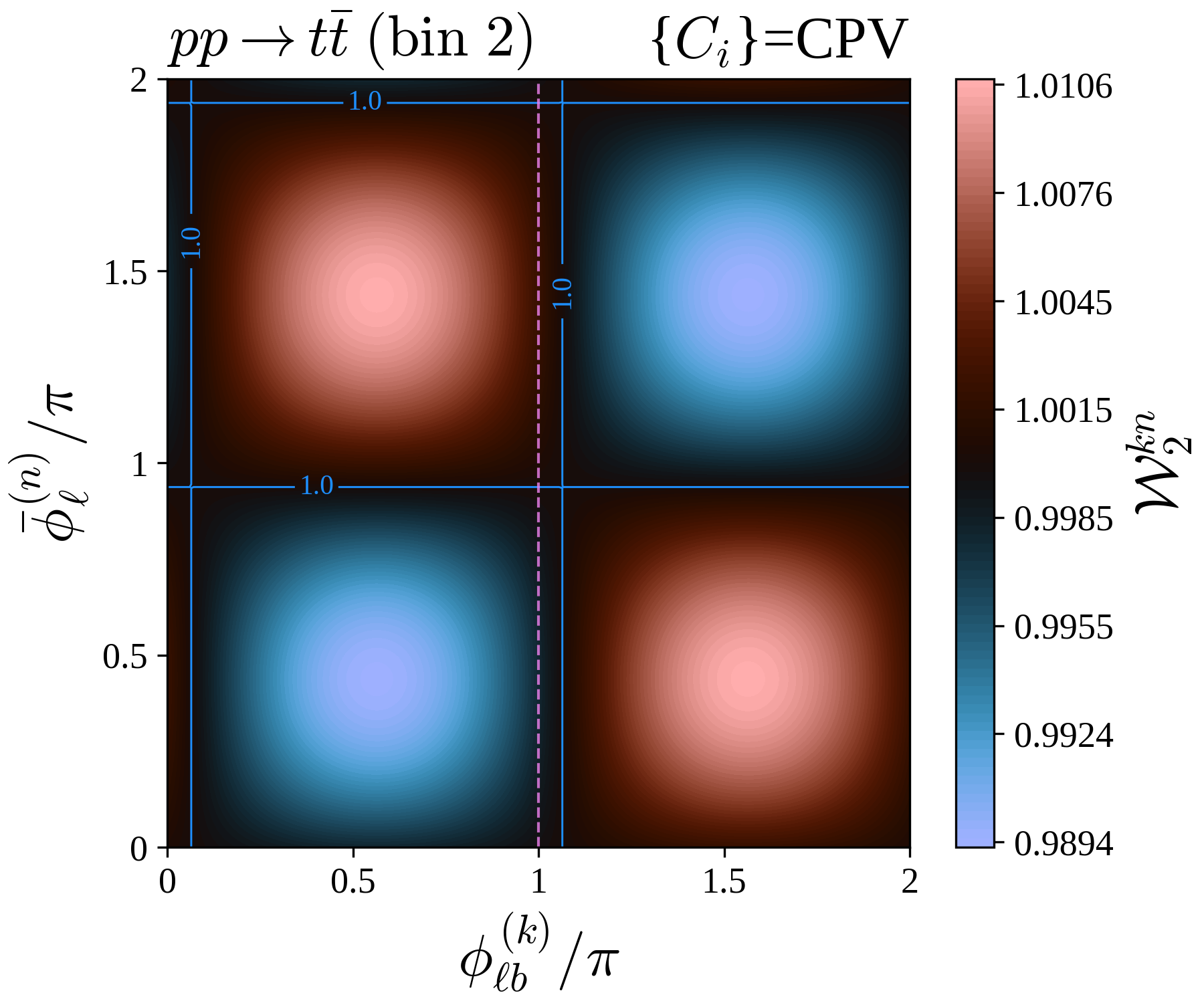}\\
     \includegraphics[width=0.445\linewidth]{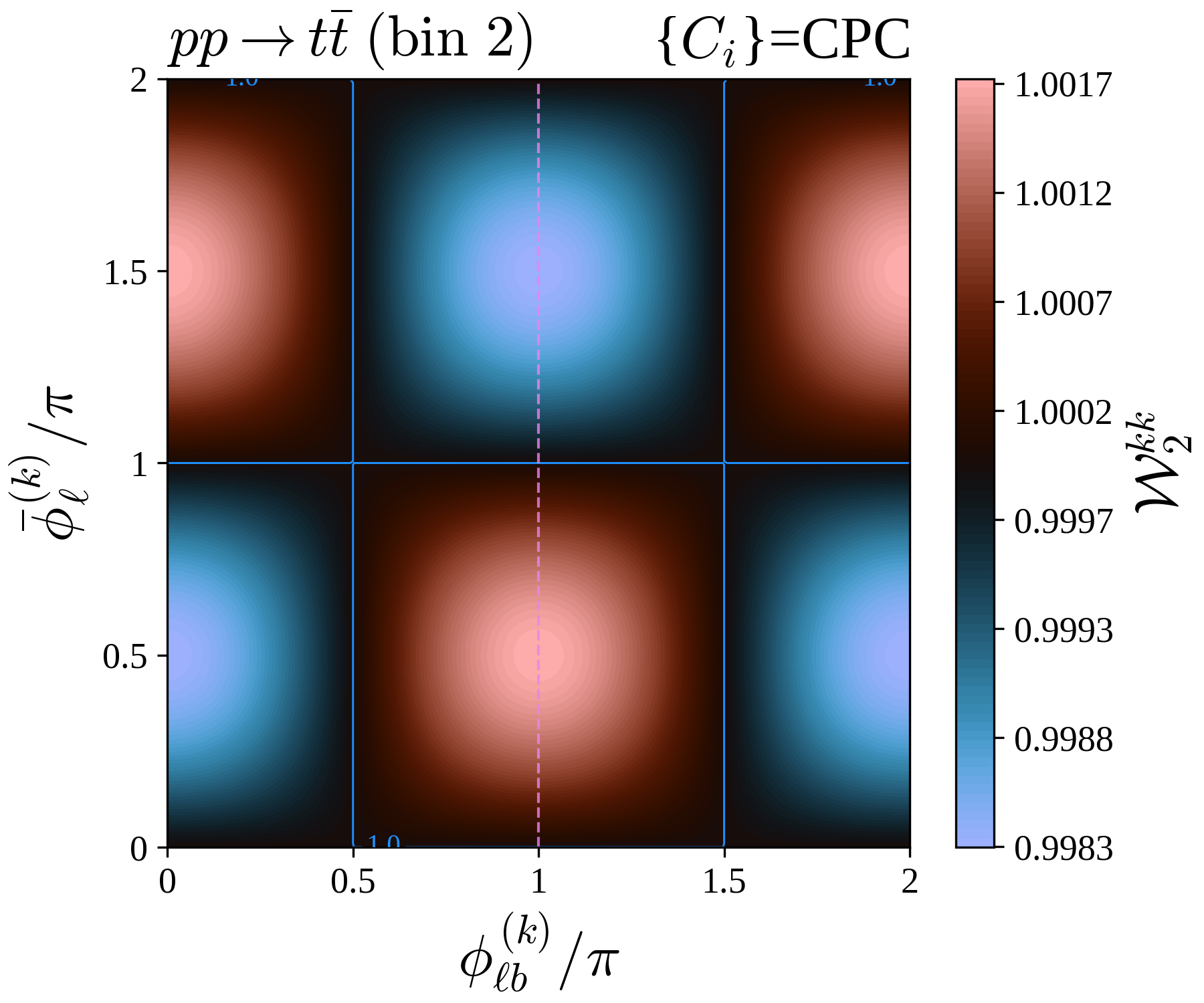}
     \includegraphics[width=0.445\linewidth]{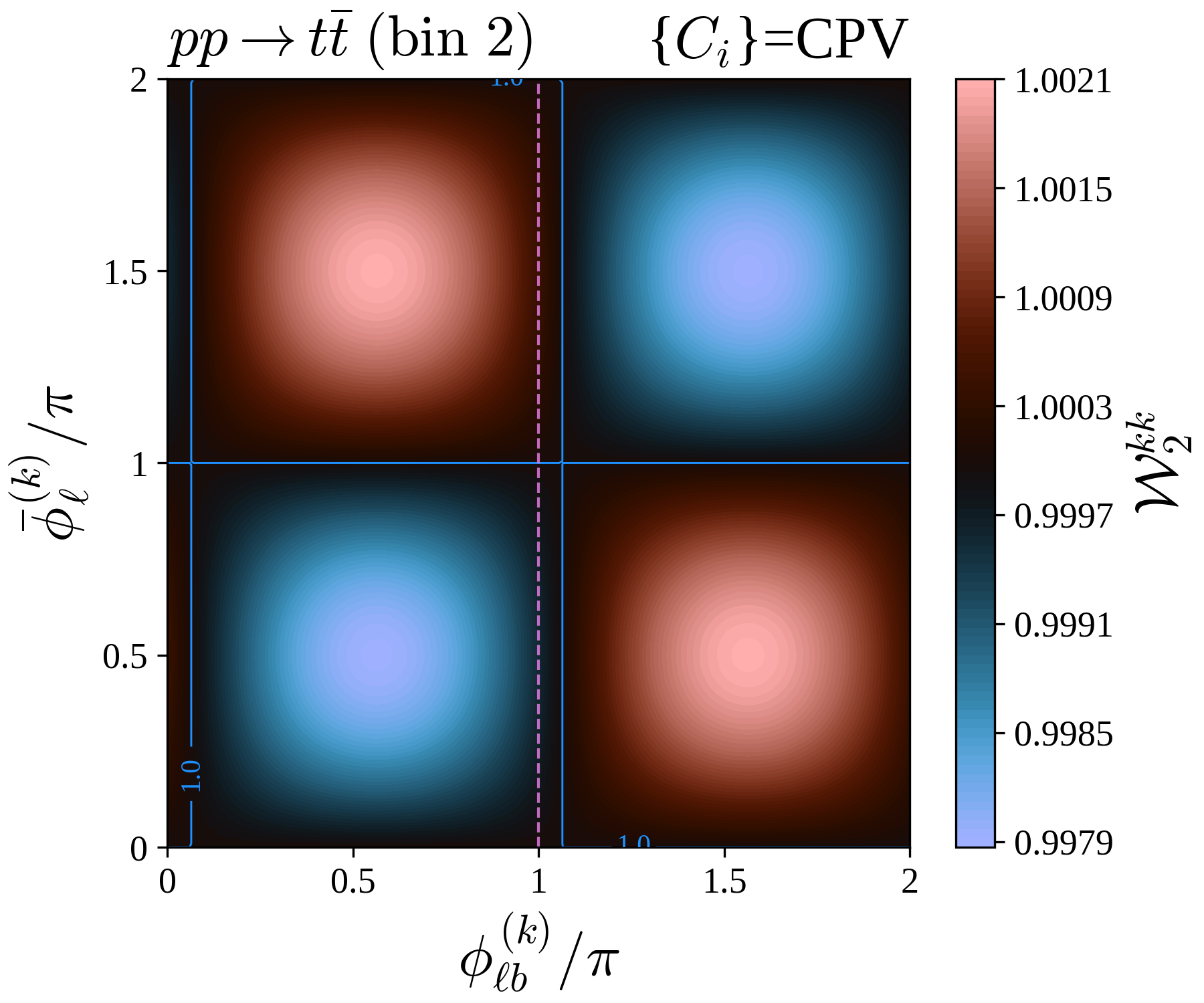} 
\caption{Normalised two-angle distribution \(\mathcal{W}_{2}^{ab}
(\phi_{\ell b}^{(a)},\bar{\phi}_{\ell}^{(b)})\),
defined in Eq.~\eqref{eq:dist2_col}, for \(pp\to t\bar t\) production in
bin~2, using the polarisation and spin-correlation coefficients extracted
from the simulations. The top and bottom rows correspond to
\((a,b)=(k,n)\) and \((a,b)=(k,k)\), respectively, while the left and
right columns show the CPC and CPV benchmark scenarios. The dashed
vertical line marks \(\phi_{\ell b}^{(k)}=\pi\), and the solid blue
contour indicates \(\mathcal{W}_{2}^{ab}=1\).}
\label{fig:dist2_pp}
\end{figure} 
In Fig.~\ref{fig:dist2_pp}, we show
\(\mathcal{W}_{2}^{ab}\) for \(pp\to t\bar t\) production in bin~2,
considering the two axis choices \((a,b)=(k,n)\) and \((a,b)=(k,k)\).
These configurations are selected to illustrate how the sensitivity of
the distribution depends on the magnitude of the relevant combination
of spin-correlation coefficients,
\begin{equation}
      C_{a j_1}\cos\bar{\phi}_{\ell}^{(b)}
    +
    C_{a j_2}\sin\bar{\phi}_{\ell}^{(b)}.  
\end{equation}
The SM prediction is not displayed because, at leading order in QCD, the
single-spin polarisations vanish and the SM decay vertex gives
\(v_c=w_s^{\rm CPV}=0\). Consequently,
\begin{equation}
    \mathcal{W}_{2}^{ab}
    \left(
        \phi_{\ell b}^{(a)},
        \bar{\phi}_{\ell}^{(b)}
    \right)
    =1
    \qquad
    \mathrm{for}\ 
   pp\to t\bar t\ \mathrm{at\  LO \ QCD},\ 
    \{C_i\} \, \mathrm{given\ by\ the \ } \mathrm{SM}.
\end{equation}

The CPC predictions remain symmetric under
\(\phi_{\ell b}^{(a)}\to2\pi-\phi_{\ell b}^{(a)}\), whereas the sine
modulation present in the CPV benchmark breaks this reflection symmetry.
The additional unit contours along the
\(\bar{\phi}_{\ell}^{(b)}\) direction correspond to the zeros of the
relevant spin-correlation combination, across which the modulation changes
sign. The choice of spin axes has a significant impact on the sensitivity of the
distribution. For the benchmark scenarios considered here, the maximum
absolute deviation from the flat SM prediction reaches approximately the
\(1\%\) level for \((a,b)=(k,n)\), whereas it remains at the
\(0.2\%\) level for \((a,b)=(k,k)\). This difference originates from the
larger magnitude of the spin-correlation combination
for the \((k,n)\) configuration. The corresponding values of the relevant
spin-correlation coefficients for the two axis choices are reported in
Tab.~\ref{tab:values_distrib_bins_2}.

The reflection decomposition introduced for distribution~1 can be applied
analogously to distribution~2. In particular, the component antisymmetric
under
\(\phi_{\ell b}^{(a)}\to2\pi-\phi_{\ell b}^{(a)}\)
isolates the CP-odd contribution proportional to \(w_s^{\rm CPV}\):
\begin{align}
    \mathcal{W}_{2,A}^{ab}
    \left(\phi_{\ell b}^{(a)},\bar{\phi}_{\ell}^{(b)}\right)
    &\equiv
    \frac{1}{2}
    \left[
        \mathcal{W}_{2}^{ab}(\phi_{\ell b}^{(a)},\bar{\phi}_{\ell}^{(b)})
        -
        \mathcal{W}_{2}^{ab}(2\pi-\phi_{\ell b}^{(a)},\bar{\phi}_{\ell}^{(b)})
    \right]
    \nonumber\\
    &=
    \left[
        B_a
        +\bar{\alpha}_{\ell}\frac{\pi}{4}
        \left(
            C_{a j_1}\cos\bar{\phi}_{\ell}^{(b)}
            +C_{a j_2}\sin\bar{\phi}_{\ell}^{(b)}
        \right)
    \right]
   \textcolor{teal} {\frac{w_s^{\rm CPV}}{4\pi}\sin\phi_{\ell b}^{(a)}} ,
   \label{eq:Asymm2}
\end{align}
For \(pp\to t\bar t\) production at LO, \(B_a=0\), so the sensitivity is
entirely controlled by the relevant combination of spin-correlation
coefficients and can be maximised by focusing on regions exhibiting large spin-correlations.

\begin{figure}[t]
    \centering
    \includegraphics[width=0.325\linewidth]{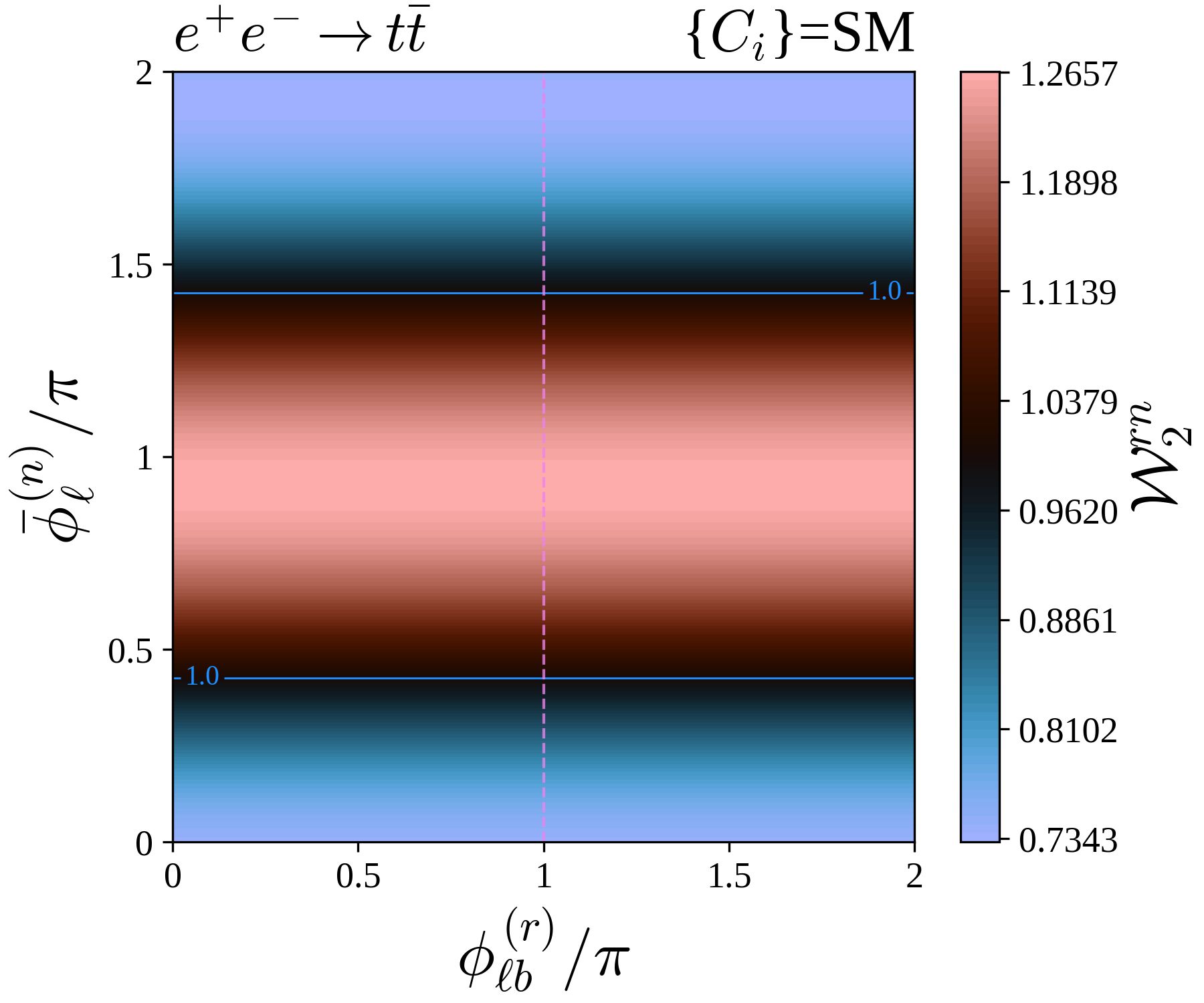}\includegraphics[width=0.325\linewidth]{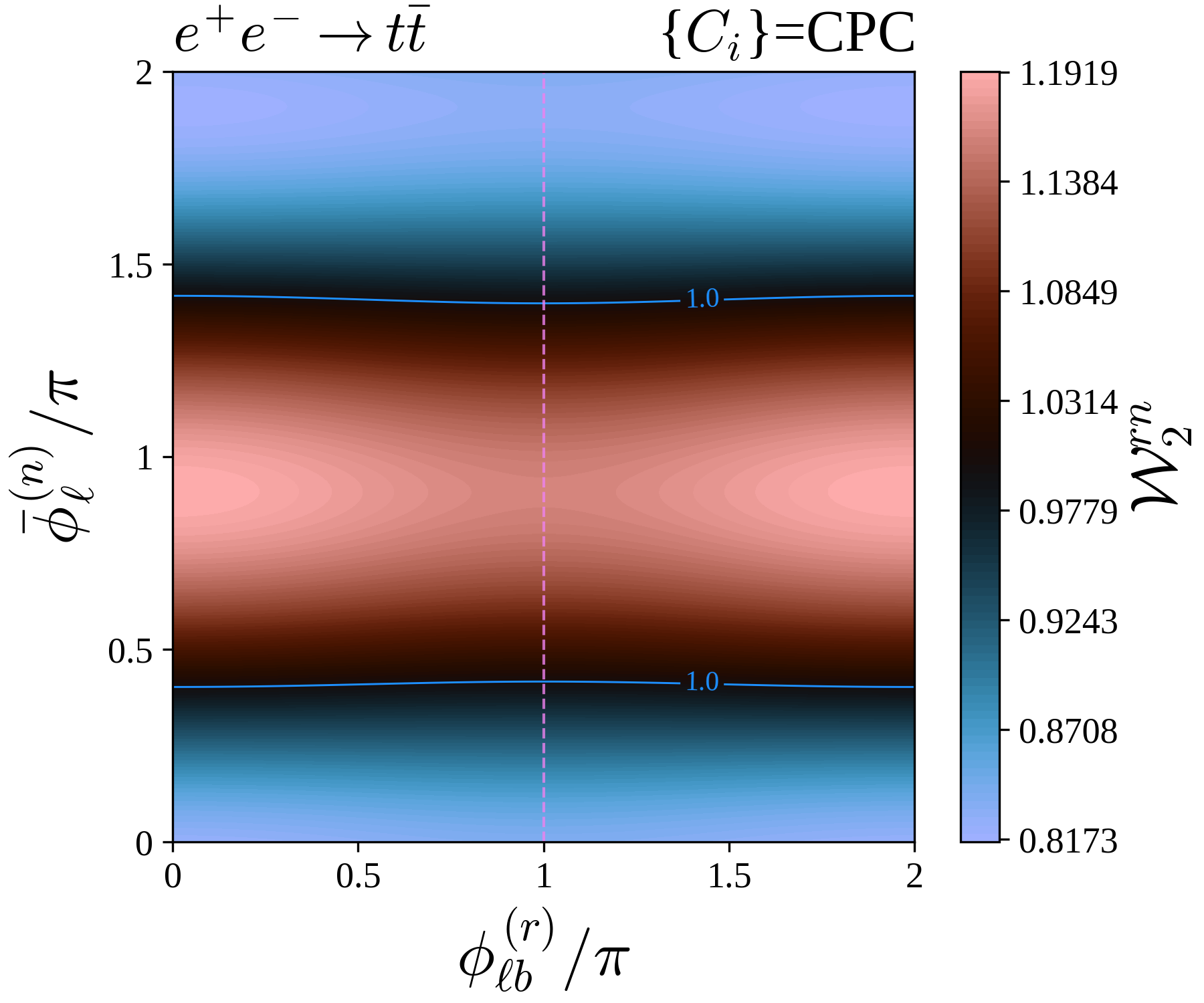} 
    \includegraphics[width=0.325\linewidth]{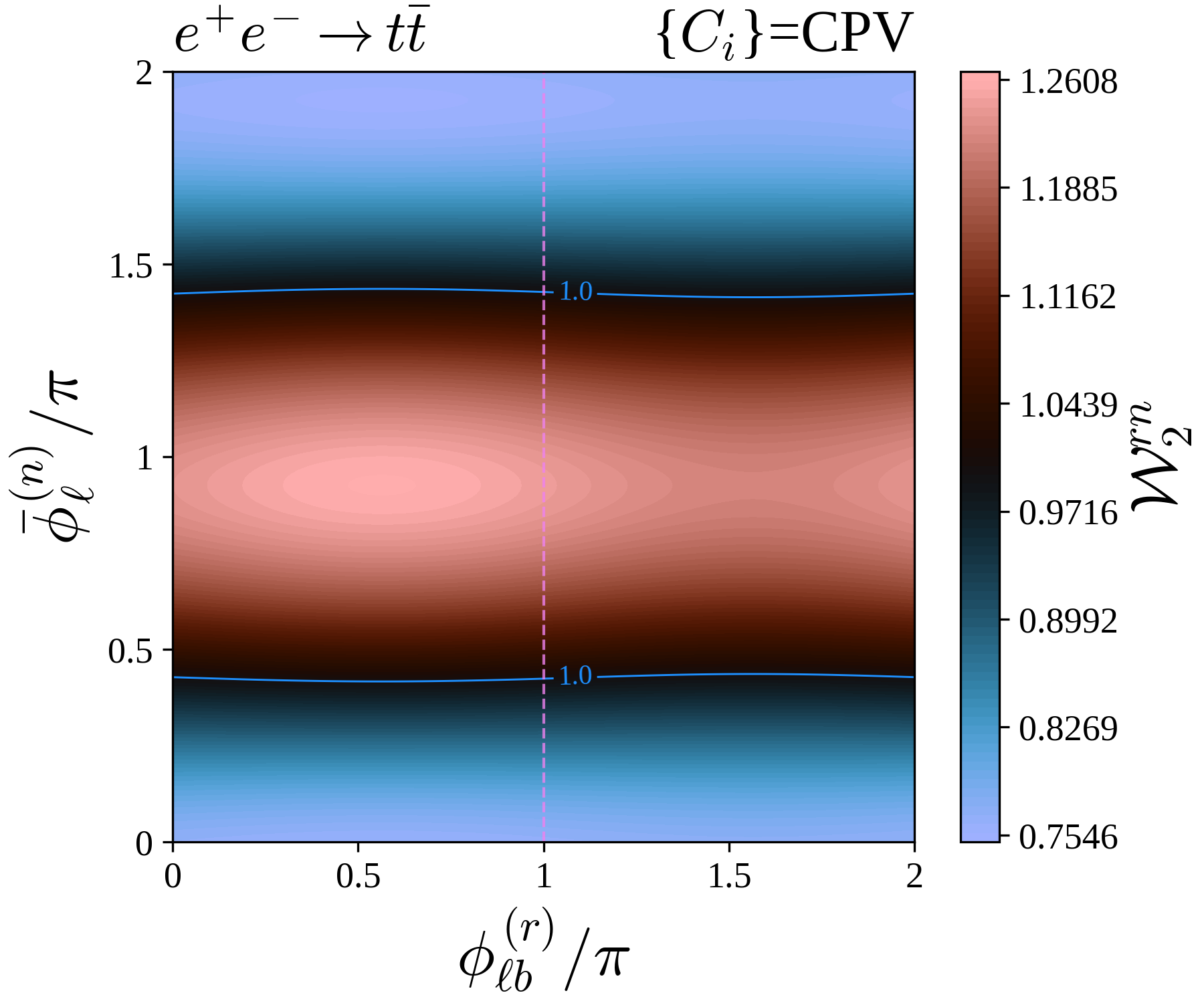}
    \caption{Normalised distribution \(\mathcal{W}_{2}^{rn}
    (\phi_{\ell b}^{(r)},\bar{\phi}_{\ell}^{(n)})\), defined in Eq.~\eqref{eq:dist2_col}, for \(e^+e^-\to t\bar t\) production at
\(\sqrt{s}=365~\mathrm{GeV}\). The panels show the SM, CPC, and CPV benchmark scenarios, defined in
Eqs.~\eqref{eq:wc_set_SM}, \eqref{eq:wc_set_CPC},
and~\eqref{eq:wc_set_CPV}, respectively. The dashed vertical
line marks \(\phi_{\ell b}^{(r)}=\pi\), while the solid blue contour
indicates \(\mathcal{W}_{2}^{rn}=1\).
}
    \label{fig:dist2_ee_rn}
\end{figure}

Figure~\ref{fig:dist2_ee_rn} shows the corresponding distribution for
\(e^+e^-\to t\bar t\) production at
\(\sqrt{s}=365~\mathrm{GeV}\), with the axis choice
\((a,b)=(r,n)\). This configuration is selected because the polarisation
and spin-correlation coefficients entering the distribution are
comparatively large. In contrast to \(pp\to t\bar t\) production at
leading order in QCD, the top and antitop quarks are individually
polarised, and the SM prediction therefore already exhibits a non-trivial
dependence on \(\bar{\phi}_{\ell}^{(n)}\). For
\(v_c=w_s^{\rm CPV}=0\), it is given by
\begin{equation}
    \mathcal{W}_{2,\mathrm{SM}}^{rn}
    \left(
        \phi_{\ell b}^{(r)},
        \bar{\phi}_{\ell}^{(n)}
    \right)
    =
    1+
    \bar{\alpha}_{\ell}^{\rm SM}\frac{\pi}{4}
    \left[
        \bar{B}_{j_1}\cos\bar{\phi}_{\ell}^{(n)}
        +
        \bar{B}_{j_2}\sin\bar{\phi}_{\ell}^{(n)}
    \right],
\end{equation}
which is independent of \(\phi_{\ell b}^{(r)}\).

The comparison of the three panels in Fig.~\ref{fig:dist2_ee_rn} shows that this polarisation-induced
contribution determines the dominant shape of the distribution. A
modified \(Wtb\) vertex changes its magnitude through the replacement
\(\bar{\alpha}_{\ell}^{\rm SM}\to\bar{\alpha}_{\ell}\) and, in addition,
introduces cosine and sine modulations along the
\(\phi_{\ell b}^{(r)}\) direction through \(v_c\) and
\(w_s^{\rm CPV}\), respectively. These additional structures are
relatively mild in the full distribution because they are superimposed on
the larger polarisation-driven SM contribution. The CP-odd contribution can be isolated directly through the
antisymmetric combination introduced above in Eq.~\eqref{eq:Asymm2}. 

Although the symmetric component retains the CP-even cosine modulation, it
also contains the \(\phi_{\ell b}^{(a)}\)-independent SM contribution,
which can obscure the new physics effect. It is therefore useful to
subtract the complete \(\phi_{\ell b}^{(a)}\)-independent part of the
distribution. The resulting combination vanishes identically for an SM
decay vertex, while retaining both the CP-even contribution proportional
to \(v_c\) and the CP-odd contribution proportional to
\(w_s^{\rm CPV}\). We thus define the subtracted distribution
\begin{align}
    \mathcal{W}_{9}^{ab}
    \left(
        \phi_{\ell b}^{(a)},
        \bar{\phi}_{\ell}^{(b)}
    \right)
    &\equiv
    \mathcal{W}_{2}^{ab}
    \left(
        \phi_{\ell b}^{(a)},
        \bar{\phi}_{\ell}^{(b)}
    \right)
    -
    \mathcal{W}_{2}^{b}
    \left(
        \bar{\phi}_{\ell}^{(b)}
    \right)\nonumber\\
    &=
    \left[
        B_a
        +
        \bar{\alpha}_{\ell}\frac{\pi}{4}
        \left(
            C_{a j_1}\cos\bar{\phi}_{\ell}^{(b)}
            +
            C_{a j_2}\sin\bar{\phi}_{\ell}^{(b)}
        \right)
    \right]
    \left[
        \textcolor{Bittersweet}{
        \frac{v_c}{4\pi}\cos\phi_{\ell b}^{(a)}}
        +
        \textcolor{teal}{
        \frac{w_s^{\rm CPV}}{4\pi}
        \sin\phi_{\ell b}^{(a)}}
    \right].
    \label{eq:distSUB_2}
\end{align}
Here,
\(\mathcal{W}_{2}^{b}(\bar{\phi}_{\ell}^{(b)})\) is the corresponding
single-angle projection, resulting in the same distribution as $\bar{6}$, defined in Eq.~\eqref{eq:dist6s}. The subtraction removes the complete
\(\phi_{\ell b}^{(a)}\)-independent contribution, including the
polarisation term and its modification through
\(\bar{\alpha}_{\ell}\). Consequently,
\(\mathcal{W}_{9}^{ab}\) vanishes identically for a SM decay vertex,
even when the top and antitop quarks are individually polarised. 
\begin{figure}[t]
    \centering
    \includegraphics[width=0.445\linewidth]{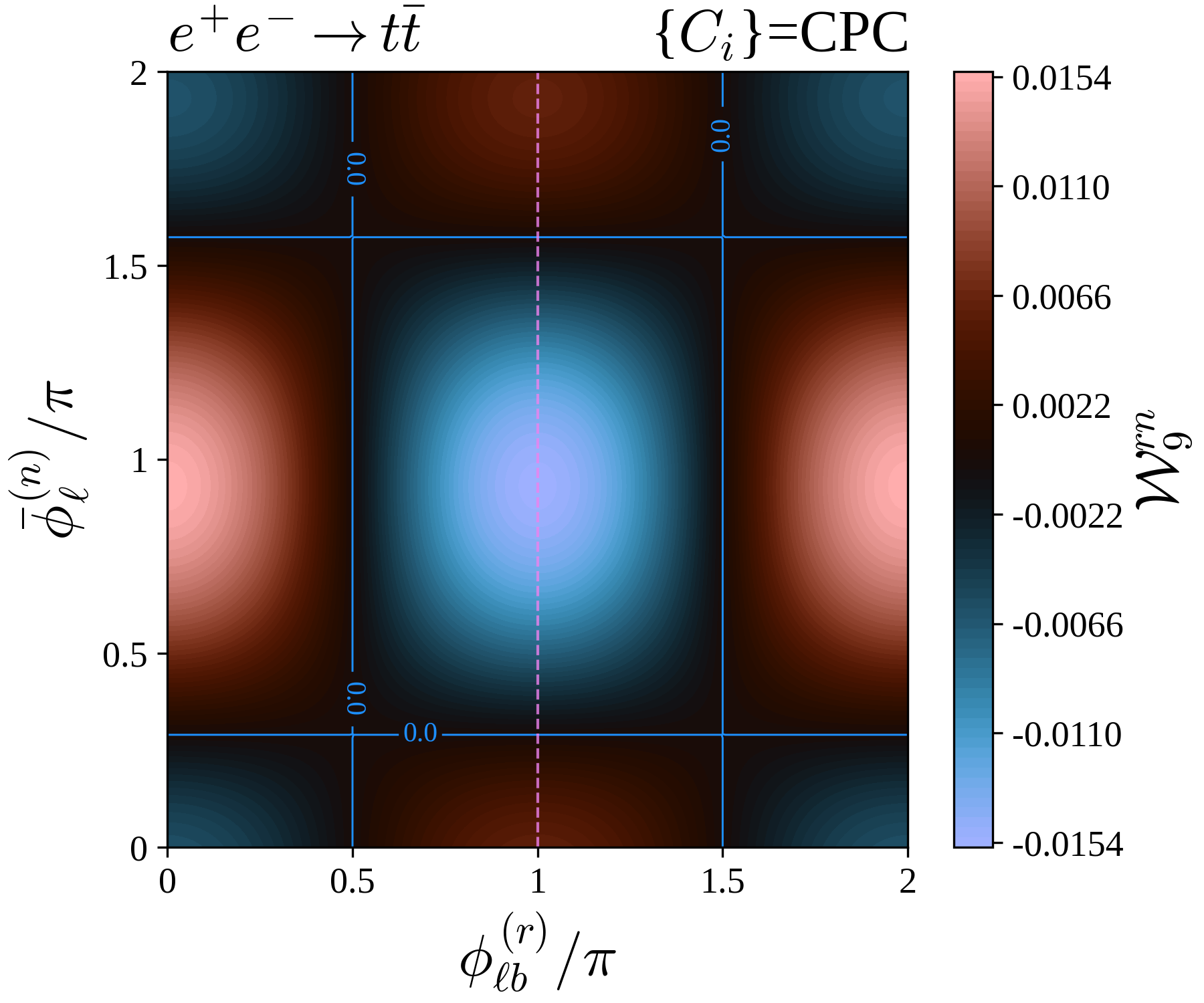}
    \includegraphics[width=0.445\linewidth]{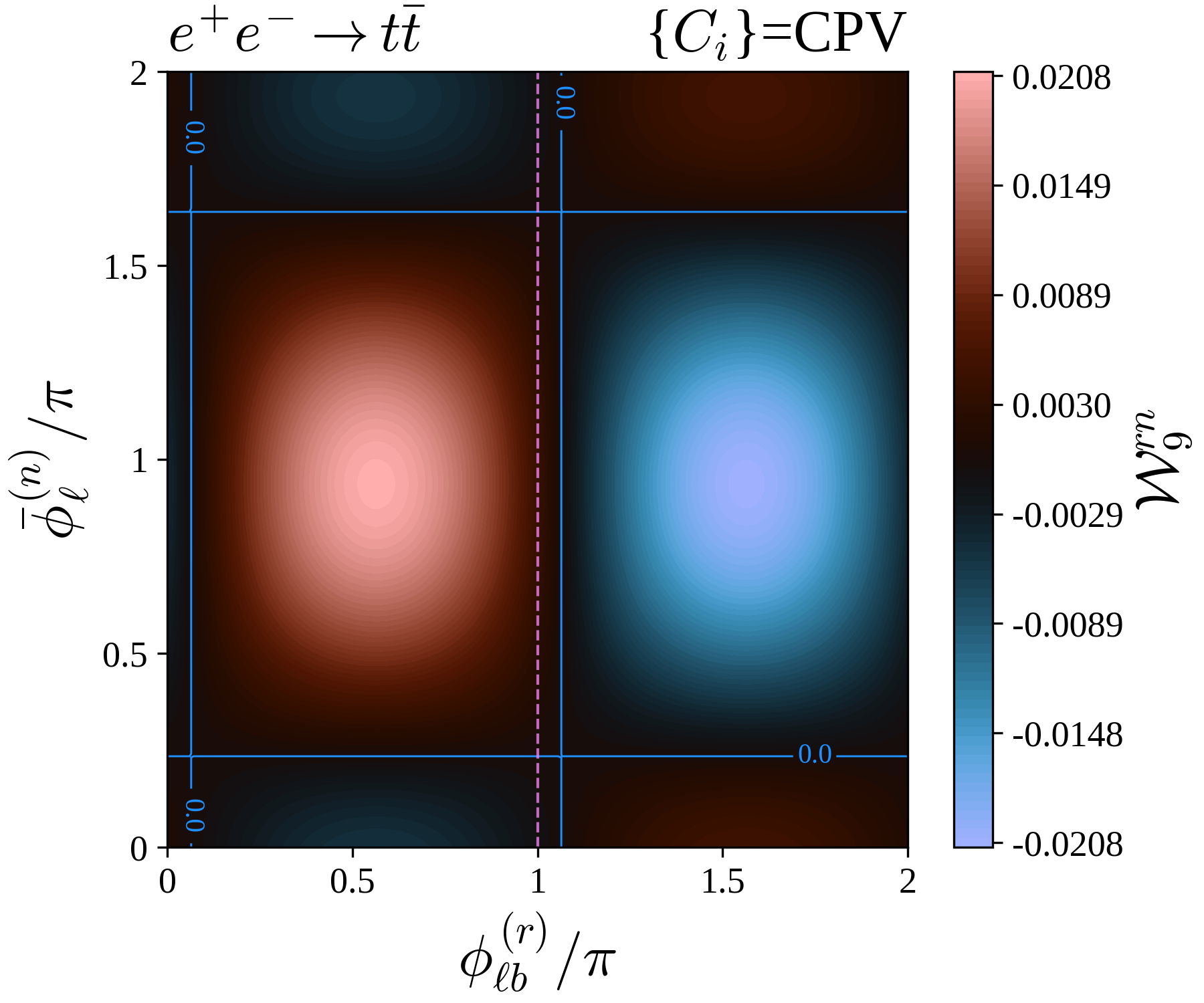}
\caption{Subtracted distribution \(\mathcal{W}_{9}^{rn}(
        \phi_{\ell b}^{(r)},
        \bar{\phi}_{\ell}^{(n)})\), defined in Eq.~\eqref{eq:distSUB_2}, for
\(e^+e^-\to t\bar t\) production at
\(\sqrt{s}=365~\mathrm{GeV}\). The left and right panels correspond to the CPC and CPV
benchmark scenarios, respectively. The dashed vertical
line marks \(\phi_{\ell b}^{(r)}=\pi\), while the solid blue contour
indicates $\mathcal{W}_{9}^{rn}=0$.}
\label{fig:distSUBTRACTION_2}
\end{figure}
Figure~\ref{fig:distSUBTRACTION_2} shows
\(\mathcal{W}_{9}^{rn}\) for the \(e^+e^-\to t\bar t\) production at 365 GeV in the CPC and CPV benchmark scenarios. The
subtraction exposes absolute deviations from zero at the
\(1.5\text{--}2\%\) level and makes the reflection properties along
\(\phi_{\ell b}^{(r)}\) more apparent. The CPC prediction remains
symmetric under
\(\phi_{\ell b}^{(r)}\to2\pi-\phi_{\ell b}^{(r)}\), whereas the sine
modulation in the CPV scenario breaks this symmetry. Equivalently, the
component antisymmetric under this reflection isolates the contribution
proportional to \(w_s^{\rm CPV}\). The subtracted distribution  $\mathcal{W}_{9}^{ab}$ therefore provides a direct probe of
modifications of the \(Wtb\) vertex: its symmetric component isolates the
CP-even cosine modulation, whereas its antisymmetric component isolates
the CP-odd sine modulation.
\subsubsection{Lepton-bottom azimuthal-angle distributions}\label{sect:dist3_discussion}

The third distribution in Eq.~\eqref{eq:dist3} also belongs to this class
of observables. In this case, the same type of angular variable is selected
from the two decay phase spaces, namely the relative lepton-bottom
azimuthal angle. The normalised distribution
takes the form
\begin{align}
\mathcal{W}_3^{ab}(\phi_{\ell b}^{(a)},\bar{\phi}_{\ell b}^{(b)})&= 1+ B_a\Bigl[\textcolor{Bittersweet}{\frac{v_c}{4\pi} \cos\phi^{(a)}_{\ell b}}+  \textcolor{teal}{\frac{w_s^{\rm CPV}}{4\pi}\sin\phi^{(a)}_{\ell b}}\Bigr]  + \bar{B}_b\Bigl[\textcolor{Bittersweet}{\frac{\bar{v}_c}{4\pi} \cos\bphi^{(b)}_{\ell b}}+  \textcolor{teal}{\frac{\bar{w}_s^{\rm CPV}}{4\pi}\sin\bphi^{(b)}_{\ell b}}\Bigr] \nonumber\\
&+C_{ab}\Bigl[\textcolor{Bittersweet}{\frac{v_c}{4\pi} \cos\phi^{(a)}_{\ell b}}+ \textcolor{teal}{ \frac{w_s^{\rm CPV}}{4\pi}\sin\phi^{(a)}_{\ell b}}\Bigr]\Bigl[\textcolor{Bittersweet}{\frac{\bar{v}_c}{4\pi} \cos\bphi^{(b)}_{\ell b}}+  \textcolor{teal}{\frac{\bar{w}_s^{\rm CPV}}{4\pi}\sin\bphi^{(b)}_{\ell b}}\Bigr]\label{eq:dist3_col}
\end{align}
This distribution is particularly interesting because its SM prediction is
flat,
\begin{equation}
    \mathcal{W}_{3,\rm SM}^{ab}
    \left(
        \phi_{\ell b}^{(a)},
        \bar{\phi}_{\ell b}^{(b)}
    \right)
    =1,
\end{equation}
so that any non-trivial angular dependence directly signals a modification
of the decay vertex. 

\begin{figure}[t]
    \centering
    \includegraphics[width=0.445\linewidth]{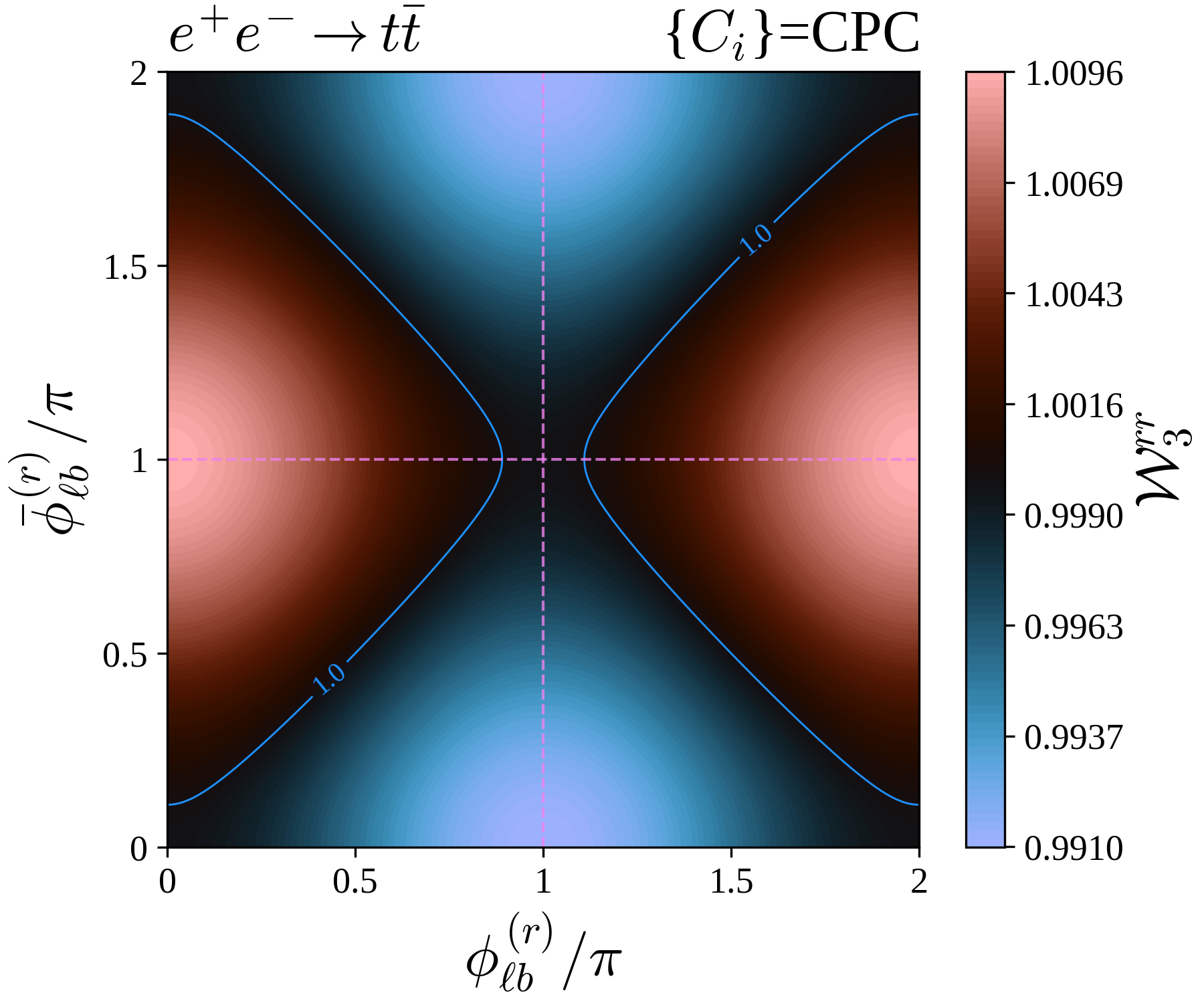} 
    \includegraphics[width=0.445\linewidth]{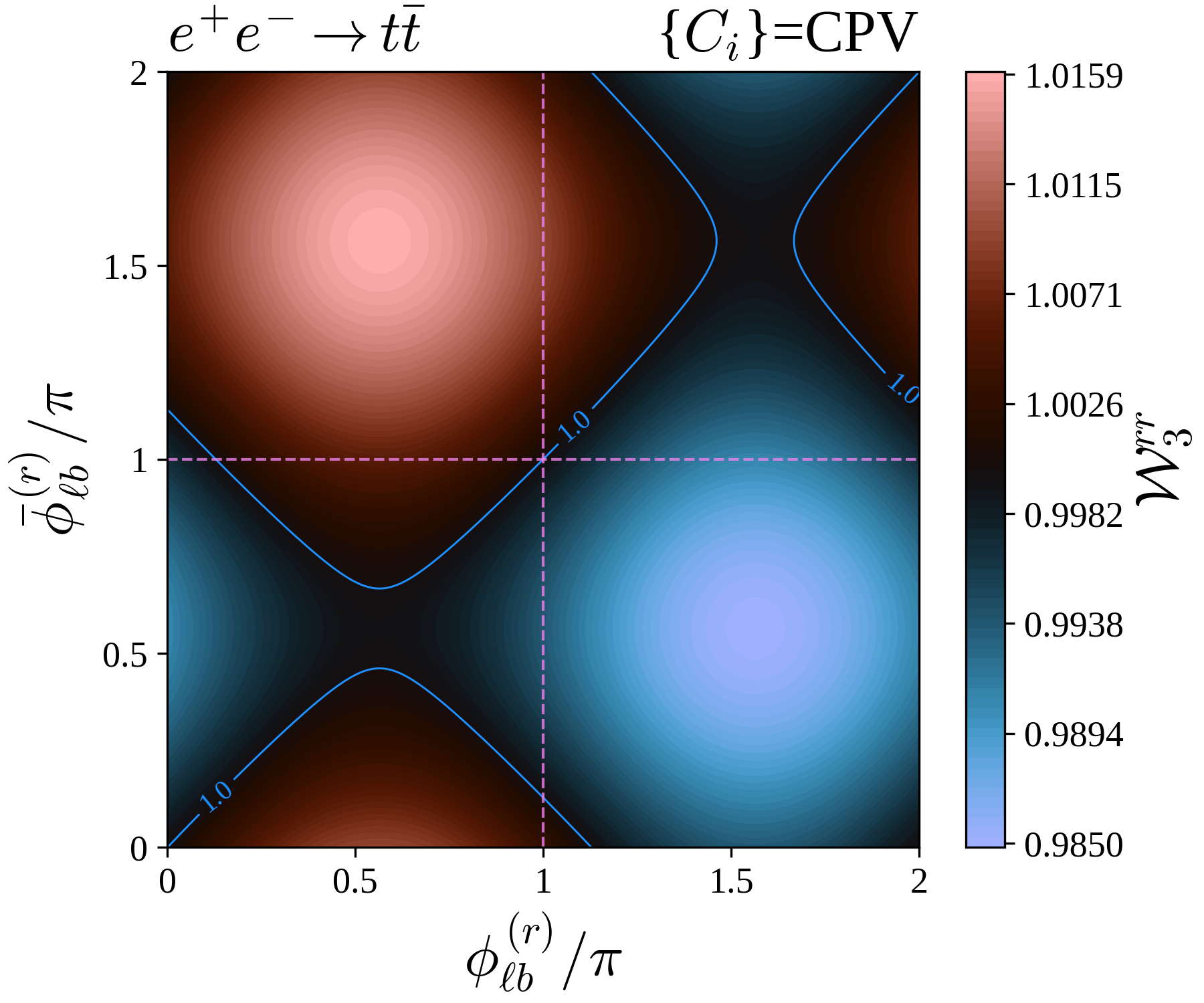}
    \caption{Normalised two-angle distribution
\(\mathcal{W}_{3}^{rr}
(\phi_{\ell b}^{(r)},\bar{\phi}_{\ell b}^{(r)})\),
defined in Eq.~\eqref{eq:dist3_col}, for
\(e^+e^-\to t\bar t\) production at
\(\sqrt{s}=365~\mathrm{GeV}\). The left and right panels
show the CPC and CPV benchmark scenarios, respectively. The dashed lines
mark \(\phi_{\ell b}^{(r)}=\pi\) and
\(\bar{\phi}_{\ell b}^{(r)}=\pi\), while the solid blue contours indicate
\(\mathcal{W}_{3}^{rr}=1\).}
    \label{fig:dist3_ee_rr}
\end{figure}

If either the top or the antitop is individually
polarised, the terms proportional to \(B_a\) or \(\bar{B}_b\) generate a
contribution that is linear in the corresponding new physics decay
coefficients and therefore starts at
\(\mathcal{O}(\Lambda^{-2})\).
The situation is different for \(pp\to t\bar t\) production at leading
order in QCD, where \(B_a=\bar{B}_b=0\). In this case, the only non-trivial
contribution is the term proportional to the spin-correlation coefficient
\(C_{ab}\). In this respect, $\mathcal{W}_3^{ab}(\phi_{\ell b}^{(a)},\bar{\phi}_{\ell b}^{(b)})$ differs significantly from
distributions discussed in sections~\ref{sect:dist1_discussion} and \ref{sect:dist2_discussion}, for which the spin-correlation
terms retain linear sensitivity to modifications of a single decay
vertex. In $\mathcal{W}_3^{ab}$, the spin-correlation term is instead bilinear
in the azimuthal coefficients of the top and antitop decays and therefore
requires non-zero new physics contributions in both decay matrices. Since
each lepton-bottom azimuthal coefficient starts at
\(\mathcal{O}(\Lambda^{-2})\), the resulting deviation from the flat SM
prediction begins at \(\mathcal{O}(\Lambda^{-4})\) and is therefore
quadratically suppressed.
 \begin{figure}[t]
    \centering
        \includegraphics[width=0.445\linewidth]{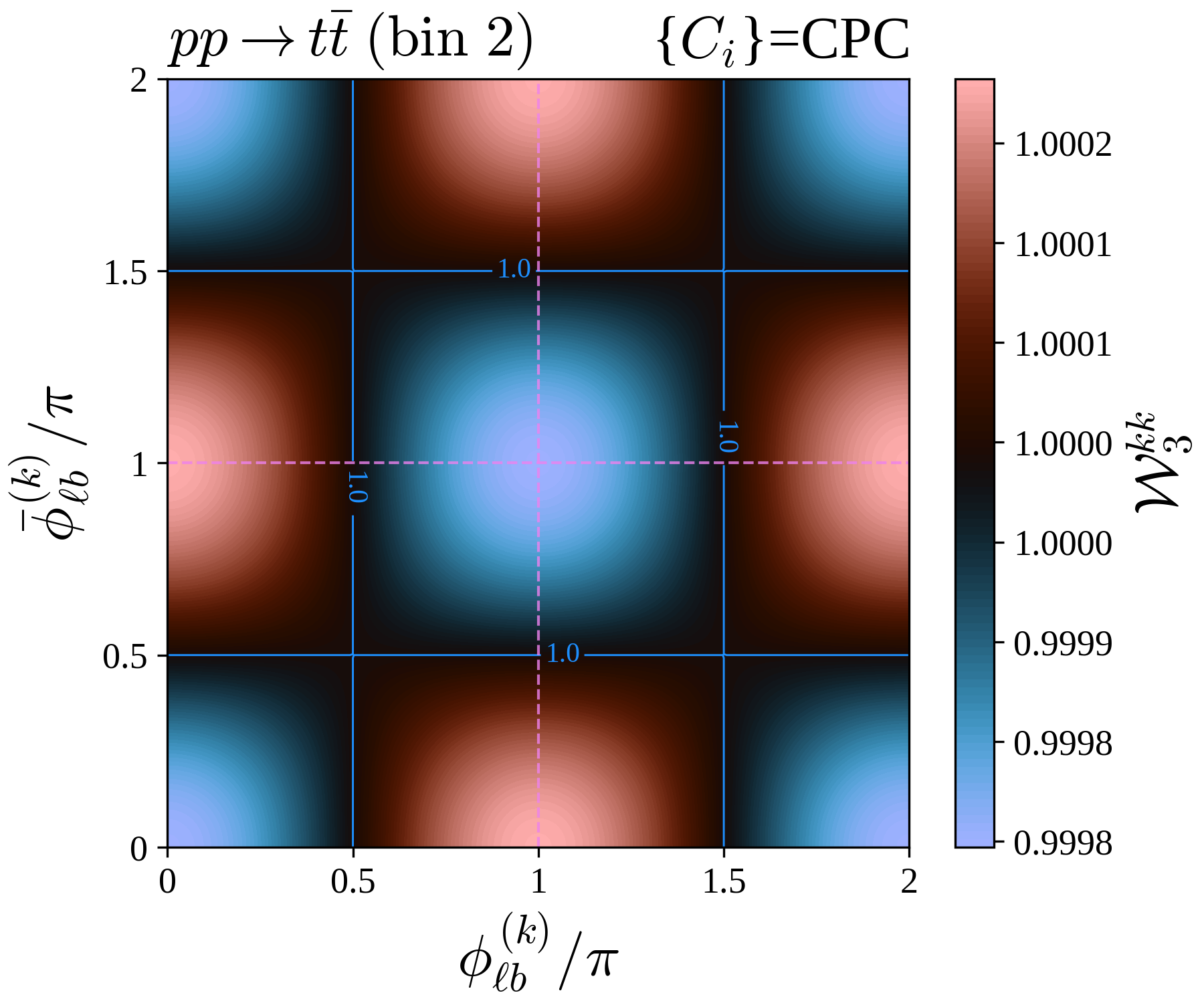}
    \includegraphics[width=0.445\linewidth]{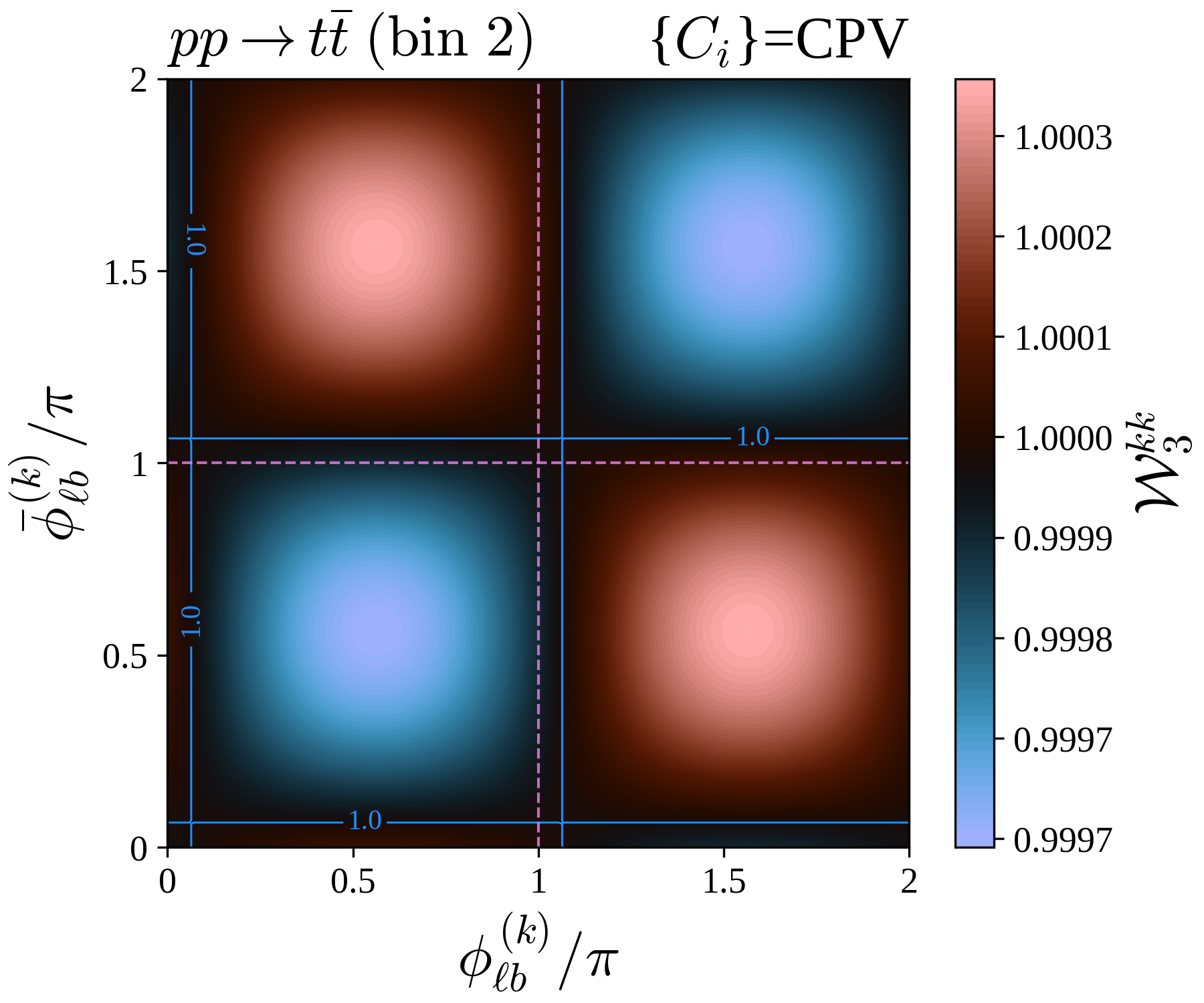} \\
       \includegraphics[width=0.445\linewidth]{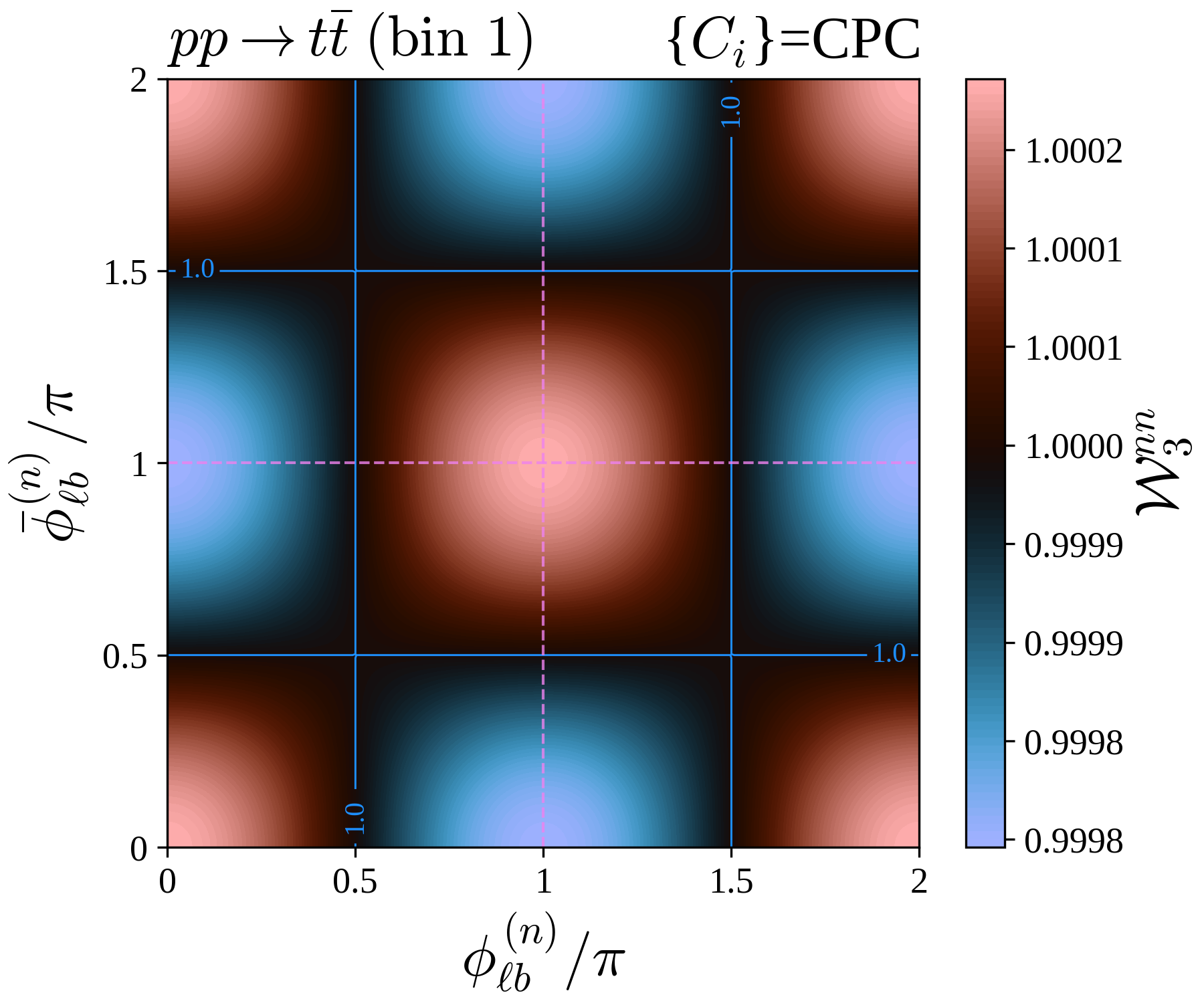}
    \includegraphics[width=0.445\linewidth]{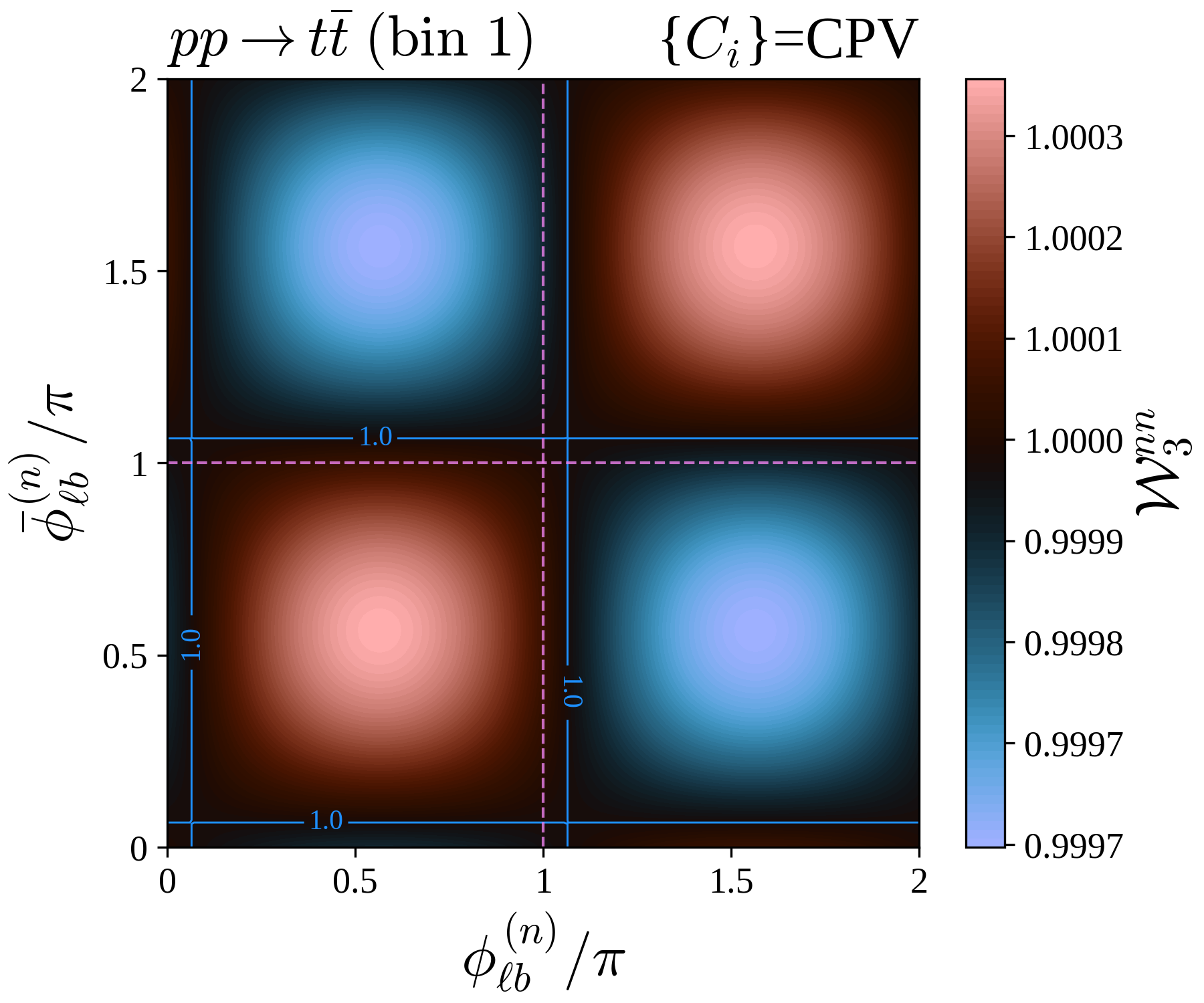}
    
\caption{Normalised two-angle distribution
\(\mathcal{W}_{3}^{ab}
(\phi_{\ell b}^{(a)},\bar{\phi}_{\ell b}^{(b)})\),
defined in Eq.~\eqref{eq:dist3_col}, for \(pp\to t\bar t\) production,
using the polarisation and spin-correlation coefficients extracted from
the simulations. The top and bottom rows correspond to
\((a,b)=(k,k)\) in bin~2 and \((a,b)=(n,n)\) in bin~1, respectively,
while the left and right columns show the CPC and CPV benchmark scenarios.
The dashed lines mark \(\phi_{\ell b}^{(a)}=\pi\) and
\(\bar{\phi}_{\ell b}^{(b)}=\pi\), while the solid blue contours indicate
\(\mathcal{W}_{3}^{ab}=1\).}
\label{fig:dist3_pp}
\end{figure}  
 Figures~\ref{fig:dist3_pp} and~\ref{fig:dist3_ee_rr} show the behaviour of
distribution~\eqref{eq:dist3_col} for \(pp\to t\bar t\) and
\(e^+e^-\to t\bar t\) production, respectively. For the LHC scenario, we
consider \((a,b)=(k,k)\) in bin~2~\eqref{eq:bin2} and
\((a,b)=(n,n)\) in bin~1~\eqref{eq:bin1}, while for the lepton-collider
scenario we choose \((a,b)=(r,r)\). In both figures, the left and right
panels correspond to the CPC and CPV benchmark scenarios, respectively.
The SM predictions are not displayed, as they exhibit a flat behaviour for both production processes.

The CPC distributions are symmetric under the two independent reflections
\begin{equation}
    \phi_{\ell b}^{(a)}
    \longrightarrow
    2\pi-\phi_{\ell b}^{(a)},
    \qquad
    \bar{\phi}_{\ell b}^{(b)}
    \longrightarrow
    2\pi-\bar{\phi}_{\ell b}^{(b)},
\end{equation}
corresponding to reflections about
\(\phi_{\ell b}^{(a)}=\pi\) and
\(\bar{\phi}_{\ell b}^{(b)}=\pi\), respectively. The sine-dependent
terms generated in the CPV scenario break these separate reflection
symmetries, producing the distinct patterns visible in the right panels
of both figures. More explicitly, the independent reflection decomposition with respect to
\(\phi_{\ell b}^{(a)}=\pi\) and
\(\bar{\phi}_{\ell b}^{(b)}=\pi\) separates the different CP-sensitive
structures. The component odd in \(\phi_{\ell b}^{(a)}\) and even in
\(\bar{\phi}_{\ell b}^{(b)}\) selects the terms proportional to
\(w_s^{\rm CPV}\), while the component even in
\(\phi_{\ell b}^{(a)}\) and odd in
\(\bar{\phi}_{\ell b}^{(b)}\) selects those proportional to
\(\bar w_s^{\rm CPV}\). The component odd in both variables isolates the
double-sine contribution proportional to
\(w_s^{\rm CPV}\bar w_s^{\rm CPV}\). Distribution~\eqref{eq:dist3_col} therefore provides simultaneous access to the lepton-bottom azimuthal structures generated in the top and antitop decay vertices.

For \(pp\to t\bar t\) production at leading order in QCD, the vanishing
single-spin polarisations, \(B_a=\bar{B}_b=0\), imply that the entire
non-trivial structure of \(\mathcal{W}_3^{ab}\) originates from the term
proportional to \(C_{ab}\). 

Although the CPC and CPV benchmarks can be distinguished through their
different angular patterns, their numerical deviations from the flat SM
prediction remain very small. The maximum absolute deviations are at the
level of a few \(10^{-4}\), corresponding to approximately
\(0.02\text{--}0.03\%\). These effects are roughly two orders of
magnitude smaller than those obtained from distributions
\(\mathcal{W}_{1}^{ab}
(\phi_{\ell b}^{(a)},\cos\bar{\theta}_{\ell}^{(b)})\) and
\(\mathcal{W}_{2}^{ab}
(\phi_{\ell b}^{(a)},\bar{\phi}_{\ell}^{(b)})\) in
Sects.~\ref{sect:dist1_discussion} and
\ref{sect:dist2_discussion}, and should therefore be regarded primarily
as an illustration of the angular structure generated by the benchmark
interactions. 

This suppression follows directly from the quadratic dependence of the
spin-correlation contribution on the modified top and antitop decay
coefficients:
\begin{equation}
    C_{ab}\,x_i\,\bar{x}_j
    =
    \mathcal{O}\left(\Lambda^{-4}\right),
    \qquad
    x_i\in
    \left\{v_c,w_s^{\rm CPV}\right\},
    \qquad
    \bar{x}_j\in
    \left\{\bar v_c,\bar w_s^{\rm CPV}\right\}.
\end{equation}
This conclusion is specific to the leading-order production density
matrix. Radiative QCD and electroweak corrections can generate non-zero
single-spin polarisations, thereby activating the contributions
\begin{equation}
    B_a\,x_i,
    \qquad
    \bar B_b\,\bar x_j,
\end{equation}
which are linear in the anomalous decay coefficients and therefore start
at \(\mathcal{O}(\Lambda^{-2})\). Although the higher-order
polarisation coefficients are expected to be small, the corresponding
terms could be numerically relevant relative to the
\(\mathcal{O}(\Lambda^{-4})\) spin-correlation contribution retained
here. A quantitative assessment would require the production density
matrix to be evaluated consistently beyond leading order. The reflection
properties used to separate the cosine and sine modulations are,
however, unaffected.

The similar magnitudes and opposite orientations of the distributions for
the two axis choices follow from the approximate relation $ C_{nn}^{\mathrm{bin\,1}}
    \simeq
    -C_{kk}^{\mathrm{bin\,2}}$.
For an individually unpolarised \(t\bar t\) sample, distributions
\(\mathcal{W}_{1}^{ab}
\) and
\(\mathcal{W}_{2}^{ab}
\) are therefore more
sensitive probes of a modified decay interaction, since they retain
contributions linear in the new physics decay coefficients.
The situation changes substantially for
\(e^+e^-\to t\bar t\) production: the non-vanishing individual
polarisations generate the terms
\begin{equation}
    B_a\,x_i,
    \qquad
    \bar{B}_b\,\bar{x}_j,
\end{equation}
which start already at
\(\mathcal{O}(\Lambda^{-2})\). Consequently, the distribution
\(\mathcal{W}_{3}^{rr}\) acquires considerably larger deviations from the
SM prediction, reaching approximately \(0.9\%\) in the CPC benchmark and
\(1.5\%\) in the CPV benchmark. The reflection properties nevertheless
remain unchanged: the CPC prediction is symmetric about both azimuthal
axes, whereas the CPV contribution breaks the corresponding separate
symmetries.

The deviations obtained for \(\mathcal{W}_{3}^{rr}\) are of the same
percent-level magnitude as those found for the subtracted distributions
\(\mathcal{W}_{8}^{rr}\) and \(\mathcal{W}_{9}^{rn}\), but no subtraction
is required in this case. With a SM prediction  exactly flat, the interpretation of 
CP-even and CP-odd harmonics is straightforward, although resolving percent-level deviations still requires adequate
experimental precision. This contrasts with the full distributions shown in
Figs.~\ref{fig:dist1_ee_rr} and~\ref{fig:dist2_ee_rn}, where the
new physics modulations are superimposed on a larger
polarisation-induced SM contribution.

The three classes of distributions discussed above therefore provide
complementary probes of modified top-quark decay interactions. For
unpolarised \(pp\to t\bar t\) production, distributions \(\mathcal{W}_{1}^{ab}\) and \(\mathcal{W}_{2}^{ab}\) are
generally more sensitive because their spin-correlation contributions
retain terms linear in the new physics decay coefficients, whereas
distribution \(\mathcal{W}_{3}^{ab}\) starts only at
\(\mathcal{O}(\Lambda^{-4})\). Nevertheless, distribution \(\mathcal{W}_{3}^{ab}\) probes the lepton-bottom
azimuthal structures of the top and antitop decays simultaneously and
exhibits a distinctive pattern under the two independent angular
reflections.

For \(e^+e^-\to t\bar t\) production, the non-vanishing single-spin
polarisations give all three distributions linear sensitivity at
\(\mathcal{O}(\Lambda^{-2})\), and the corresponding deviations can be
of a similar numerical order. Their sensitivity is nevertheless
distributed differently over the angular phase space and is controlled by
different combinations of polarisation and spin-correlation coefficients.

In particular, distributions \(\mathcal{W}_{1}^{ab}\) and
\(\mathcal{W}_{2}^{ab}\) contain sizeable polarisation-induced SM
contributions, while distribution \(\mathcal{W}_{3}^{ab}\) remains constant in
the SM. As already commented above, a flat baseline makes the origin of any fitted harmonic more
transparent at parton level. Yet, per se, it does not imply greater
experimental sensitivity, which is controlled by the size of the deviation
and the associated uncertainties. A combined analysis of the three
distributions is therefore important both for increasing sensitivity and
for distinguishing the CP-even and CP-odd structures responsible for a
possible deviation.
\subsection{Single-variable  distributions sensitive to CP violation in decay}
\label{sec:additional-projections}

Further single-variable observables can be constructed from the
two-angle distributions derived in the previous subsection. One
possibility is to integrate over one of the two angular variables,
yielding a one-angle distribution associated with a single decay chain.
Alternatively, the two angles can be combined into a single variable,
thereby retaining information on the correlations between the top and
antitop decays.

One-angle distributions that preserve their SM functional form, including
those employed in the tomographic reconstruction and the single-lepton
azimuthal distributions defined in
Eqs.~\eqref{eq:dist6} -\eqref{eq:dist6s}, have already been discussed in
Sec.~\ref{subsect:BSM_decay}. In the following, we instead focus mainly on
single-variable projections whose angular dependence is generated by
modifications of the \(Wtb\) decay vertex and which therefore provide
direct sensitivity to CP-even and CP-odd decay interactions.

Starting with the single-angle observables, the distributions in the
relative lepton--bottom azimuthal angles
\(\phi_{\ell b}^{(a)}\) and
\(\bar{\phi}_{\ell b}^{(b)}\), defined in
Eqs.~\eqref{eq:dist7} and~\eqref{eq:dist7s}, acquire a non-trivial angular
dependence only in the presence of a modified \(Wtb\) vertex. They can be
written as:
\begin{align}
    \mathcal{W}_7^a
    \left(\phi_{\ell b}^{(a)}\right)
    &=
    1
    +B_a
    \left[
        \textcolor{Bittersweet}{
        \frac{v_c}{4\pi}\cos\phi_{\ell b}^{(a)}}
        +
        \textcolor{teal}{
        \frac{w_s^{\rm CPV}}{4\pi}
        \sin\phi_{\ell b}^{(a)}}
    \right],
    \label{eq:dist7_col}
    \\
    \mathcal{W}_{\bar{7}}^b
    \left(\bar{\phi}_{\ell b}^{(b)}\right)
    &=
    1
    +\bar{B}_b
    \left[
        \textcolor{Bittersweet}{
        \frac{\bar v_c}{4\pi}
        \cos\bar{\phi}_{\ell b}^{(b)}}
        +
        \textcolor{teal}{
        \frac{\bar w_s^{\rm CPV}}{4\pi}
        \sin\bar{\phi}_{\ell b}^{(b)}}
    \right].
    \label{eq:dist7s_col}
\end{align}
The cosine modulation is controlled by the CP-even coefficients
\(v_c\) and \(\bar v_c\), whereas the sine modulation is generated by the
CP-odd coefficients \(w_s^{\rm CPV}\) and
\(\bar w_s^{\rm CPV}\). Since these terms are multiplied by the
single-spin polarisations \(B_a\) and \(\bar{B}_b\), the distributions retain
sensitivity to the modified decay vertex only when the corresponding top
or antitop quark is individually polarised. In the absence of
single-particle polarisation, both distributions reduce to the flat
prediction
\begin{equation}
    \mathcal{W}_7^a
    =
    \mathcal{W}_{\bar{7}}^b
    =
    1,
\end{equation}
independently of the values of the decay coefficients.

\begin{figure}[t]
    \centering
    \includegraphics[width=0.7\linewidth]{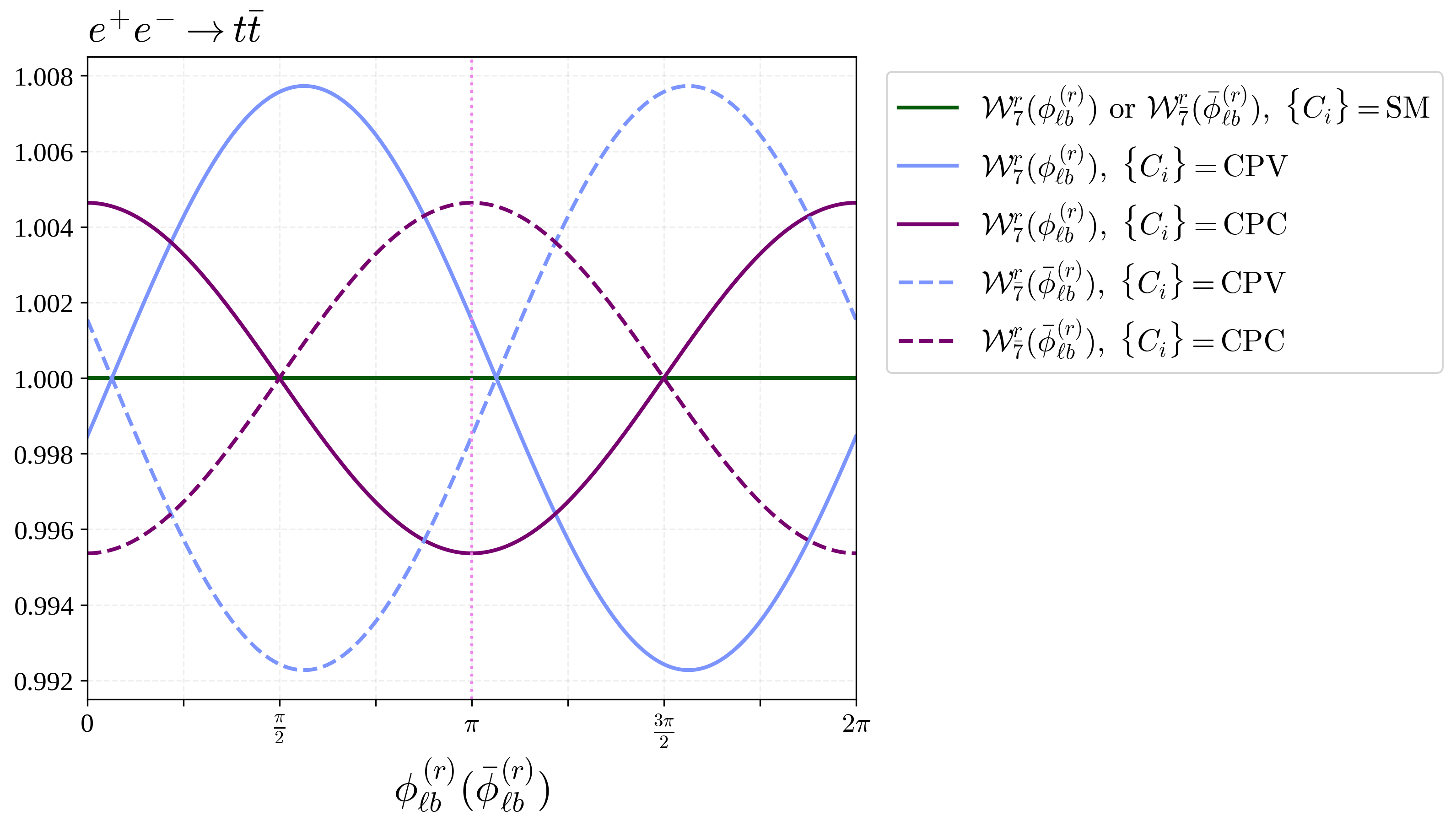}
    \caption{Single-angle distributions
\(\mathcal{W}_7^r(\phi_{\ell b}^{(r)})\) and
\(\mathcal{W}_{\bar{7}}^r(\bar{\phi}_{\ell b}^{(r)})\) for
\(e^+e^-\to t\bar t\) production at
\(\sqrt{s}=365~\mathrm{GeV}\). The panel shows the
SM, CPC, and CPV benchmark scenarios. The dotted vertical lines mark
\(\phi_{\ell b}^{(r)}=\pi\) and
\(\bar{\phi}_{\ell b}^{(r)}=\pi\), respectively.}
    \label{fig:dist7_ee}
\end{figure}

Figure~\ref{fig:dist7_ee} shows
\(\mathcal{W}_7^r(\phi_{\ell b}^{(r)})\) and
\(\mathcal{W}_{\bar{7}}^r(\bar{\phi}_{\ell b}^{(r)})\) for the SM, CPC, and
CPV benchmark scenarios in \(e^+e^-\to t\bar t\) production at
\(\sqrt{s}=365~\mathrm{GeV}\). In the CPC scenario, the distributions are
symmetric under the reflections
\[
    \phi_{\ell b}^{(r)}
    \to 2\pi-\phi_{\ell b}^{(r)},
    \qquad
    \bar{\phi}_{\ell b}^{(r)}
    \to 2\pi-\bar{\phi}_{\ell b}^{(r)},
\]
whereas the CP-odd sine modulations present in the CPV scenario break
these symmetries.

The maximum absolute deviation from the SM prediction is approximately
\(0.8\%\) in the CPV scenario and \(0.5\%\) in the CPC scenario. The
sensitivity is therefore lower than that obtained from some of the
two-angle observables involving the same spin axis, such as
\(\mathcal{W}_8^{rr}\). 
The distributions
\(\mathcal{W}_7^r\) and \(\mathcal{W}_{\bar{7}}^r\) provide
one-dimensional projections using an angle from a single decay chain. They
may be useful for limited event samples, where a two-dimensional distribution
would be sparsely populated. This simplification, however, comes at the cost of integrating out part of the spin-correlation information.

We underline that, since \(O_{\phi Q}^{3}\) and \(O_{tW}\) modify both the
production and decay stages, for the CPV scenario the presence of an imaginary component for \(C_{tW}\) can induce a net asymmetry between the polarisation vectors $\Delta\mathbf{B}\neq0$, as reported in
Tab.~\ref{tab:spin-pol-corr-ee-fig}. In particular, for the components entering the distributions in Fig.~\ref{fig:dist7_ee}, we find: \begin{equation}
    \left|B_r-\bar{B}_r\right|
    \sim 10^{-5},
    \qquad
    \{C_i\}=\mathrm{CPV}.
\end{equation} Therefore, any difference in the amplitude of the two distributions induced by $\Delta B_r$ is numerically negligible and not visible in the scale of the figure. 

The angular
patterns seen in Fig.~\ref{fig:dist7_ee} instead originate from the opposite sign of the decay
coefficients in the charge-conjugate top and antitop channels, see Eq.~\eqref{eq:rel_tandtbar_decaycoeff}. Consequently, both the CP-even cosine modulation and the CP-odd sine
modulation reverse sign between
\(\mathcal{W}_7^r\) and \(\mathcal{W}_{\bar{7}}^r\).

 We finally consider single-variable distributions constructed from
algebraic combinations of two azimuthal angles associated with the two
decay chains. As for the combinations of polar angles
\(\xi^{(ab)}\) and \(\xi_-^{(ab)}\), defined in
Eqs.~\eqref{eq:diff-xi-SM} and~\eqref{eq:diff-xi_MINUS-SM},
respectively, the projection from two angular variables onto a single
combined variable removes the terms proportional to the individual
polarisations. The resulting distributions nevertheless retain a linear
dependence on the relevant spin-correlation coefficients.

We define the sum and difference of the two azimuthal angles as
\begin{equation}
    \phi_{n,m}^{\pm(a,b)}
    \equiv
    \left[
        \phi_n^{(a)}
        \pm
        \bar{\phi}_m^{(b)}
    \right]_{2\pi},
    \qquad
    n,m\in\{\ell,\ell b\},
    \label{eq:combined_azimuthal_angles}
\end{equation}
where the symbol \([\cdots]_{2\pi}\) denotes the mapping of the resulting
angle onto the interval \([0,2\pi)\). Starting from
Eqs.~\eqref{eq:dist2}--\eqref{eq:dist3} and~\eqref{eq:dist5}, one obtains,
among others, the following single-variable distributions:
\begin{align}
    \frac{2\pi}{\sigma}\frac{d\sigma}{d\phi^{\pm(a,b)}_{\ell b,\ell }}&=1+ \frac{\bar{\alpha}_\ell}{32}\left[ (\textcolor{Bittersweet}{v_c}C_{aj_1}\mp \textcolor{teal}{w^{\rm CPV}_s}C_{aj_2})\cos\phi^{\pm(a,b)}_{\ell b,\ell }+ ( \textcolor{teal}{w^{\rm CPV}_s}C_{aj_1}\pm \textcolor{Bittersweet}{v_c}C_{aj_2})  \sin \phi^{\pm(a,b)}_{\ell b,\ell } \right]\,,
    \label{eq:dist2abpm}\\
   \frac{2\pi}{\sigma}\frac{d\sigma}{d\phi^{\pm(a,b)}_{\ell b,\ell b}}&=1+ \frac{C_{ab}}{32\pi^2}\left[ (\textcolor{Bittersweet}{v_c\bar{v}_c}\mp\textcolor{teal}{ w_s^{\rm CPV} \bar{w}_s^{\rm CPV}})\cos\phi^{\pm(a,b)}_{\ell b,\ell b}+ (\textcolor{Bittersweet}{\bar{v}_c}\textcolor{teal}{w_s^{\rm CPV}}\pm \textcolor{Bittersweet}{v_c} \textcolor{teal}{\bar{w}_s^{\rm CPV}})  \sin\phi^{\pm(a,b)}_{\ell b,\ell b}  \right]\,,
    \label{eq:dist3pm}\\
    \frac{2\pi}{\sigma}\frac{d\sigma}{d\phi^{\pm(a,b)}_{\ell,\ell}}&=1+ \alpha_\ell\bar{\alpha}_\ell\frac{\pi^2}{32}\left[(C_{i_1j_1}\mp C_{i_2j_2})\cos\phi^{\pm(a,b)}_{\ell,\ell}+( C_{i_2j_1}\pm C_{i_1j_2})\sin\phi^{\pm(a,b)}_{\ell,\ell}\right]\,.
    \label{eq:dist5pm}
\end{align}
Further non-equivalent distributions can be obtained by exchanging the
roles of the top and antitop decay chains. We do not show separate numerical
plots for these combined-angle observables because they are one-dimensional
projections of the two-angle distributions already illustrated in
Sects.~\ref{sect:dist2_discussion} and~\ref{sect:dist3_discussion} and do not
introduce qualitatively new harmonic structures. The versions obtained by
exchanging the two decay chains are related by the corresponding relabelling
of the top and antitop angular variables.

The distributions in Eqs.~\eqref{eq:dist2abpm}
and~\eqref{eq:dist3pm} are flat for SM decay vertex and remain
sensitive to modified decay interactions even when the individual
top and antitop polarisations vanish. Distribution~\eqref{eq:dist2abpm}
depends linearly on \(v_c\) and \(w_s^{\rm CPV}\), and therefore receives
new physics contributions starting at
\(\mathcal{O}(\Lambda^{-2})\). By contrast,
distribution~\eqref{eq:dist3pm} is bilinear in the decay coefficients of
the top and antitop channels and consequently starts at
\(\mathcal{O}(\Lambda^{-4})\).

For generic nonzero values of both \(C_{aj_1}\) and \(C_{aj_2}\), the
cosine and sine harmonics in Eq.~\eqref{eq:dist2abpm} contain mixtures of
the CP-even coefficient \(v_c\) and the CP-odd coefficient
\(w_s^{\rm CPV}\). Consequently, the observation of a cosine or sine
modulation in either
\(\phi_{\ell b,\ell}^{+(a,b)}\) or
\(\phi_{\ell b,\ell}^{-(a,b)}\) does not, by itself, provide a direct
classification of the CP nature of the decay interaction.
If the relevant spin-correlation coefficients are known, the cosine and
sine coefficients extracted from either angular distribution are
sufficient to disentangle \(v_c\) and \(w_s^{\rm CPV}\). The distribution
in the other combined angle then provides a complementary determination
and a consistency check.

If \(C_{aj_1}\) and \(C_{aj_2}\) are not known independently, the two
distributions can instead be analyzed simultaneously. Denoting by
\(A_\pm\) and \(D_\pm\) the cosine and sine coefficients extracted from
independent harmonic fits to the distributions in
\(\phi_{\ell b,\ell}^{+(a,b)}\) and
\(\phi_{\ell b,\ell}^{-(a,b)}\), one finds
\begin{align}
    A_++A_- &\propto
    \textcolor{Bittersweet}{v_c}\,C_{aj_1},
    &
    A_--A_+ &\propto
    \textcolor{teal}{w_s^{\rm CPV}}\,C_{aj_2},
    \\
    D_++D_- &\propto
    \textcolor{teal}{w_s^{\rm CPV}}\,C_{aj_1},
    &
    D_+-D_- &\propto
    \textcolor{Bittersweet}{v_c}\,C_{aj_2}.
\end{align}
These combinations separately determine the products of the decay
coefficients with the corresponding spin-correlation coefficients.
Moreover, whenever the relevant denominators are nonzero, the ratios of
the corresponding combinations determine
\(w_s^{\rm CPV}/v_c\). The relative size of the CP-odd and CP-even decay
contributions can therefore be extracted without knowing the overall
magnitude of the spin-correlations. Their absolute values, however,
remain degenerate with \(C_{aj_1}\) and \(C_{aj_2}\) and require either
an independent determination of the spin-correlation coefficients or a
simultaneous fit including additional observables.

The interpretation becomes particularly transparent when one of the two
spin-correlation coefficients vanishes. For \(C_{aj_2}=0\), the cosine
and sine modulations directly probe \(v_c\) and \(w_s^{\rm CPV}\),
respectively. For \(C_{aj_1}=0\), this correspondence is reversed: the
cosine modulation probes \(w_s^{\rm CPV}\), while the sine modulation
probes \(v_c\). Thus, the CP nature of a given harmonic is determined by
the structure of the relevant spin-correlations rather than by the
trigonometric function alone. If both coefficients vanish, the
distribution is flat and loses sensitivity to the modified decay vertex.

Using the relations between the barred and unbarred decay coefficients
given in Eq.~\eqref{eq:rel_tandtbar_decaycoeff},
the two distributions in Eq.~\eqref{eq:dist3pm} acquire different
harmonic structures. For the distribution in
\(\phi_{\ell b,\ell b}^{-(a,b)}\), the sine coefficient vanishes
identically, independently of the presence of CP-odd decay interactions.
Its angular dependence is therefore entirely described by a cosine
modulation proportional to
\(-[v_c^2+(w_s^{\rm CPV})^2]\).

By contrast, the distribution in
\(\phi_{\ell b,\ell b}^{+(a,b)}\) contains a cosine modulation
proportional to
\((w_s^{\rm CPV})^2-v_c^2\) and a sine modulation proportional to
\(-2v_cw_s^{\rm CPV}\). The latter vanishes in the CPC limit. Hence, a
breaking of the reflection symmetry,
\begin{equation}
    \phi_{\ell b,\ell b}^{+(a,b)}
    \longrightarrow
    2\pi-\phi_{\ell b,\ell b}^{+(a,b)}\,,
\end{equation}
provides a direct indication of a CP-odd contribution to the modified
decay vertex. The absence of such an asymmetry does not, however, exclude
CP violation, since the sine coefficient also vanishes when \(v_c=0\). Provided that \(C_{ab}\) is known and non-zero, the cosine coefficients of
the two distributions can be combined to determine
\(v_c^2\) and \((w_s^{\rm CPV})^2\), and therefore
\(\lvert v_c\rvert\) and \(\lvert w_s^{\rm CPV}\rvert\). The relative
sign of \(v_c\) and \(w_s^{\rm CPV}\) is encoded in the sine coefficient
of the
\(\phi_{\ell b,\ell b}^{+(a,b)}\)-dependent distribution.

Finally, we consider the distributions in Eq.~\eqref{eq:dist5pm}, which
generalise the sum-and-difference distributions in the charged-lepton
azimuthal angles introduced in Ref.~\cite{Baumgart:2012ay}. Unlike the
previous combined-azimuthal observables, these distributions are already
non-flat in the SM. Modified decay interactions enter only through the overall
factor \(\alpha_\ell\bar\alpha_\ell\), which rescales the modulation
amplitudes without changing the harmonic structure. Because the harmonic
coefficients involve specific combinations of the spin-correlation matrix,
these observables also probe the production spin structure and its CP
properties.

First, for a given choice of the relevant angular variable, the distribution becomes flat only
if the two corresponding combinations vanish simultaneously:
\begin{equation}
    C_{i_1j_1}\mp C_{i_2j_2}=0,
    \qquad
    C_{i_2j_1}\pm C_{i_1j_2}=0.
\end{equation}
Additionally, an explicit dependence on the symmetric and antisymmetric components of the spin-correlation matrix is displayed for the same-axes combinations $(a,a)$. For example, using the symmetric and antisymmetric parts of the correlations matrix, 
the distribution constructed with the difference between the two charged-lepton azimuthal angles can be written as: 
\begin{align}
    \mathcal{W}^{aa}_{10}\left(\phi_{\ell,\ell}^{-(a,a)}\right)\equiv \frac{2\pi}{\sigma}
    \frac{d\sigma}
    {d\phi_{\ell,\ell}^{-(a,a)}}
    &=
    1+
    \alpha_\ell\bar{\alpha}_\ell\frac{\pi^2}{32}
    \Bigl[
        \left(
            C_{i_1i_1}+C_{i_2i_2}
        \right)
        \cos\phi_{\ell,\ell}^{-(a,a)}
        +2C_{i_2i_1}^{A}
        \sin\phi_{\ell,\ell}^{-(a,a)}
    \Bigr].
    \label{eq:dist5m_CA}
\end{align}
Within the symmetry classification of Ref.~\cite{Lamba:2026jfm},
\(C_{i_2i_1}^{A}\neq0\) is a CP-violation marker in the common spin basis
when CP-related kinematic configurations are compared. The contribution can
be extracted from the antisymmetric angular combination: 
\begin{equation}
    \mathcal{W}_{10,A}^{aa}
    \left(\phi_{\ell,\ell}^{-(a,a)}\right)
    \equiv
    \frac{1}{2}
    \left[
        \mathcal{W}^{aa}_{10}\left(\phi_{\ell,\ell}^{-(a,a)}\right)
        -
        \mathcal{W}^{aa}_{10}\left(2\pi -\phi_{\ell,\ell}^{-(a,a)}\right)    \right]= \alpha_\ell\bar{\alpha}_\ell\frac{\pi^2}{16}C_{i_2i_1}^{A}\sin\phi_{\ell,\ell}^{-(a,a)}. 
\end{equation}
Therefore, the distribution $\mathcal{W}^{aa}_{10}\left(\phi_{\ell,\ell}^{-(a,a)}\right)$ and its combination $\mathcal{W}_{10,A}^{aa}
\left(\phi_{\ell,\ell}^{-(a,a)}\right)$ constitute complementary observables with respect to the distribution in Eq.~\eqref{eq:diff-xi_MINUS-SM} for the extraction of the antisymmetric part of the spin-correlation matrix and the eventual detection of CP violation. Analogously, the $\phi_{\ell,\ell}^{+(a,a)}$--distribution is proportional to the symmetric combination $C_{i_2i_1}^{S}$ and to $C_{i_1i_1}-C_{i_2i_2}$ in its sine and cosine modulation, respectively. 
\subsection{Combined strategy for separating CP violation in production and decay}
\label{sec:strategy}

The results obtained in this and the previous section
suggest a two-stage analysis. Polar-angle observables reconstruct the
production density matrix, while observables involving the relative
\(b\)--lepton azimuthal angles test the decay vertices. Their combination
separates CP-odd effects generated in production from those generated in
the top or antitop decay.

The polar-angle distributions in
Eqs.~\eqref{eq:diff-single-polar-angle-SM}--\eqref{eq:diff-xi-SM}
retain their tomographic form in the presence of the anomalous \(Wtb\)
interactions considered here. They determine the combinations
\begin{equation}
    \alpha_\ell B_a,
    \qquad
    \bar\alpha_\ell \bar B_b,
    \qquad
    \alpha_\ell\bar\alpha_\ell C_{ab}.
\end{equation}
Within the present decay framework,
\(\bar\alpha_\ell=-\alpha_\ell\), and \(\alpha_\ell\) depends only on
CP-even combinations of the anomalous couplings. Consequently, assuming
the SM analysing powers when the decay vertex is modified rescales the
inferred values of \(\mathbf B\), \(\bar{\mathbf B}\), and \(\mathbf C\),
but does not by itself generate a polarisation difference or an
antisymmetric spin-correlation matrix. In this case, however, the analysing powers should
nevertheless be included as fit parameters  in a
precision reconstruction.

After comparing CP-related phase-space configurations and expressing the
two spins in the common spin basis, CP invariance of the production
density matrix requires
\begin{equation}
    \Delta\mathbf B
    =0,
    \qquad
    \mathbf C^A
    =0.
\end{equation}
The distribution in \(\xi_-^{(ab)}\),
Eq.~\eqref{eq:diff-xi_MINUS-SM}, directly probes \(C^A_{ab}\).
The charged-lepton azimuthal distribution in
Eq.~\eqref{eq:dist5m_CA} provides a complementary projection of the
same antisymmetric spin-correlation structure. A statistically
significant non-zero value of either \(\Delta\mathbf B\) or
\(\mathbf C^A\) therefore identifies CP violation in production within
the assumptions of the factorised analysis.

Anomalous decay interactions are instead identified through the relative
azimuthal angles \(\phi_{\ell b}^{(a)}\) and
\(\bar\phi_{\ell b}^{(b)}\). For a distribution
\(\mathcal W(\phi_{\ell b}^{(a)},\ldots)\), the reflection-odd component
\begin{equation}
    \mathcal W_A(\phi_{\ell b}^{(a)},\ldots)
    =
    \frac{1}{2}
    \left[
        \mathcal W(\phi_{\ell b}^{(a)},\ldots)
        -
        \mathcal W(2\pi-\phi_{\ell b}^{(a)},\ldots)
    \right]
\end{equation}
isolates the contribution proportional to \(w_s^{\rm CPV}\), whereas
the corresponding reflection-even modulation probes \(v_c\). Thus, in
the tree-level setup without absorptive phases, a non-zero
reflection-odd component diagnoses CP violation in the decay vertex. 

The preferred decay-sensitive observables depend on the production spin
state. When
\(\mathbf B=\bar{\mathbf B}=0\) but \(\mathbf C\neq0\), the mixed
distributions in Eqs.~\eqref{eq:dist1_col} and~\eqref{eq:dist2_col},
or their subtracted forms in
Eqs.~\eqref{eq:distSUB_1} and~\eqref{eq:distSUB_2}, retain
\(\mathcal O(\Lambda^{-2})\) sensitivity through the spin-correlations.
In the same limit, the single-decay distributions
\(\mathcal W_7\) and \(\mathcal W_{\bar 7}\) are flat, while
\(\mathcal W_3\), Eq.~\eqref{eq:dist3_col}, starts at
\(\mathcal O(\Lambda^{-4})\). When non-zero single-spin polarisations are
present, \(\mathcal W_7\), \(\mathcal W_{\bar 7}\), and
\(\mathcal W_3\) also acquire contributions linear in the anomalous decay
coefficients.

A consistent analysis should therefore fit the production parameters
\(\mathbf B\), \(\bar{\mathbf B}\), and \(\mathbf C\) together with the
decay parameters \(\alpha_\ell\), \(v_c\), and
\(w_s^{\rm CPV}\), using the barred coefficients fixed by the
charge-conjugation relations when appropriate. Subject to the assumptions above and barring accidental cancellations, a
violation of \(\Delta\mathbf B=0\) or \(\mathbf C^A=0\), in the absence
of a decay-side reflection-odd modulation, points to CP violation in
production. Conversely, a reflection-odd \(b\)--lepton azimuthal
modulation, with the production CP relations satisfied, points to CP
violation in decay. The simultaneous observation of both structures
indicates contributions from both stages.
A non-flat \(\mathcal W_3\), \(\mathcal W_7\), or
\(\mathcal W_{\bar 7}\) signals an anomalous decay interaction and
therefore motivates fitting \(\alpha_\ell\), but does not by itself imply
a sizeable departure of \(\alpha_\ell\) from its SM value.


\section{Conclusions}
\label{sec:conclusions}

We have developed a quantum-tomographic framework for the study of CP violation in top-quark pair production and decay. The starting point is the factorisation, in the narrow-width approximation, of the full process
\begin{equation}
    I\to t\bar t\to b\ell^+\nu\,\bar b\ell^-\bar\nu\,, 
\end{equation}
into a production density matrix and two decay density matrices. This factorised structure allows the Fano--Bloch coefficients of the
\(t\bar t\) quantum state, namely the polarisation vectors
\(\mathbf B\) and \(\bar{\mathbf B}\) and the spin-correlation matrix
\(\mathbf C\), to be related directly to angular distributions of the
decay products.

We have first reviewed the standard tomographic reconstruction for SM top
decays, where the charged leptons act as optimal spin analysers, and then
extended the analysis to a general anomalous \(Wtb\) vertex. The polar-angle
distributions retain the same functional form as in the SM, with anomalous
decay effects entering through possible shifts of the spin-analysing powers
\(\alpha_\ell\) and \(\bar\alpha_\ell\). They therefore retain a direct
tomographic interpretation, provided that the analysing powers are fitted
together with \(\mathbf B\), \(\bar{\mathbf B}\), and \(\mathbf C\).
Fixing the analysing powers to their SM values in the presence of anomalous
decays would instead bias the extracted normalisations.

The anomalous decay vertex also generates new angular structures involving
the relative azimuthal angles between the charged lepton and the \(b\) jet.
In particular, CP violation in the decay produces sine modulations in
\(\phi_{\ell b}\) and \(\bar\phi_{\ell b}\), controlled by
\(w_s^{\rm CPV}\) and \(\bar w_s^{\rm CPV}\). These terms vanish in the SM
and, within the factorised framework considered here, provide a clean probe
of CP-odd contributions to the decay vertex.

On this basis, we have devised a combined strategy for separating
production and decay CP violation through their distinct angular
signatures. Polar-angle tomography reconstructs the production density
matrix and tests its CP properties through
\(\Delta\mathbf{B}\) and \(\mathbf{C}^{A}\), while the reflection-odd
components of the relative \(b\)--lepton azimuthal distributions probe
CP violation in the decay vertices through \(w_s^{\rm CPV}\) and
\(\bar w_s^{\rm CPV}\). Combining these complementary classes of observables provides, within the
factorised framework considered here, a systematic way to distinguish
CP-odd effects originating in production, in decay, or in both.

The numerical examples presented here are intended to illustrate the
information contained in the different angular structures rather than to
provide a complete sensitivity analysis. They are based on parton-level
predictions and benchmark choices of the anomalous couplings; a realistic
assessment will require detector effects, backgrounds, event reconstruction,
and correlated experimental and theoretical uncertainties.

The HL-LHC provides the immediate opportunity to pursue this programme.
Recent measurements of top-quark spin entanglement have already demonstrated
that information on the \(t\bar t\) quantum state can be extracted from
dileptonic angular distributions. The larger data set expected at the HL-LHC
should enable more differential measurements, including analyses in
phase-space regions optimised for specific polarisation and spin-correlation
components. Important theoretical extensions include higher-order QCD and
electroweak corrections and a simultaneous fit of 
Wilson coefficients in production and in decay.  A future lepton collider would provide a complementary
environment, with cleaner kinematics and non-vanishing single-top
polarisations facilitating the joint analysis of production and decay effects.


\section*{Acknowledgements}
F.M. and O.M. thank the CERN Theory Department for hospitality when the final part of this work was performed. O.M. has received partial support from the European Union’s Horizon 2020 research and innovation programme under the Marie Skłodowska-Curie Staff Exchange grant agreement No 101086085-ASYMMETRY. P.L. and F.M. are partially supported by the F.R.S.–FNRS (Belgian Fund for Scientific Research) through the IISN convention 4.4517.08, ``Theory of fundamental interactions''. P.L. also acknowledges the support by the Fonds Special de Recherche at UCLouvain.  E.V. is supported by the European Research Council (ERC) under the European Union’s Horizon 2020 research and innovation programme (Grant agreement No. 949451).


\appendix

\section{Three-body phase space for top and antitop decays}\label{app:3-body-phase-space}

Here we present an explicit derivation of the three-body phase space for
top-quark decay in the rest frame of the decaying particle. The five
independent phase-space degrees of freedom are parametrised in terms of two
invariant masses and three angular variables. The derivation for the
antitop-quark decay proceeds analogously, with the corresponding final-state
variables treated as an independent set.

The same three-dimensional Cartesian basis is used in the rest frames of
both decaying particles and is chosen to coincide with the helicity basis
defined in the top-quark rest frame. In constructing the angular variables,
we take the charged lepton as the spin analyser, while the neutrino
four-momentum is eliminated using momentum conservation. This choice is
made for convenience, however the roles of the final-state momenta can be
interchanged if a different set of kinematic or angular dependence is
desired.
The Lorentz-invariant three-body phase-space measure takes the standard form
\begin{equation}
  d\Phi_3
    =  (2\pi)^4\delta^4(p_t-\sum_ip_i)\prod_{i}\frac{d^3\vec{p_i}}{(2\pi)^32E_i}\,. 
    \label{eq:threebody}
\end{equation}
Using the on-shell identity
\begin{equation}
    \frac{d^3\vec p_a}{2E_a}
    =
    d^4p_a\,
    \delta\!\left(p_a^2-m_a^2\right)
    \Theta\!\left(p_a^0\right),
\end{equation}
the neutrino phase-space measure can be expressed as a four-dimensional
integral. Performing the \(d^4p_n\) integration using the
four-momentum-conserving delta function then gives
\begin{equation}
    d\Phi_3
    =
    \frac{1}{(2\pi)^5}
    \frac{d^3\vec p_\ell}{2E_\ell}
    \frac{d^3\vec p_b}{2E_b}\,
    \delta\!\left[\left(p_t-p_b-p_\ell\right)^2\right]
    \Theta\!\left(p_t^0-p_\ell^0-p_b^0\right),
    \label{eq:app-phase-space}
\end{equation}
where the neutrino has been taken to be massless.

We now work in the top-quark rest frame. For a massless charged lepton and
neglecting terms of \(\mathcal{O}(m_b^2)\), the momentum-space measures may
be written as
\begin{equation}
    d^3\vec p_\ell\,d^3\vec p_b
    =
    E_\ell^2\,dE_\ell\,d\Omega_\ell\,
    E_b^2\,dE_b\,d\Omega_{\ell b}.
    \label{eq:3mometa}
\end{equation}
Here,
\(\Omega_\ell=\{\theta_\ell,\phi_\ell\}\) denotes the usual pair of
spherical angles specifying the direction of the charged lepton in the
chosen Cartesian basis. The charged lepton therefore serves as the spin
analyser. The direction of the \(b\) quark is then defined relative to that
of the charged lepton through
\(\Omega_{\ell b}=\{\theta_{\ell b},\phi_{\ell b}\}\), where
\(\theta_{\ell b}\) and \(\phi_{\ell b}\) specify the orientation of
\(\hat p_b\) in the rotated frame whose polar axis is aligned with
\(\hat p_\ell\). The explicit expressions for \(\hat p_\ell\) and
\(\hat p_b\) are given in Eq.~\eqref{eq:4-momenta-in-t-rest}.
With these definitions, the remaining delta function in
Eq.~\eqref{eq:app-phase-space} takes the form
\begin{equation}
    \delta\!\left[
        m_t^2
        -2m_tE_\ell
        -2m_tE_b
        +2E_\ell E_b
        \left(1-\cos\theta_{\ell b}\right)
    \right].
    \label{eq:delta-costhetalb}
\end{equation}
We use this constraint to perform the integration over
\(d\cos\theta_{\ell b}\). Requiring the corresponding solution to lie
within the physical range 
 yields
\begin{equation}
    \frac{m_t-2E_b}{2}
    \leq E_\ell
    \leq \frac{m_t}{2}.
\end{equation}
The positive-energy condition inherited from
Eq.~\eqref{eq:app-phase-space},
\begin{equation}
    \Theta\!\left(m_t-E_\ell-E_b\right),
\end{equation}
further restricts the kinematically allowed region.
To replace the two energies by invariant masses, we define
\begin{align}
    m_{12}^2
    &=
    (p_\ell+p_n)^2
    =
    (p_t-p_b)^2
    \equiv q^2,
    \\
    m_{23}^2
    &=
    (p_n+p_b)^2
    =
    (p_t-p_\ell)^2.
\end{align}
In the top-quark rest frame, these relations give
\begin{equation}
    E_\ell
    =
    \frac{m_t^2-m_{23}^2}{2m_t},
    \qquad
    E_b
    =
    \frac{m_t^2-q^2}{2m_t}.
    \label{eq:energies-invmasses}
\end{equation}
The corresponding Jacobian is
\begin{equation}
    dE_\ell\,dE_b
    =
    \frac{1}{4m_t^2}\,
    dm_{23}^2\,dq^2.
\end{equation}
The integration limits for \(m_{23}^2\) follow directly from the allowed
range of \(E_\ell\), while those for \(q^2\) are determined by the
kinematically accessible invariant mass of the lepton--neutrino system.
The resulting three-body phase-space measure can therefore be written as
\begin{equation}
   d\Phi_3=  \frac{1}{(2\pi)^5}\frac{1}{32m_{t}^2} dm^2_{23}dq^2d\Omega_\ell d\phi_{\ell b}  \quad \textnormal{with :  } \begin{cases}
     \cos\theta_{\ell b}= \frac{m_t^4-m_{23}^2 \left(m_t^2+q^2\right)-m_t^2
   q^2}{\left(m_{23}^2-m_t^2\right) \left(m_t^2-q^2\right)}\\
      0  \leq m_{23}^2 \leq m_{t}^2-q^2 \\
      0\leq q^2\leq (m_t-m_b)^2
      \label{eq:phasespace_t}
     \end{cases}\,. 
\end{equation}
All kinematic limits are given for massless charged leptons and neutrinos,
with terms of \(\mathcal{O}(m_b^2)\) neglected.
The phase space for the antitop-quark decay can be obtained analogously.
    \\
If the NWA is subsequently applied to the squared
decay amplitudes, the squared \(W\)-boson propagators generate contributions
proportional to
\(\delta(q^2-m_W^2)\) and
\(\delta(\bar q^2-m_W^2)\).
The integrations over \(dq^2\) and \(d\bar q^2\) then fix
$q^2=\bar q^2=m_W^2$,
reducing each three-body phase space to four independent variables.

Finally, all quantities entering our calculation are expressed in terms of
scalar products and contractions with the Levi--Civita tensor. Scalar
products are Lorentz invariant, while Levi--Civita contractions are
invariant under proper orthochronous Lorentz transformations. They may
therefore be evaluated equivalently in the top- or antitop-quark rest frame
and in the \(t\bar t\) zero-momentum frame.

The polar axis used to define the spherical angles
\(\theta_\ell\) and \(\phi_\ell\) may be chosen arbitrarily among the three
axes of the Cartesian basis. Since the direction of the \(b\) quark is
specified relative to that of the charged lepton, the angles
\(\theta_{\ell b}\) and \(\phi_{\ell b}\) also depend, in principle, on
this choice. Momentum conservation, however, fixes
\(\theta_{\ell b}\) uniquely in terms of the invariant masses, and this
angle is subsequently removed by the delta-function integration. The only
remaining dependence on the reference axis therefore enters through the
three angular variables \(\theta_\ell\), \(\phi_\ell\), and
\(\phi_{\ell b}\).
According to this parametrisation, the independent four-momenta in the
top-quark rest frame are

 \be \begin{cases}
&p_t=(m_{t}, 0)\\[2mm]
&p_\ell= E_\ell(m^2_{23}) (1, \hat{p_\ell}),\qquad \hat{p_\ell}= \mathcal{R}(\theta^{(a)}_\ell,\phi^{(a)}_\ell)\hat{a} \\[2mm]
    &p_b= E_b(q^2) (1, \hat{p_b}),\qquad  \hat{p_b}= \mathcal{R}(\theta^{(a)}_\ell,\phi^{(a)}_\ell)\mathcal{R}(\theta^{(a)}_{\ell b},\phi^{(a)}_{\ell b})\hat{a}.\\[2mm]
 \end{cases}
 \label{eq:4-momenta-in-t-rest}
\ee
The rotation operator $\mathcal{R}$ is defined as
\begin{equation}
    \mathcal{R}(\theta^{(a)}_u, \phi^{(a)}_u)= \mathcal{R}_a(\phi^{(a)}_u)\mathcal{R}_{i_2}(\theta^{(a)}_u)  \textnormal{    with    } \hat{i}_1\wedge \hat{i}_2= \hat{a}.
\end{equation}
where $u=\ell,\ell b$, and $\mathcal{R}_v(\omega)$, with $v=x,y,z$, denotes the standard rotation matrix through an angle $w$ about the axis $\hat{v}$.

The antitop-quark phase space is parametrised analogously in the antitop rest
frame. We denote by \(\hat b\) the polar axis used to define the direction of
the charged antilepton.\footnote{The unit vector \(\hat b\) should not be
confused with the direction of the bottom quark. It denotes one of the three
axes of the chosen Cartesian basis.} The Cartesian frame is taken to have the
same orientation as that used for the top-quark decay. The corresponding
four-momenta are
 \be \begin{cases}
&\bp_t=(m_{t},0)\\[2mm]
&\bp_{\ell}= \bar{E}_{\ell}(\bar{m}^2_{23}) (1, \hat{\bp}_{\ell}),\qquad \hat{\bp}_{\ell}= \mathcal{R}(\btheta^{(b)}_{\ell},\bphi^{(b)}_{\ell})\hat{b} \\[2mm]
    &\bp_{b}= \bar{E}_{b}(\bar{q}^2) (1, \hat{\bp}_{b}),\qquad  \hat{\bp}_{b}= \mathcal{R}(\btheta^{(b)}_{\ell},\bphi^{(b)}_{\ell})\mathcal{R}(\btheta^{(b)}_{\ell b},\bphi^{(b)}_{\ell b})\hat{b}.\\[2mm]
 \end{cases}
\ee


\section{Analytical expressions for $ \alpha_\ell, \ v_c $ and $w_s^{\rm CPV}$} \label{app:analytic-expressions}

\subsection{Analytical expression of $\alpha_\ell$} \label{app:alpha_analytical}
In the following, we present the expression for the spin-analysing power of the charged-lepton produced in the \( t\) decay, assuming \(V_{tb}=1\). We write it in the form
\begin{equation}
\alpha_\ell
=
\frac{\mathrm{N}(\alpha_\ell)}
{\mathrm{D}(\alpha_\ell)},
\label{eq:sap_ND}
\end{equation}
where \(\mathrm{N}(\alpha_\ell)\) and \(\mathrm{D}(\alpha_\ell)\) denote the corresponding numerator and denominator, respectively. For the charged-lepton produced in the \(\bar t\) decay, the spin-analysing power is given by $\bar{\alpha}_\ell=-\alpha_\ell$.
The numerator of the charged-lepton spin-analysing power is given by
\begin{flalign}
    \mathrm{N}(\alpha_\ell)&=  \left[ 4 g^2 \left(m_t^6-3 m_t^2 m_W^4+2 m_W^6\right) \right] \nonumber\\
    &+\frac{1}{\mathbf{\Lambda^2}}\Bigg[C^3_{\phi Q}   \left(
   8 g^2 v^2 (m_t^6-3 m_t^2 m_W^4+2 m_W^6) \right)\nonumber\\
   &+ \mathrm{Re}(C_{\phi tb})   \left( 24 g^2 m_b m_t m_W^2 v^2 (m_W^2-m_t^2) \right) \nonumber \\
   &  
     +    \mathrm{Re}(C_{bW})   \left( 48 \sqrt{2} g m_b m_W^2 v (m_W^4-m_t^4) \right)  \nonumber\\
     &+  \mathrm{Re}(C_{tW})   \left( 48 \sqrt{2} g m_t m_W^2 v
   \left(m_t^2-m_W^2\right)^2 \right)\Bigg]\nonumber\\ 
&  + \frac{1}{\mathbf{\Lambda^4}} \Bigg[(C^3_{\phi Q})^2\left( 4 g^2 v^4 \left(m_t^6-3 m_t^2 m_W^4+2 m_W^6\right) \right)\nonumber\\
&+  \mathrm{Re}( C^3_{\phi Q}C_{\phi tb})\left( 24 g^2 m_b m_t m_W^2 v^4 \left(m_W^2-m_t^2\right)\right) \nonumber\\   
   &  + \mathrm{Re}( C^3_{\phi Q} C_{bW})\left( 48 \sqrt{2} g m_b m_W^2 v^3 \left(m_W^4-m_t^4\right) \right) \nonumber\\
   &+  \mathrm{Re}(C^3_{\phi Q} C_{tW})\left( 48 \sqrt{2} g m_t m_W^2 v^3
   \left(m_t^2-m_W^2\right)^2\right)\nonumber \\    
   &  -|C_{\phi tb}|^2\left( g^2 v^4 \left(m_t^6-12 m_t^4 m_W^2+9 m_t^2 m_W^4+24 m_t^2 m_W^4 \log (m_t/m_W)+2 m_W^6\right) \right) \nonumber\\    
   &  + \mathrm{Re}(C_{\phi tb} C_{bW}^*)\left( 24 \sqrt{2} g m_t m_W^2 v^3 \left(m_t^2-m_W^2\right)^2 \right)\nonumber\\
   &+ \mathrm{Re}(C_{\phi tb} C_{tW})\left( 24 \sqrt{2} g m_b m_W^2
   v^3 \left(m_t^4-8 m_t^2 m_W^2 \log (m_t/m_W)-m_W^4\right) \right) \nonumber\\    
   &  +|C_{bW}|^2\left(32 m_W^2 v^2 \left(2 m_t^6-3 m_t^4
   m_W^2+m_W^6\right) \right)\nonumber\\
   &- \mathrm{Re}(C_{bW} C_{tW})\left( 384 m_b m_t m_W^4 v^2 \left(m_t^2-m_W^2\right) \right) \nonumber\\    
   &  -|C_{tW}|^2\left( 32
   m_W^2 v^2 \left(2 m_t^6+9 m_t^4 m_W^2-24 m_t^4 m_W^2 \log (m_t/m_W)-12 m_t^2 m_W^4+m_W^6\right) \right) \Bigg] && 
\end{flalign}
The corresponding denominator is given by
\begin{flalign}
    \mathrm{D}(\alpha_\ell)&=  \left[ 4 g^2 \left(m_t^6-3 m_t^2 m_W^4+2 m_W^6\right) \right] \nonumber\\
   &+ \frac{1}{\mathbf{\Lambda^2}} \Bigg[C^3_{\phi Q}   \left( 8
   g^2 v^2 \left(m_t^6-3 m_t^2 m_W^4+2 m_W^6\right) \right) \nonumber\\
   &+   \mathrm{Re}(C_{\phi tb})   \left( 24 g^2 m_b m_t m_W^2 v^2 \left(m_W^2-m_t^2\right)\right) \nonumber\\   
   &+    \mathrm{Re}(C_{bW})   \left( 48 \sqrt{2} g m_b m_W^2 v \left(m_W^4-m_t^4\right) \right) \nonumber\\
   &+    \mathrm{Re}(C_{tW})   \left(48 \sqrt{2} g m_t m_W^2 v
   \left(m_t^2-m_W^2\right)^2 \right)\Bigg] \nonumber\\   
   &+ \frac{1}{\mathbf{\Lambda^4}} \Bigg[(C^3_{\phi Q})^2\left(4 g^2 v^4 \left(m_t^6-3 m_t^2 m_W^4+2 m_W^6\right) \right) \nonumber\\
   &+  \mathrm{Re}( C^3_{\phi Q}C_{\phi tb})\left( 24 g^2 m_b m_t m_W^2 v^4 \left(m_W^2-m_t^2\right) \right) \nonumber\\    
   &+   \mathrm{Re}(C^3_{\phi Q}  C_{bW} )\left( 48 \sqrt{2} g m_b m_W^2 v^3 \left(m_W^4-m_t^4\right) \right) \nonumber\\
   &+  \mathrm{Re}(C^3_{\phi Q} C_{tW})\left( 48 \sqrt{2} g m_t m_W^2 v^3
   \left(m_t^2-m_W^2\right)^2\right)\nonumber \\   
   &+|C_{\phi tb}|^2\left( g^2 v^4 \left(m_t^6-3 m_t^2 m_W^4+2 m_W^6\right) \right) \nonumber\\
   &+\mathrm{Re}(C_{\phi tb} C_{bW}^*)\left( 24
   \sqrt{2} g m_t m_W^2 v^3 \left(m_t^2-m_W^2\right)^2\right) \nonumber\\  
   &+ \mathrm{Re}(C_{\phi tb} C_{tW})\left( 24 \sqrt{2} g m_b m_W^2 v^3 \left(m_W^4-m_t^4\right) \right) \nonumber\\
   &+(|C_{bW}|^2+|C_{tW}|^2)\left( 32 m_W^2 v^2 \left(2
   m_t^6-3 m_t^4 m_W^2+m_W^6\right)\right)\nonumber\\   
   &- \mathrm{Re}(C_{bW} C_{tW})\left( 384 m_b m_t m_W^4 v^2 \left(m_t^2-m_W^2\right) \right)\Bigg]  &&
\end{flalign}
Our result agrees with Ref.~\cite{Aguilar-Saavedra:2010ljg}, apart from a single term, which may be typographical errors in that reference.

Although both \(\mathrm{N}(\alpha_\ell)\) and \(\mathrm{D}(\alpha_\ell)\) contain terms linear in the Wilson coefficients, these contributions cancel in the ratio when Eq.~\eqref{eq:sap_ND} is consistently expanded in inverse powers of the EFT scale. Consequently, the first deviation from the SM prediction, \(\alpha_\ell^{\mathrm{SM}}=1\), arises at quadratic order, namely at \(\mathcal{O}(\Lambda^{-4})\). Expanding the ratio, we obtain
\begin{flalign}
   \notag \alpha_\ell&=1\,- \frac{1}{\mathbf{\Lambda^4}}\Bigg[ \frac{\mathrm{Re}(C_{\phi tb}C_{tW})}{gf_0}\left(12 \sqrt{2} m_b m_W^2 v^3 \left(-m_t^4+4 m_t^2 m_W^2 \log \left(m_t/m_W\right)+m_W^4\right)\right)\nonumber\\
   &-\frac{|C_{\phi tb}|^2}{2f_0} \left(v^4 \left(m_t^6-6 m_t^4 m_W^2+3 m_t^2 m_W^4+12 m_t^2 m_W^4 \log \left(m_t/m_W\right)+2 m_W^6\right)\right)\nonumber  \\
    & -\frac{|C_{tW}|^2}{g^2f_0}\left(16m_W^2 v^2 \left(2 m_t^6+3 m_t^4 m_W^2+12 m_t^4 m_W^2 \log \left(m_W/m_t\right)-6 m_t^2 m_W^4+m_W^6\right)\right)\Bigg] \nonumber\\
   & +\frac{1}{\mathbf{\Lambda^6}}\left(\frac{v^3}{ g^3 \left(m_t^2-m_W^2\right)^3 \left(m_t^2+2 m_W^2\right)^2} \right)\nonumber\\
   &*\Bigg[
   \left(-\mathrm{Re}(C_{bW})
   \left(6 \sqrt{2} m_b m_W^2 
   \left(m_t^2+m_W^2\right)
   \right)
   -\mathrm{Re}(C_{\phi tb})
   \left(3 g m_b m_t m_W^2 v
   \right)
   \right)\nonumber\\
   &+(m_t^2-m_W^2)  \left(C^3_{\phi Q}\left( g v \left(m_t^2+2 m_W^2\right)\right)+\mathrm{Re}(C_{tW})\left(6 \sqrt{2} m_t m_W^2 \right)\right)\Bigg) \nonumber\\
& *\notag\Bigg(-\mathrm{Re}\left(C_{\phi tb} C_{tW}\right)\left(24 \sqrt{2} g m_b m_W^2 v  \left(m_t^4-4 m_t^2 m_W^2 \log \left(m_t/m_W\right)-m_W^4\right)\right)\\
&\notag +|C_{\phi tb}|^2 \left(g^2 v^2
    \left(m_t^6-6 m_t^4 m_W^2+3 m_t^2 m_W^4+12 m_t^2 m_W^4 \log \left(m_t/m_W\right)+2 m_W^6\right)\right)\\
& + |C_{tW}|^2 \left(32m_W^2 \left(2 m_t^6+3
   m_t^4 m_W^2+12 m_t^4 m_W^2 \log \left(m_W/m_t\right)-6 m_t^2 m_W^4+m_W^6\right)\right)\Bigg)\Bigg]\nonumber\\
   &+ O\left(\frac{1}{\mathbf{\Lambda^8}}\right)\,. &&\label{eq:sap_expanded}
\end{flalign}
where $f_0=\left(m_t^6-3 m_t^2 m_W^4+2 m_W^6\right)$
Retaining the \(\mathcal{O}(\Lambda^{-6})\) contribution is phenomenologically relevant. For instance, terms proportional to \(\lvert C_{tW}\rvert^2\) and to \(\lvert C_{tW}\rvert^2\mathrm{Re}(C_{tW})\), which enter at \(\mathcal{O}(\Lambda^{-4})\) and \(\mathcal{O}(\Lambda^{-6})\), respectively, can be numerically comparable over the range of Wilson coefficients considered here. This also explains why the curves \(\alpha_\ell(\mathrm{Re}(C_{tW}))\) and \(\alpha_\ell(\mathrm{Im}(C_{tW}))\) in Fig.~\ref{fig:sap} do not coincide.
\\
\subsection{Analytical expressions of $v_c, w_s^{\rm CPV}$}
\label{App:vcwcpv}

We write $v_c$ in the form
\begin{align}
    v_c=\frac{\mathrm{N}(v_c)}{\mathrm{D}(v_c)} ,
\end{align}
where \(\mathrm{N}(v_c)\) and \(\mathrm{D}(v_c)\) denote the corresponding numerator and denominator, respectively. 
The numerator of $v_c$ is given by
\begin{flalign}
    \mathrm{N}(v_c)&= \frac{1}{\mathbf{\Lambda^2}} \Bigg[ \mathrm{Re}(C_{\phi tb})   \left( 6 \pi ^3 g^2 m_b m_W v^2 (m_t-m_W)
   (m_t+m_W)^2 \right)\nonumber \\
   &-   \mathrm{Re}(C_{tW})   \left( 12 \sqrt{2} \pi ^3 g m_W v (m_t-m_W)^2 (m_t+m_W)^3 \right) \Bigg]\nonumber\\
   & +\frac{1}{\mathbf{\Lambda^4}}\Bigg[ \mathrm{Re}( C^3_{\phi Q}C_{\phi tb})\left(
  6 \pi ^3 g^2 m_b m_W v^4 (m_t-m_W) (m_t+m_W)^2 \right)\nonumber \\
   &-\mathrm{Re}(C^3_{\phi Q} C_{tW})\left( 12\sqrt{2} \pi ^3 g m_W v^3 (m_t-m_W)^2
   (m_t+m_W)^3\right) \nonumber\\   
   &-|C_{\phi tb}|^2\left( 3 \pi ^3 g^2 m_t m_W v^4 (m_t-m_W)^2 (m_t+3 m_W) \right)\nonumber\\
   &- \mathrm{Re}(C_{\phi tb} C_{bW}^*)\left( 6
   \sqrt{2} \pi ^3 g m_W v^3 (m_t-m_W)^2 (m_t+m_W)^3\right) \nonumber\\    &+ \mathrm{Re}(C_{\phi tb} C_{tW})\left( 96 \sqrt{2} \pi ^3 g m_b m_t m_W^3 v^3 (m_t-m_W) \right) \nonumber\\
   &+ \mathrm{Re}(C_{bW} C_{tW})\left( 96 \pi ^3 m_b m_W^3 v^2 (m_t-m_W) (m_t+m_W)^2 \right) \nonumber\\    &-|C_{tW}|^2\left( 96 \pi ^3 m_t m_W^3 v^2
   (m_t-m_W)^2 (3 m_t+m_W) \right) \Bigg] \label{eq:vcnum}&&
\end{flalign}
The corresponding denominator is given by
\begin{flalign}
    \mathrm{D}(v_c)&=    \left[ 8 g^2 (m_t+m_W)^2 \left(m_t^3-m_t^2 m_W+2 m_t m_W^2-2 m_W^3\right) \right] \nonumber\\
    &+ \frac{1}{\mathbf{\Lambda^2}}\Bigg[ C^3_{\phi Q}   \left( 16 g^2 v^2 (m_t+m_W)^2 \left(m_t^3-m_t^2 m_W+2 m_t m_W^2-2 m_W^3\right) \right) \nonumber\\   
    &-   \mathrm{Re}(C_{\phi tb})   \left( 48 g^2 m_b m_t m_W^2 v^2 (m_t+m_W) \right) \nonumber\\
  &- \mathrm{Re}(C_{bW})   \left( 96 \sqrt{2} g m_b m_W^2 v
   \left(m_t^3+m_t^2 m_W+m_t m_W^2+m_W^3\right) \right) \nonumber\\   
   &+ \mathrm{Re}(C_{tW})   \left((96 \sqrt{2} g m_t m_W^2 v (m_t-m_W) (m_t+m_W)^2 \right) \Bigg]\nonumber\\
   &+ \frac{1}{\mathbf{\Lambda^4}}\Bigg[(C^3_{\phi Q})^2\left( 8 g^2 v^4 (m_t+m_W)^2 \left(m_t^3-m_t^2 m_W+2
   m_t m_W^2-2 m_W^3\right) \right) \nonumber\\ 
   &  -\mathrm{Re}( C^3_{\phi Q}C_{\phi tb})\left(48 g^2 m_b m_t m_W^2 v^4 (m_t+m_W) \right) \nonumber\\
   &- \mathrm{Re}(C^3_{\phi Q}C_{bW})\left( 96 \sqrt{2} g m_b m_W^2 v^3
   \left(m_t^3+m_t^2 m_W+m_t m_W^2+m_W^3\right)\right) \nonumber\\   
   &+  \mathrm{Re}(C^3_{\phi Q} C_{tW})\left( 96 \sqrt{2} g m_t m_W^2 v^3 (m_t-m_W)
   (m_t+m_W)^2 \right)\nonumber\\
   &+  |C_{\phi tb}|^2\left(2 g^2 v^4 (m_t+m_W)^2 \left(m_t^3-m_t^2 m_W+2 m_t m_W^2-2 m_W^3\right)\right) \nonumber\\    
   & +\mathrm{Re}(C_{\phi tb}C_{bW}^*)\left(48 \sqrt{2} g m_t m_W^2 v^3 (m_t-m_W) (m_t+m_W)^2
    \right)\nonumber\\
    &- \mathrm{Re}(C_{bW} C_{tW})\left( 768 m_b m_t m_W^4 v^2 (m_t+m_W) \right) \nonumber\\
    &- \mathrm{Re}(C_{\phi tb} C_{tW})\left( 48 \sqrt{2} g m_b m_W^2 v^3 \left(m_t^3+m_t^2 m_W+m_t m_W^2+m_W^3\right) \right) \nonumber\\   
   &-(|C_{bW}|^2+|C_{tW}|^2)\left( 64 m_W^2 v^2 (m_t+m_W)^2 \left(-2 m_t^3+2 m_t^2 m_W-m_t m_W^2+m_W^3\right)\right) \Bigg]\label{eq:vcdeno} &&
\end{flalign}
Similarly, we write $w_s^{\rm CPV}$ in the form
\begin{equation}
     w^{\rm CPV}_s=\frac{\mathrm{N}(w^{\rm CPV}_s)}{\mathrm{D}(w^{\rm CPV}_s)} ,
\end{equation}
where \(\mathrm{N}(w^{\rm CPV}_s)\) and \(\mathrm{D}(w^{\rm CPV}_s)\) denote the corresponding numerator and denominator, respectively. 
The numerator of $w^{\rm CPV}_s$ is given by
\begin{flalign}
    \mathrm{N}(w^{\rm CPV}_s)&= \frac{1}{\mathbf{\Lambda^2}}\Big[\mathrm{Im}\left( C_{\phi tb}^* \right) \left(6 \pi ^3 g^2 m_b m_W v^2 \left(m_W^2- m_t^2\right)\right) \nonumber\\
    &\notag+\mathrm{Im}\left(C_{tW}\right) \left(12\sqrt{2} \pi ^3 g m_W v \left(m_t^2- m_W^2\right)^2
  \right)\Big]\\
   &+\frac{1}{\mathbf{\Lambda^4}}\Big[\mathrm{Im}\left(  C^3_{\phi Q}C_{\phi tb}^*\right) \left(6 \pi ^3 g^2 m_b m_W v^4 \left(m_W^2- m_t^2\right) \right)\nonumber\\
   &\notag+\mathrm{Im} \left(  C^3_{\phi Q} C_{tW}\right)\left( 12
   \sqrt{2} \pi ^3 g m_W v^3 \left(m_t^2- m_W^2\right)^2\right) \\
   &+\mathrm{Im} \left(C_{bW}C_{\phi tb}^*\right)\left(6 \sqrt{2} \pi ^3 g m_W v^3 \left(m_t^2- m_W^2\right)^2
    \right)\nonumber\\
    &+\mathrm{Im}\left(C_{bW} C_{tW}\right) \left(96 \pi ^3 m_b m_W^3 v^2 \left(m_W^2- m_t^2\right) \right)\Big]&&
\end{flalign}
The corresponding denominator is given by
\begin{flalign}
    \mathrm{D}(w^{\rm CPV}_s)&=    \left[ -8 g^2 \left(m_t^4+m_t^2 m_W^2-2 m_W^4\right) \right] \nonumber\\
    &+ \frac{1}{\mathbf{\Lambda^2}}\Big[-C^3_{\phi Q}   \left(16
   g^2 v^2 \left(m_t^4+m_t^2 m_W^2-2 m_W^4\right) \right) +  \mathrm{Re}(C_{\phi tb})   \left( 48 g^2 m_b m_t m_W^2 v^2 \right) \nonumber\\  
   &+   \mathrm{Re}(C_{bW})
     \left(96 \sqrt{2} g m_b m_W^2 v \left(m_t^2+m_W^2\right) \right) +   \mathrm{Re}(C_{tW})   \left( 96 \sqrt{2} g m_t m_W^2 v
   \left(m_W^2-m_t^2\right)\right)\Big]\nonumber\\
   &+\frac{1}{\mathbf{\Lambda^4}}\Big[-(C^3_{\phi Q})^2\left( 8 g^2 v^4 \left(m_t^4+m_t^2 m_W^2-2 m_W^4\right)\right) +  \mathrm{Re}( C^3_{\phi Q}C_{\phi tb})\left( 48 g^2 m_b m_t m_W^2 v^4 \right) \nonumber\\
   &+ \mathrm{Re}(C^3_{\phi Q}C_{bW})\left( 96 \sqrt{2} g
   m_b m_W^2 v^3 \left(m_t^2+m_W^2\right) \right) \nonumber\\
   &+  \mathrm{Re}(C^3_{\phi Q} C_{tW})\left( 96 \sqrt{2} g m_t m_W^2 v^3 \left(m_W^2-m_t^2\right) \right) -|C_{\phi tb}|^2\left( 2 g^2 v^4 \left(m_t^4+m_t^2 m_W^2-2 m_W^4\right) \right) \nonumber\\
   &+\mathrm{Re}(C_{\phi tb} C_{bW}^*)\left( 48 \sqrt{2} g m_t m_W^2 v^3
   \left(m_W^2-m_t^2\right) \right)\nonumber\\
   &+ \mathrm{Re}(C_{\phi tb} C_{tW})\left( 48 \sqrt{2} g m_b m_W^2 v^3 \left(m_t^2+m_W^2\right) \right) + \mathrm{Re}(C_{bW} C_{tW})\left(768 m_b m_t m_W^4 v^2 \right)\nonumber\\
   &+(|C_{bW}|^2+|C_{tW}|^2)\left(64 m_W^2 v^2 \left(-2
   m_t^4+m_t^2 m_W^2+m_W^4\right) \right)   \Big] &&
\end{flalign}
From the structure of $w_s^{\mathrm{CPV}}$, and in particular from its numerator, one can directly identify its dependence on the imaginary parts of the Wilson coefficients and on their relative phases. These contributions enter, respectively, at linear and quadratic order in the $1/\Lambda^2$ expansion. Consequently, $w_s^{\mathrm{CPV}}$ provides a sensitive probe of CP-odd effects. The term linear in $\operatorname{Im}(C_{tW})$ is consistent with the result of Ref.~\cite{Zhang:2010dr}.


\section{Numerical inputs for the angular distributions}\label{app:fano_num}
This appendix collects the numerical inputs used in the angular
distributions discussed in Sec.~\ref{sec:discussion} and the
subsequent subsections. We first report the benchmark Wilson coefficient
scenarios and the corresponding decay parameters, and then provide the
polarisation and spin-correlation coefficients used for
\(pp\to t\bar t\) and \(e^+ e^-\to t\bar t\) production in the selected kinematic regions.
\subsection{Benchmark Wilson coefficient scenarios and decay parameters}
\label{app:benchmark_WCs}
The benchmark points for the Wilson coefficient assignments and the corresponding decay parameters entering the
angular distributions are collected in
Tab.~\ref{tab:benchmark_decay_parameters}. Guided by the current
experimental constraints reported in
Tab.~\ref{tab:EFT-op-limits}, we scan symmetric intervals for the real
and, where applicable, imaginary parts of each Wilson coefficient.
Throughout the analysis, the new physics scale is fixed to
\(\Lambda=1~\mathrm{TeV}\). 
\begin{table}[t]
    \centering
    \renewcommand{\arraystretch}{1.3}
    \begin{tabular}{c|c|c|c|c}
        \hline
        Scenario
        & Wilson coefficients
        & \(w_s^{\rm CPV}\)
        & \(v_c\)
        & \(\alpha_\ell\)
        \\
        \hline\hline
        SM
        &
        \(\begin{gathered}
            C_{\phi Q}^{3}=C_{\phi tb}=C_{bW}=C_{tW}=0
        \end{gathered}\)
        & \(0\)
        & \(0\)
        & \(1\)
        \\
        \hline
        CPV
        &
        \(\begin{gathered}
            C_{\phi Q}^{3}=-0.7,\qquad
            C_{\phi tb}=-2.6\,i,\\
            C_{bW}=-0.7,\qquad
            C_{tW}=-0.2\,i
        \end{gathered}\)
        & \(0.318398\)
        & \(-0.0650312\)
        & \(0.995218\)
        \\
        \hline
        CPC
        &
        \(\begin{gathered}
            C_{\phi Q}^{3}=-0.7,\qquad
            C_{\phi tb}=2.0,\\
            C_{bW}=-0.7,\qquad
            C_{tW}=-0.2
        \end{gathered}\)
        & \(0\)
        & \(0.269343\)
        & \(0.996783\)
        \\
        \hline
    \end{tabular}
    \caption{
    Benchmark Wilson coefficient scenarios and the corresponding decay
    parameters entering the angular distributions. All Wilson coefficients
    are dimensionless, and the new physics scale is fixed to
    \(\Lambda=1~\mathrm{TeV}\).
    }
    \label{tab:benchmark_decay_parameters}
\end{table}
\subsection{Polarisation and spin-correlation coefficients for
\texorpdfstring{\(pp\to t\bar t\)}{pp to ttbar} production}
\label{app:pp_spin_coefficients}

The single-particle polarisations and spin-correlation coefficients used
in the numerical evaluation of the distributions are extracted from the
leading-order Monte Carlo simulations described in
Ref.~\cite{Lamba:2026jfm}, using the kinematic regions defined in
Eqs.~\eqref{eq:bin1} and~\eqref{eq:bin2}. At this order, the individual
top and antitop polarisations vanish, while non-zero spin-correlations
remain.
Table~\ref{tab:values_distrib_bins_1} contains the coefficients entering the
distributions of Sec.~\ref{sect:dist1_discussion} and~\ref{sect:dist3_discussion}. The combinations entering the distributions in Sec.~\ref{sect:dist2_discussion}
 are reported in
Tab.~\ref{tab:values_distrib_bins_2}.

\begin{table}[H]
    \centering
    \renewcommand{\arraystretch}{1.25}
    \begin{tabular}{c|c|c|c|c}
        \hline
        Kinematic region
        & \((a,b)\)
        & \(B_a\)
        & \(\bar{B}_b\)
        & \(C_{ab}\)
        \\
        \hline\hline
        bin~1 & \((n,n)\) & \(0\) & \(0\) & \(-0.52264652\) \\
        bin~2 & \((k,k)\) & \(0\) & \(0\) & \( \phantom{-}0.51472757\) \\
        \hline
    \end{tabular}
    \caption{
    Polarisation and spin-correlation coefficients entering
    \(\mathcal{W}_{1}^{ab}\) and \(\mathcal{W}_{3}^{ab}\) for the selected
    \(pp\to t\bar t\) kinematic regions and spin-axis configurations.
    }
    \label{tab:values_distrib_bins_1}
\end{table}
\begin{table}[H]
    \centering
    \renewcommand{\arraystretch}{1.25}
    \begin{tabular}{c|c|c|c|c|c|c}
        \hline
        Kinematic region
        & \((a,b)\)
        & \(B_a\)
        & \(\bar{B}_{j_1}\)
        & \(\bar{B}_{j_2}\)
        & \(C_{a j_1}\)
        & \(C_{a j_2}\)
        \\
        \hline\hline
        bin~2
        & \((k,n)\)
        & \(0\)
        & \(0\)
        & \(0\)
        & \(0.10237656\)
        & \(0.51472757\)
        \\
        bin~2
        & \((k,k)\)
        & \(0\)
        & \(0\)
        & \(0\)
        & \(0\)
        & \(0.102377\)
        \\
        \hline
    \end{tabular}
    \caption{
    Polarisation and spin-correlation coefficients entering
    \(\mathcal{W}_{2}^{ab}\) in bin~2 for the two spin-axis configurations
    considered in Fig.~\ref{fig:dist2_pp}.
    }
    \label{tab:values_distrib_bins_2}
\end{table}
\subsection{Polarisation and spin-correlation coefficients for
\texorpdfstring{\(e^+e^-\to t\bar t\)}{e+e- to ttbar} production}
\label{app:fano_num-ee}

The polarisation and spin-correlation coefficients used for the
\(e^+e^-\to t\bar t\) distributions at
\(\sqrt{s}=365~\mathrm{GeV}\) are collected in
Tab.~\ref{tab:spin-pol-corr-ee-fig}. The values are inclusive in the
production angle \(\theta\). Their analytical expressions, together with
further details of the calculation, are presented in the companion
paper~\cite{Lamba:2026jfm}.

Among the Wilson coefficients entering the benchmark scenarios, only
\(C_{\phi Q}^{3}\) and \(C_{tW}\) modify both the production process and
the top-quark decay. Consequently, the SM, CPC, and CPV scenarios defined
in Eqs.~\eqref{eq:wc_set_CPC} and~\eqref{eq:wc_set_CPV} lead to different
values of the production polarisation and spin-correlation coefficients.
The results are evaluated using the same choice
\(\Lambda=1~\mathrm{TeV}\) adopted throughout the numerical analysis.

\begin{table}[H]
    \centering
    \renewcommand{\arraystretch}{1.25}
    \begin{tabular}{c|ccccc}
        \hline
        Scenario
        & \(B_r\)
        & \(\bar{B}_r\)
        & \(\bar{B}_k\)
        & \(C_{rr}\)
        & \(C_{rk}\)
        \\
        \hline\hline
        SM
        & \(0.329124\)
        & \(0.329124\)
        & \(-0.0784796\)
        & \(0.630828\)
        & \(-0.131923\)
        \\
        CPV
        & \(0.298852\)
        & \(0.298861\)
        & \(-0.0709490\)
        & \(0.631992\)
        & \(-0.127747\)
        \\
        CPC
        & \(0.216378\)
        & \(0.216378\)
        & \(-0.0636772\)
        & \(0.627138\)
        & \(-0.137060\)
        \\
        \hline
    \end{tabular}
    \caption{
    Polarisation and spin-correlation coefficients used in the
    \(e^+e^-\to t\bar t\) angular distributions at
    \(\sqrt{s}=365~\mathrm{GeV}\), inclusive in the production angle
    \(\theta\). The listed coefficients enter
    \(\mathcal{W}_{1}^{rr}\), \(\mathcal{W}_{2}^{rn}\), \ 
    \(\mathcal{W}_{3}^{rr}\), \ \(\mathcal{W}_{7}^{r}\) and \(\mathcal{W}_{\bar 7}^{r}\).
    }
    \label{tab:spin-pol-corr-ee-fig}
\end{table}

\bibliographystyle{JHEP}
\bibliography{main}

\end{document}